\documentclass[twocolumn]{aastex63}
\usepackage{booktabs}
\usepackage{verbatim}
\usepackage{CJKutf8}

\newcommand{\oi}{[O\,{\scriptsize I}]}

\shorttitle{CO disk size} 
\shortauthors{Long et al.}

\begin{document}
\begin{CJK*}{UTF8}{gbsn}

\title{Gas Disk Sizes from CO Line Observations: A Test of Angular Momentum Evolution}

\correspondingauthor{Feng Long}
\email{feng.long@cfa.harvard.edu}

\author[0000-0002-7607-719X]{Feng Long(龙凤)}
\affiliation{Center for Astrophysics \textbar\, Harvard \& Smithsonian, 60 Garden St., Cambridge, MA 02138, USA}

\author[0000-0003-2253-2270]{Sean M. Andrews}
\affiliation{Center for Astrophysics \textbar\, Harvard \& Smithsonian, 60 Garden St., Cambridge, MA 02138, USA}

\author[0000-0003-4853-5736]{Giovanni Rosotti}
\affiliation{School of Physics and Astronomy, University of Leicester, Leicester LE1 7RH, UK}
\affiliation{Leiden Observatory, Leiden University, P.O. Box 9513, 2300 RA Leiden, the Netherlands}

\author[0000-0001-6307-4195]{Daniel Harsono}
\affiliation{Institute of Astronomy, Department of Physics, National Tsing Hua University, Hsinchu, Taiwan}
\affiliation{Institute of Astronomy and Astrophysics, Academia Sinica, No. 1, Sec. 4, Roosevelt Road, Taipei 10617, Taiwan}
   
\author[0000-0001-8764-1780]{Paola Pinilla}
\affiliation{Max-Planck-Institut f\"{u}r Astronomie, K\"{o}nigstuhl 17, 69117, Heidelberg, Germany}
\affiliation{Mullard Space Science Laboratory, University College London, Holmbury St Mary, Dorking, Surrey RH5 6NT, UK}

\author[0000-0003-1526-7587]{David J. Wilner}
\affiliation{Center for Astrophysics \textbar\, Harvard \& Smithsonian, 60 Garden St., Cambridge, MA 02138, USA}

\author[0000-0001-8798-1347]{Karin I \"Oberg}
\affiliation{Center for Astrophysics \textbar\, Harvard \& Smithsonian, 60 Garden St., Cambridge, MA 02138, USA}

\author[0000-0003-1534-5186]{Richard Teague}
\affiliation{Center for Astrophysics \textbar\, Harvard \& Smithsonian, 60 Garden St., Cambridge, MA 02138, USA}

\author{Leon Trapman}
\affiliation{Department of Astronomy, University of Wisconsin-Madison, 475 N Charter St, Madison, WI 53706}

\author{Beno\^{i}t Tabone}
\affiliation{Universit\'{e} Paris-Saclay, CNRS, Institut d’Astrophysique Spatiale, 91405 Orsay, France}


\begin{abstract}
The size of a disk encodes important information about its evolution. Combining new Submillimeter Array (SMA) observations with archival Atacama Large Millimeter Array (ALMA) data, we analyze mm continuum and CO emission line sizes for a sample of 44 protoplanetary disks around stars with masses of 0.15--2\,$M_{\odot}$ in several nearby star-forming regions. 
Sizes measured from $^{12}$CO line emission span from 50 to 1000\,au. This range could be explained by viscous evolution models with different $\alpha$ values (mostly of $10^{-4}-10^{-3}$) and/or a spread of initial conditions. 
The CO sizes for most disks are also consistent with MHD wind models that directly remove disk angular momentum, but very large initial disk sizes would be required to account for the very extended CO disks in the sample. As no CO size evolution is observed across stellar ages of 0.5--20\,Myr in this sample, determining the dominant mechanism of disk evolution will require a more complete sample for both younger and more evolved systems. 
We find that the CO emission is universally more extended than the continuum emission by an average factor of $2.9\pm1.2$. 
The ratio of the CO to continuum sizes does not show any trend with stellar mass, mm continuum luminosity, or the properties of substructures. 
The GO Tau disk has the most extended CO emission in this sample, with an extreme CO to continuum size ratio of 7.6. 
Seven additional disks in the sample show high size ratios ($\gtrsim4$) that we interpret as clear signs of substantial radial drift.

\end{abstract}


\section{Introduction} \label{sec:intro}
Circumstellar disks are the cradles of young planets. Their physical structures and chemical conditions largely determine the properties of the resulting planetary systems. Detailed characterization of the gas and dust distributions in disks is therefore fundamental for developing theoretical models of planet formation (e.g., \citealt{Benz2014prpl}, \citealt{MorbidelliRaymond2016}). These disk structures are closely linked to how disks evolve. However, the physical mechanisms that drive one key component of disk evolution -- the angular momentum transport -- are still debated (e.g., \citealt{Turner2014}).

In the classical viscous evolution model, mass accretion onto the central star requires the outward transport of angular momentum \citep{Lynden-Bell1974, Hartmann1998}. Consequently, the gas density distribution becomes more radially extended over time. 
In the alternative magneto-hydrodynamical (MHD) disk wind model, angular momentum is transported in the wind, moving vertically away from the disk \citep{Blandford1982, Ferreira1997, Gressel2015, Bai2016}. That behavior can result in a more radially compact gas density distribution (e.g., \citealt{ Lesur2021, YangBai2021arXiv}). Comparing disk sizes in a large sample that spans a range of evolutionary stages could help differentiate between the two scenarios.

Measurements of disk sizes are available from observations with (sub-)millimeter interferometers, which can spatially resolve disks in nearby ($\lesssim$200\,pc) star-forming regions. The (sub-)mm continuum emission from dust has been resolved in $\sim$200 disks (e.g., \citealt{Tripathi2017, Andrews2018_Lmm, Hendler2020}). However, these continuum observations primarily trace the mm-sized particles that have aerodynamically decoupled from the gas and migrated inwards toward any pressure maxima \citep{Weidenschilling1977}. The extent of the continuum emission is therefore a diagnostic of the evolution of disk solids (e.g., \citealt{Testi2014, Birnstiel2014}). Observations of the molecular gas reservoir in disks are the key probes of the disk evolution tied to angular momentum transport. The most common tracer of the gas disk is the bright CO line emission. However, as the gas emission usually emerges in a narrow velocity range, detecting these lines is more challenging, and systematic CO disk surveys are still limited (e.g., \citealt{Ansdell2016, Long2017, Boyden2020}).

The size of the bright $^{12}$CO line emission is found to be 
larger than that of the mm continuum emission in disks \citep[e.g.,][]{Panic2009, Andrews2012, Ansdell2018}. 
This size discrepancy has been widely interpreted as the consequence of dust evolution, associated with the growth and inward migration of solids \citep{Birnstiel2014, Trapman2019}. But some of the discrepancy can be attributed to the different optical depths of the size tracers \citep{Hughes2008, Facchini2017}. How well the CO-to-continuum size ratio probes the relative spatial distribution of the gas and solid reservoir masses, how it varies with stellar and disk properties, and how the inferred behaviors would relate to fundamental physical processes, are still open questions.

In this work, we combine new SMA CO line observations with archival ALMA data to build a collective view of CO and mm continuum disk sizes. 
With a sample of 44 targets across a wide range of stellar and disk properties, we explore how the CO and mm continuum sizes can help improve our understanding of disk evolution processes. We introduce the sample and the data in Section~\ref{sec:obs}, and the method for size measurement in Section~\ref{sec:measure}. The sizes and their relationships with stellar/disk properties are summarized in Section~\ref{sec:results}. In Section~\ref{sec:diss}, we discuss the role of the CO emission sizes in differentiating between turbulent viscosity and MHD disk wind models and the CO-to-continuum size ratio in probing the dust disk evolution. Finally, we summarize the main findings in Section~\ref{sec:sum}.

\section{The Sample and Data} \label{sec:obs}

\subsection{Three Disks with New SMA Observations}

We observed DL Tau, GO Tau, and UZ Tau\footnote{The UZ Tau system contains four stellar components, including the spectroscopic binary UZ Tau Eab and the close binary UZ Tau Wab. The East and West components are separated by $3\farcs6$. In this work we are only interested in the UZ Tau E system and refer to it as UZ Tau for short throughout the paper.} with eight 6\,m antennas of the SMA \citep{Ho2004} in both extended and subcompact configurations. These targets were observed in two shared tracks in the extended configuration on 2019 November 10 and 11, and one shared track in the subcompact configuration on 2019 November 19. The dual-sideband receivers were both tuned to the same local oscillator (LO) frequency of 225.5\,GHz to maximize the sensitivity in the $^{12}$CO $J=2-1$ line. The SWARM correlator was configured with 16384 channels in each of the four spectral chunks per sideband, with a spectral resolution of 140\,kHz (or 0.18\,km\,s$^{-1}$ velocity resolution at 230.538\,GHz) and covering the spectral ranges of 213.4--221.6\,GHz and 229.4--237.6\,GHz. The science targets were observed in an alternating sequence with the gain calibrators 3C111 and 0510+180. The bandpass calibrator 3C454.3 and flux calibrators MWC349A and Titan were observed at the beginning of each track.

The raw visibility data were calibrated using the IDL-based \textsc{MIR} software package\footnote{\url{https://lweb.cfa.harvard.edu/~cqi/mircook.html}}, following the standard SMA data reduction procedures. After the initial flagging of data with abnormal phase and amplitude variations, the bandpass response was determined with the observation of the bright quasar 3C454.3. The amplitude scale was set based on the frequent monitoring of MWC349A and Titan, with a typical systematic uncertainty of $\sim10\%$.  The repeated observations of 3C111 and 0510+180 provided the complex gain response of the system, which were then applied to the science targets. The calibrated visibility data\footnote{Self-calibration was not applied for these data sets where peak signal-to-noise is only $\sim$5--10$\sigma$ in individual channels.} for individual spectral chunks were imported into \textsc{CASA} for further imaging. Using \textsc{CASA} version 5.6.0 \citep{McMullin2007}, the continuum baseline was first subtracted with the \texttt{uvcontsub} task in the spectral chunks covering the $^{12}$CO $J=2-1$ line. The continuum-subtracted visibility data from both receivers and antenna configurations of individual disks were then combined using the \texttt{concat} task. The $^{12}$CO image cubes with channel widths of 0.25\,km\,s$^{-1}$ were produced using the \texttt{tclean} task with natural weighting to maximize the sensitivity, resulting in a typical beam size of $1\farcs3\times0\farcs9$ with an rms noise level of $\sim$80\,mJy\,beam$^{-1}$ per channel ($\sim$5\,K). Keplerian masks were applied in the CLEAN process, which were generated based on stellar and disk parameters from previous observations for each disk \citep{Long2018} and designed to encompass the observed emission in individual channels.

\subsection{Archival Data} 
Besides the three disks with new SMA observations, we collated a sample of disks for which spatially resolved $^{12}$CO observations are available from the ALMA archive\footnote{V4046 Sgr is the one exception, for which $^{12}$CO data were taken with the SMA \citep{Rosenfeld2012}.} for a feasible measurement of the CO disk size. The selected sample includes most well-studied systems and spans a wide range in stellar and disk properties. 

The ALMA Large Project DSHARP (Disk Substructures at High Angular Resolution Project, \citealt{Andrews2018_dsharp}) observed twenty disks in the 1.3\,mm continuum emission and $^{12}$CO $J=2-1$ lines at $\sim0\farcs03$ resolution. We selected the twelve disks with considerably low cloud contamination in the $^{12}$CO emission. These disks are mostly located in the Lupus, Oph, and Upper Sco star-forming regions. The image products from the project website\footnote{ \url{https://bulk.cv.nrao.edu/almadata/lp/DSHARP/}} were used in this study.

The ALMA Lupus survey reported gas disk sizes for 22 disks based on $^{12}$CO $J=2-1$ observations at $\sim0\farcs25$ resolution \citep{Ansdell2018}. 
As five disks (GW Lup, IM Lup, MY Lup, Sz 129, and HT Lup\footnote{The triple system, HT Lup, with A-B components only separated by $0\farcs1$, is excluded.}) have higher quality data from DSHARP, we included the remaining 17 Lupus disks, and directly adopted the size measurements in \citet{Ansdell2018} for the full sample analysis in this work.

Six disks around low mass stars (0.1--0.2\,$M_\odot$) in the Taurus star-forming region were included based on the recent work of \citet{Kurtovic2021}, which targeted the 890\,$\mu$m continuum emission and $^{12}$CO $J=3-2$ line at $\sim0\farcs1$ resolution. DM Tau is historically known to host an extended CO gas disk \citep{Guilloteau1998, Pietu2007}. The ALMA data with $\sim0\farcs3$ resolution by \citet{Flaherty2020} were used for DM Tau in our analysis. 
The $^{12}$CO $J=2-1$ emission from the FP Tau disk was recently observed in an ALMA Chemistry program with $\sim0\farcs2$ resolution \citep{Pegues2021}. We also included the measurements for CX Tau, which was reported as the first example with very high gas-to-dust disk size ratios ($R_{\rm CO}/R_{\rm mm}\sim4$, \citealt{Facchini2019}).

Three additional sources from other star-forming regions with available $^{12}$CO data were also included: TW Hya \citep{Huang2018_CO},  V4046 Sgr \citep{Rosenfeld2012}, and J11004022-7619280 \citep{Pegues2021}. Data images for the three sources and the above Taurus disks were kindly shared by authors of these references.

\subsection{Host Star Properties}
The full sample includes 12 disks from Taurus, 21 disks from Lupus, 5 disks from Oph, and 6 disks from an assortment of five other regions. Two sources (UZ Tau and V4046 Sgr) are  spectroscopic binaries and host circumbinary disks. In addition, UZ Tau and Sz~123~A have stellar companions with moderate separations ($3\farcs6$ and $1\farcs7$, respectively, \citealt{Kraus2009, Ghez1997}). No additional stellar companions were reported for the remaining sample to our best knowledge \citep[e.g.,][]{Kraus2012, Zurlo2020}. Therefore, the effect of tidal truncation is minimal for this sample.

Target distances ($d$) were estimated based on Gaia DR2 parallax measurements \citep{Gaia2018}. 
The stellar properties for individual disks were adopted from previous studies (accounted for updated Gaia distances) and summarized in Table~\ref{tab:source_prop}. Though different evolutionary models were used to derive stellar parameters, no significant differences were observed. 
Specifically, for Taurus targets, effective temperature ($T_{\rm eff}$) and luminosity ($L_{*}$ scaled by $d^2$) taken from the optical spectral survey of \citet{HH2014} were used to calculate stellar mass and age using \citet{baraffe2015} and the non-magnetic \citet{feiden2016} models of pre-main sequence stellar evolution \citep{Long2019}. Stellar masses and ages for most Lupus disks and the DSHARP sample were estimated based on the MIST models using literature values of $T_{\rm eff}$ and $L_{*}$ \citep{Andrews2018_Lmm, Andrews2018_dsharp}. 
For the few remaining targets, stellar properties were adopted from the corresponding references summarized in Table~\ref{tab:source_prop}. The listed stellar masses for UZ Tau and V4046 Sgr are the total mass of the two stellar components, as derived from the gas disk rotation \citep{Rosenfeld2012, Czekala2019}. We also updated the stellar masses for a few systems when dynamical stellar mass measurements are available (see Table~\ref{tab:source_prop}). 

\begin{figure}[!th]
\centering
    \includegraphics[width=0.9\linewidth]{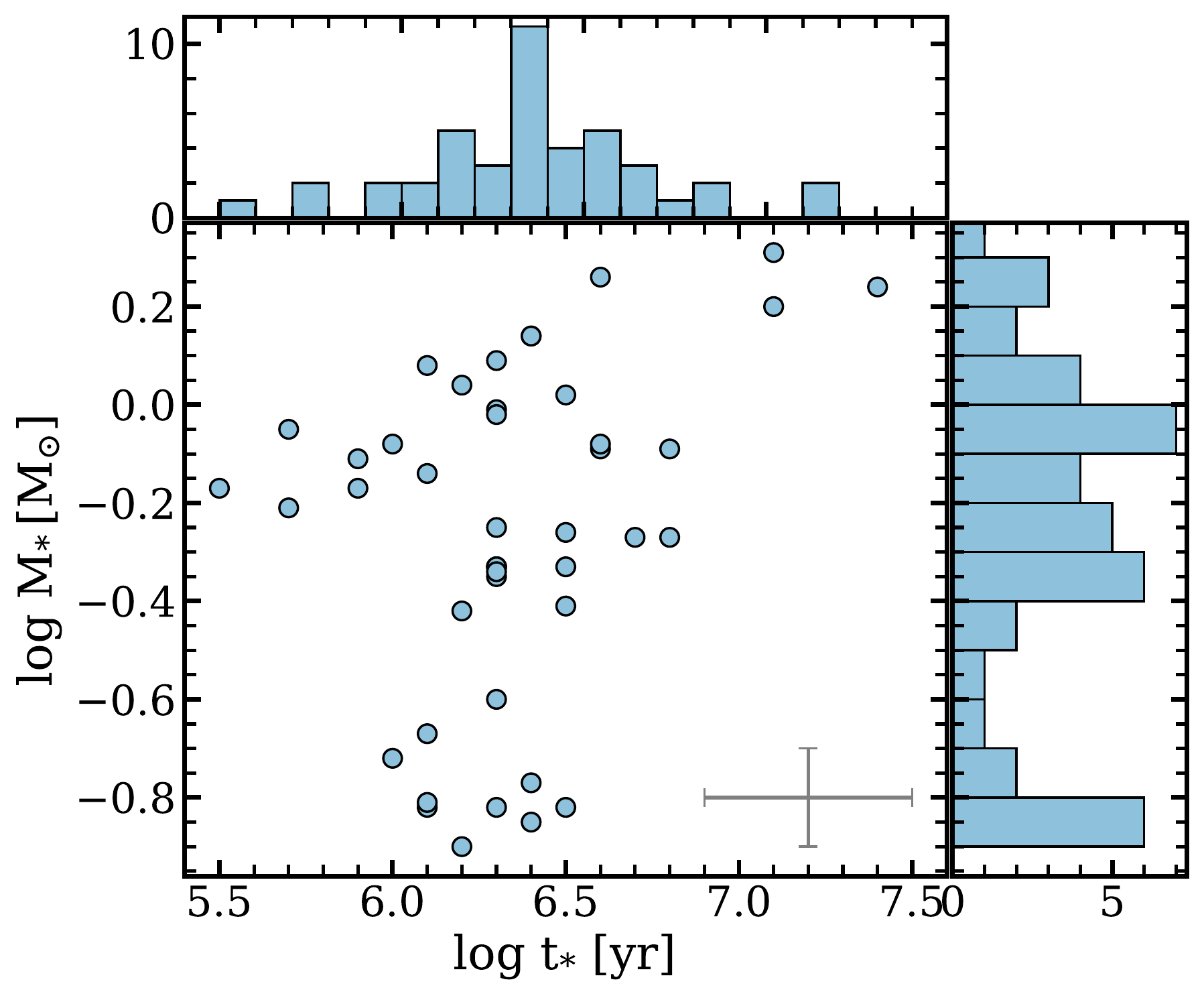} \\ 
\caption{The stellar age and mass distribution for the sample of 44 disks. Most systems have ages around 2--3\,Myr. Though the stellar mass distribution is more uniform, this sample lacks low mass host stars at younger and more evolved stages. A typical errorbar in log($M_{*}$) and log($t_*$) is shown in the lower right corner and corresponds to $\pm$0.1 and $\pm$0.3\,dex, respectively.
\label{fig:age-mass}}
\end{figure}

This sample covers a wide range of young star properties, with stellar masses of 0.15--2\,$M_{\odot}$ (spectral types from M5 to A1) and spanning more than two orders of magnitude in stellar luminosities. The stellar ages range from 0.5 to 20\,Myr, while the average age is about 2--3\,Myr as most of the disks are located in the Taurus and Lupus regions. As shown in Figure~\ref{fig:age-mass}, this sample is particularly incomplete for lower mass host stars towards younger and more evolved systems.

\begin{deluxetable*}{clllllllccl}
\tabletypesize{\scriptsize}
\tablecaption{Host Stellar Properties\label{tab:source_prop}}
\tablewidth{0pt}
\tablehead{
\colhead{Index} & \colhead{Name} & \colhead{2MASS} & \colhead{Region} & \colhead{d} & \colhead{SpTy} & \colhead{$L_*$} & \colhead{$T_{\rm eff}$} & \colhead{$log(M_*)$} & \colhead{$log(t_*)$} &  \colhead{ref} \\ 
\colhead{} & \colhead{} & \colhead{} & \colhead{} & \colhead{(pc)} & \colhead{} & \colhead{($L_\odot$)} & \colhead{(K)} & \colhead{($M_\odot$)} &\colhead{(Myr)} & \colhead{}  \\
} 
\colnumbers
\startdata
  1  &     CX Tau & J04144786$+$2648110 &   Taurus &  128 &  M2.5 &  0.25 & 3488 & -0.42$\pm$0.02 &    6.2$\pm$0.3  &  1,2*  \\
   2  &     DL Tau & J04333906$+$2520382 &   Taurus &  159 &  K5.5 &  0.65 & 4277 &  0.02$\pm$0.01 &    6.5$\pm$0.3  &  1,2*  \\
   3  &     DM Tau & J04334871$+$1810099 &   Taurus &  145 &    M3 &  0.14 & 3415 & -0.26$\pm$0.02 &    6.5$\pm$0.2  &  1,2*  \\
   4  &     GO Tau & J04430309$+$2520187 &   Taurus &  144 &  M2.3 &  0.21 & 3516 & -0.35$\pm$0.01 &    6.3$\pm$0.3  &  1,2*  \\
   5  &     UZ Tau & J04324303$+$2552311 &   Taurus &  131 &  M1.9 &  0.83 & 3574 &  0.08$\pm$0.01 &    6.1$\pm$0.3  &  1,3*  \\
   6  &     FP Tau & J04144730$+$2646264 &   Taurus &  128 &  M2.6 &  0.16 & 3273 & -0.41$\pm$0.02 &    6.5$\pm$0.3  &  1,4*  \\
   7  &     CIDA 1 & J04141760$+$2806096 &   Taurus &  136 &  M4.5 &   0.2 & 3200 & -0.72$\pm$0.06 &    6.0$\pm$0.2  &  1  \\
   8  &     CIDA 7 & J04422101$+$2520343 &   Taurus &  136 &  M5.1 &  0.08 & 3111 & -0.82$\pm$0.08 &    6.3$\pm$0.2  &  1  \\
   9  &      MHO 6 & J04322210$+$1827426 &   Taurus &  142 &    M5 &  0.06 & 3125 & -0.77$\pm$0.07 &    6.4$\pm$0.2  &  1  \\
  10  &     J0415 & J04155799$+$2746175 &   Taurus &  135 &  M5.2 &  0.05 & 3098 & -0.82$\pm$0.08 &    6.5$\pm$0.2  &  1  \\
  11  &     J0420 & J04202555$+$2700355 &   Taurus &  170 & M5.25 &  0.07 & 3091 & -0.85$\pm$0.08 &    6.4$\pm$0.2  &  1  \\
  12  &     J0433 & J04334465$+$2615005 &   Taurus &  173 &  M5.2 &  0.12 & 3098 & -0.82$\pm$0.05 &    6.1$\pm$0.2  &  1  \\
  13  &      Sz 65 & J15392776$-$3446171 &    Lupus &  155 &  K7   &  0.87 & 4060 & -0.14$\pm$0.11 &    6.1$\pm$0.4  &  5  \\
  14  &      Sz 73 & J15475693$-$3514346 &    Lupus &  156 &  K7   &  0.46 & 4060 & -0.09$\pm$0.09 &    6.6$\pm$0.4  &  5  \\
  15  &      Sz 75 & J15491210$-$3539051 &    Lupus &  151 &  K6   &  1.45 & 4205 & -0.11$\pm$0.11 &    5.9$\pm$0.4  &  5  \\
  16  &      Sz 76 & J15493074$-$3549514 &    Lupus &  159 &  M4   &  0.11 & 3270 & -0.60$\pm$0.05 &    6.3		    &  6  \\
  17  & J15560210 & J15560210$-$3655282 &    Lupus &  158 &  M1   &  0.22 & 3705 & -0.33$\pm$0.10 &    6.3		    &  6  \\
  18  &      Sz 84 & J15580252$-$3736026 &    Lupus &  152 &  M5   &  0.12 & 3125 & -0.81$\pm$0.08 &    6.1$\pm$0.4  &  5  \\
  19  &     RY Lup & J15592838$-$4021513 &    Lupus &  158 &  K2   &  1.91 & 4900 &  0.14$\pm$0.06 &    6.4$\pm$0.3  &  5  \\
  20  & J16000236 & J16000236$-$4222145 &    Lupus &  164 &  M4   &  0.17 & 3270 & -0.67$\pm$0.08 &    6.1$\pm$0.4  &  5  \\
  21  &     EX Lup & J16030548$-$4018254 &    Lupus &  157 &  M0   &  0.75 & 3850 & -0.25$\pm$0.09 &    6.3		    &  6  \\
  22  &     Sz 133 & J16032939$-$4140018 &    Lupus &  153 &  K2   &  1.05 & 4350 & -0.01$\pm$0.09 &    6.3$\pm$0.4  &  5  \\
  23  &      Sz 91 & J16071159$-$3903475 &    Lupus &  159 &  M1   &  0.19 & 3705 & -0.27$\pm$0.11 &    6.8$\pm$0.4  &  5  \\
  24  &      Sz 98 & J16082249$-$3904464 &    Lupus &  156 &  K7   &  1.51 & 4060 & -0.21$\pm$0.11 &    5.7$\pm$0.4  &  5  \\
  25  &     Sz 100 & J16082576$-$3906011 &    Lupus &  136 &  M5.5 &  0.08 & 3057 & -0.90$\pm$0.08 &    6.2$\pm$0.5  &  5  \\
  26  & J16083070 & J16083070$-$3828268 &    Lupus &  155 &  K2   &  1.82 & 4900 &  0.14$\pm$0.07 &    6.4$\pm$0.3  &  5  \\
  27  &  V1094 Sco & J16083617$-$3923024 &    Lupus &  153 &  K6   &  1.12 & 4205 & -0.33$\pm$0.11 &    6.3		    &  6  \\
  28  &     Sz 111 & J16085468$-$3937431 &    Lupus &  157 &  M1   &  0.21 & 3705 & -0.27$\pm$0.11 &    6.7$\pm$0.4  &  5  \\
  29  &    Sz 123A & J16105158$-$3853137 &    Lupus &  162 &  M1   &  0.13 & 3705 & -0.33$\pm$0.09 &    6.3		    &  6  \\
  30  &     GW Lup & J15464473$-$3430354 &    Lupus &  155 &  M1.5 &  0.33 & 3631 & -0.34$\pm$0.13 &    6.3$\pm$0.4  &  7  \\
  31  &     IM Lup & J15560921$-$3756057 &    Lupus &  158 &    K5 &  2.57 & 4266 & -0.05$\pm$0.11 &    5.7$\pm$0.4  &  7  \\
  32  &     MY Lup & J16004452$-$4155310 &    Lupus &  156 &    K0 &  0.87 & 5129 &  0.09$\pm$0.08 &    6.3 &  7  \\
  33  &     Sz 129 & J15591647$-$4157102 &    Lupus &  161 &    K7 &  0.44 & 4074 & -0.08$\pm$0.09 &    6.6$\pm$0.4  &  7  \\
  34  &     AS 209 & J16491530$-$1422087 &      Oph &  121 &    K5 &  1.41 & 4266 & -0.08$\pm$0.12 &    6.0$\pm$0.4  &  7  \\
  35  &       SR 4 & J16255615$-$2420481 &      Oph &  134 &    K7 &  1.17 & 4074 & -0.17$\pm$0.12 &    5.9$\pm$0.4  &  7  \\
  36  &    DoAr 25 & J16262367$-$2443138 &      Oph &  138 &    K5 &  0.95 & 4266 & -0.02$\pm$0.11 &    6.3$\pm$0.4  &  7  \\
  37  &    DoAr 33 & J16273901$-$2358187 &      Oph &  139 &    K4 &  1.51 & 4467 &  0.04$\pm$0.12 &    6.2$\pm$0.4  &  7  \\
  38  &    WaOph 6 & J16484562$-$1416359 &      Oph &  123 &    K6 &  2.88 & 4169 & -0.17$\pm$0.13 &    5.5$\pm$0.5  &  7  \\
  39  &  HD 142666 & J15564002$-$2201400 &   UppSco &  148 &    A8 &  9.12 & 7586 &  0.20$\pm$0.02 &    7.1$\pm$0.3  &  7  \\
  40  &  HD 143006 & J15583692$-$2257153 &   UppSco &  165 &    G7 &   3.8 & 5623 &  0.26$\pm$0.06 &    6.6$\pm$0.3  &  7  \\
  41  &  HD 163296 & J17562128$-$2157218 &  isolate &  101 &    A1 & 16.98 & 9333 &  0.31$\pm$0.04 &    7.1$\pm$0.6  &  7  \\
  42  &     J1100 & J11004022$-$7619280 &     ChaI &  191 &    M4 &   0.1 & 3270 & -0.33$\pm$0.05 &    6.5		    &  4,8*  \\
  43  &     TW Hya & J11015191$-$3442170 &      TWA &   60 &  M0.5 &  0.34 & 4070 & -0.09$\pm$0.10 &    6.8$\pm$0.4  &  1,5  \\
  44  &  V4046 Sgr & J18141047$-$3247344 & $\beta$ Pic & 72.5 & K5,K7 &  0.86 & 4350 &  0.24$\pm$0.02 &    7.4$\pm$0.1  &  9,10*  \\
\enddata
\tablecomments{Gaia DR2 distances are adopted for individual disks \citep{Gaia2018}, with typical uncertainty of 1\,pc. Stellar masses for the two spectroscopic binaries, UZ Tau and V4046 Sgr, are the total mass of the two components, derived from disk CO rotation. Sources with dynamical stellar mass adopted are marked with $*$ for their corresponding references.
Lupus disks without individually calculated stellar ages take the average age of 2\,Myr for Lupus Clouds. The stellar age for MY Lup also takes 2\,Myr since the disk is inclined and flared enough to extinct the host, which overestimates the stellar age \citep{Andrews2018_Lmm}.}

\tablerefs{1=\citet{HH2014}, 2=\citet{Simon2019}, 3=\citet{Czekala2019},  4=\citet{Pegues2021}, 5=\citet{Andrews2018_Lmm}, 6=\citet{Ansdell2018}, 7=\citet{Andrews2018_dsharp}, 8=\citet{Pascucci2016}, 9=\citet{Rosenfeld2012}, 10=\citet{HH2015}}
\end{deluxetable*}


\section{Size Measurements} \label{sec:measure}
The most general definition of the size of an object is taken as the distance from its center to the outer edge. For astrophysical objects like disks, which fade around the edges and merge into the interstellar environment, the above definition is not applicable due to observational limitations. For practical purposes, \citet{Tripathi2017} introduced the disk effective radius ($R_{\rm eff}$) concept, the radius that encompasses a fraction ($x$) of its total flux. For a constructed cumulative intensity profile, $f_\nu(r) = 2\pi \int_0^r I_\nu(r')r'dr'$, $R_{\rm eff}$ is taken as $f_\nu(R_{\rm eff})=xF_{\nu}$, where $F_{\nu}=f_\nu(\infty)$ is the total flux.
This effective radius definition has been widely applied in disk studies using the intensity profiles inferred from modeling the continuum visibilities \citep{Andrews2018_Lmm, Long2019, Hendler2020}.
Following a similar concept, some other works use increasing elliptical apertures in the image plane to construct a cumulative flux curve and determine the effective radius \citep{Ansdell2018, Huang2018_ring}.

In this paper, we define the size for a given disk tracer as the 90\% effective radius based on the derived azimuthally averaged radial profiles from \textit{deprojected} data images.
The choice of $x=0.9$ ensures the inclusion of most of the disk area. Meanwhile, this choice is consistent with the measurements of the 17 Lupus disks from the ALMA Lupus Survey \citep{Ansdell2018}, which enables a joint analysis by directly adopting their results. We also tabulate the measurements with $x=0.68, 0.95$ in Appendix Table~\ref{tab:sizes_apx}.  

\begin{figure*}[!th]
\centering
    \includegraphics[width=0.9\linewidth]{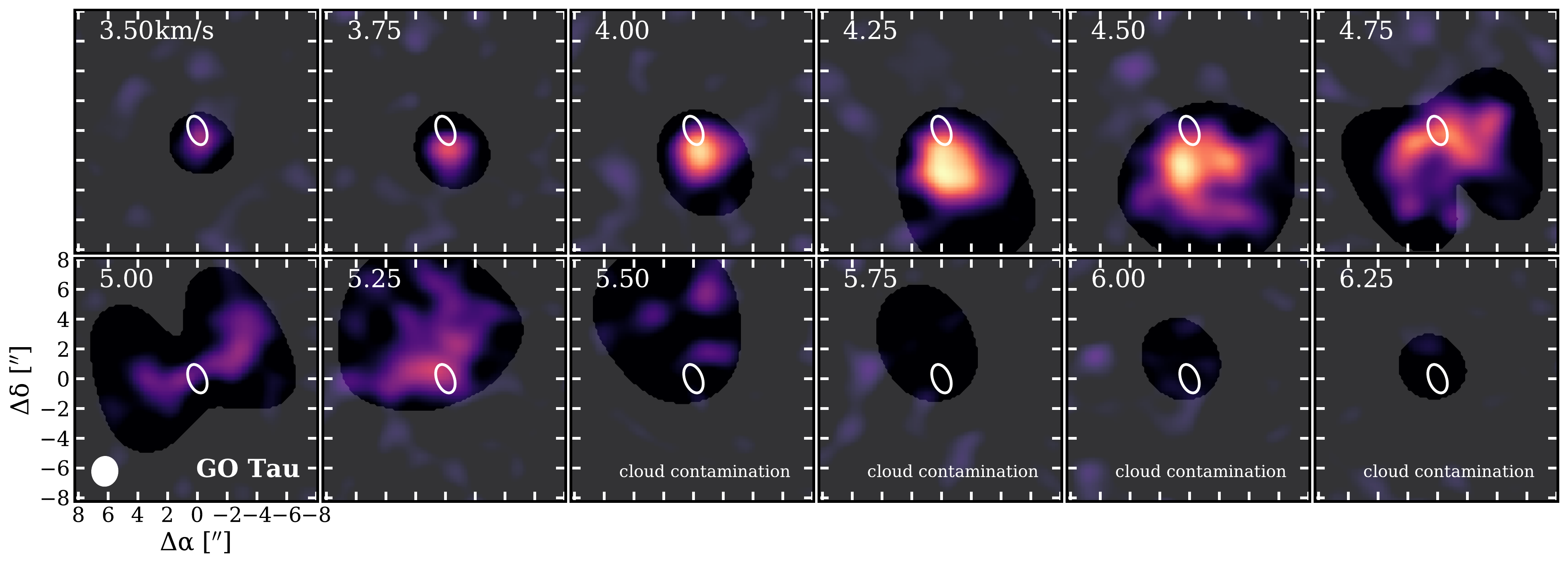} \\ 
    \includegraphics[width=0.9\linewidth]{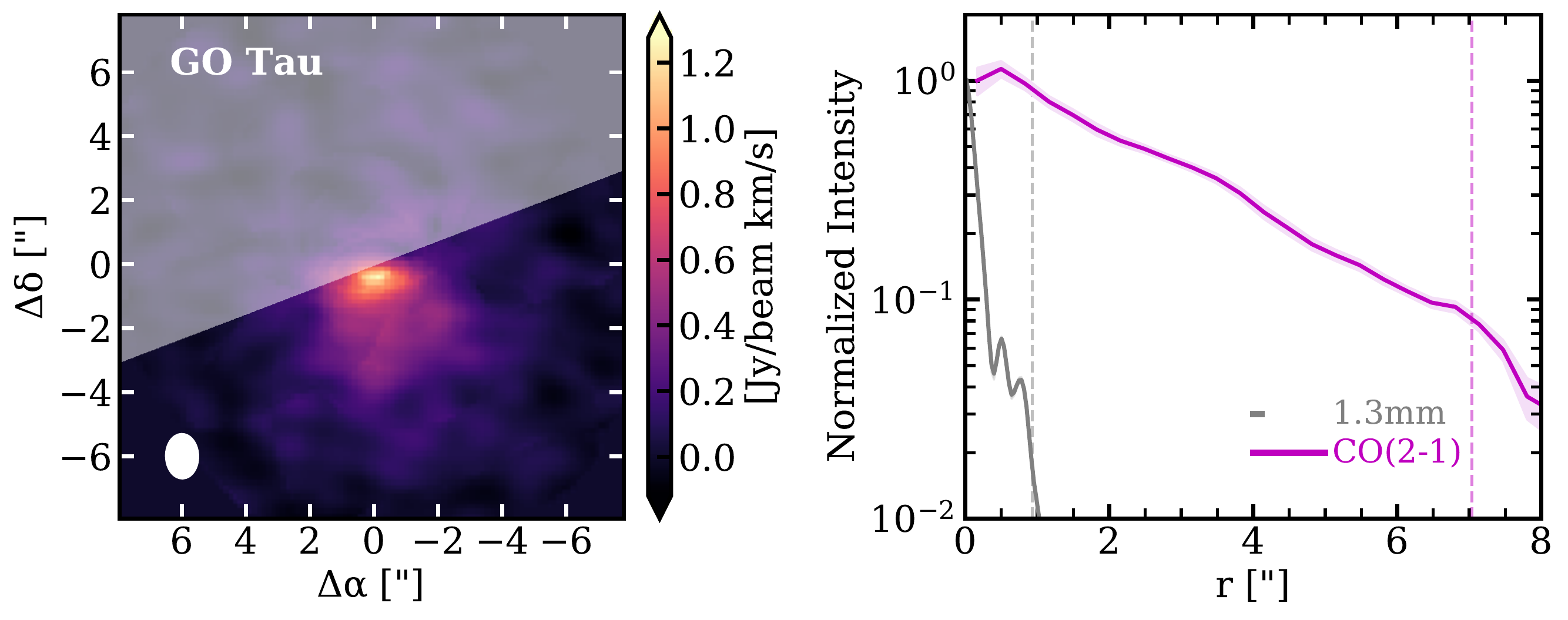}
\caption{\textbf{Top:} Channel maps of the $^{12}$CO $J=2-1$ emission for GO Tau from SMA observations (tapered to 2$''$ resolution for illustration purpose here). The Keplerian mask is shown in each channel, representing regions with expected emission. The white contour indicates $R_{\rm mm}$ as measured from the 1.3\,mm continuum emission \citep{Long2018}. \textbf{Bottom Left:} The moment-zero map, created by integrating over the velocity channels and regions marked out by the Keplerian mask. The redshifted NE side of the disk is further masked out as white-shades where severe cloud absorption presents. \textbf{Bottom Right:} The azimuthally averaged CO and mm continuum radial profiles, from unmasked regions of the moment-zero map and 1.3\,mm image, respectively. $R_{\rm mm}$ and $R_{\rm CO}$ are marked as vertical dashed lines. The synthesized beam sizes are indicated by the horizontal bars. \label{fig:gotau}}
\end{figure*}

\begin{figure*}[!th]
\centering
    \includegraphics[width=\linewidth]{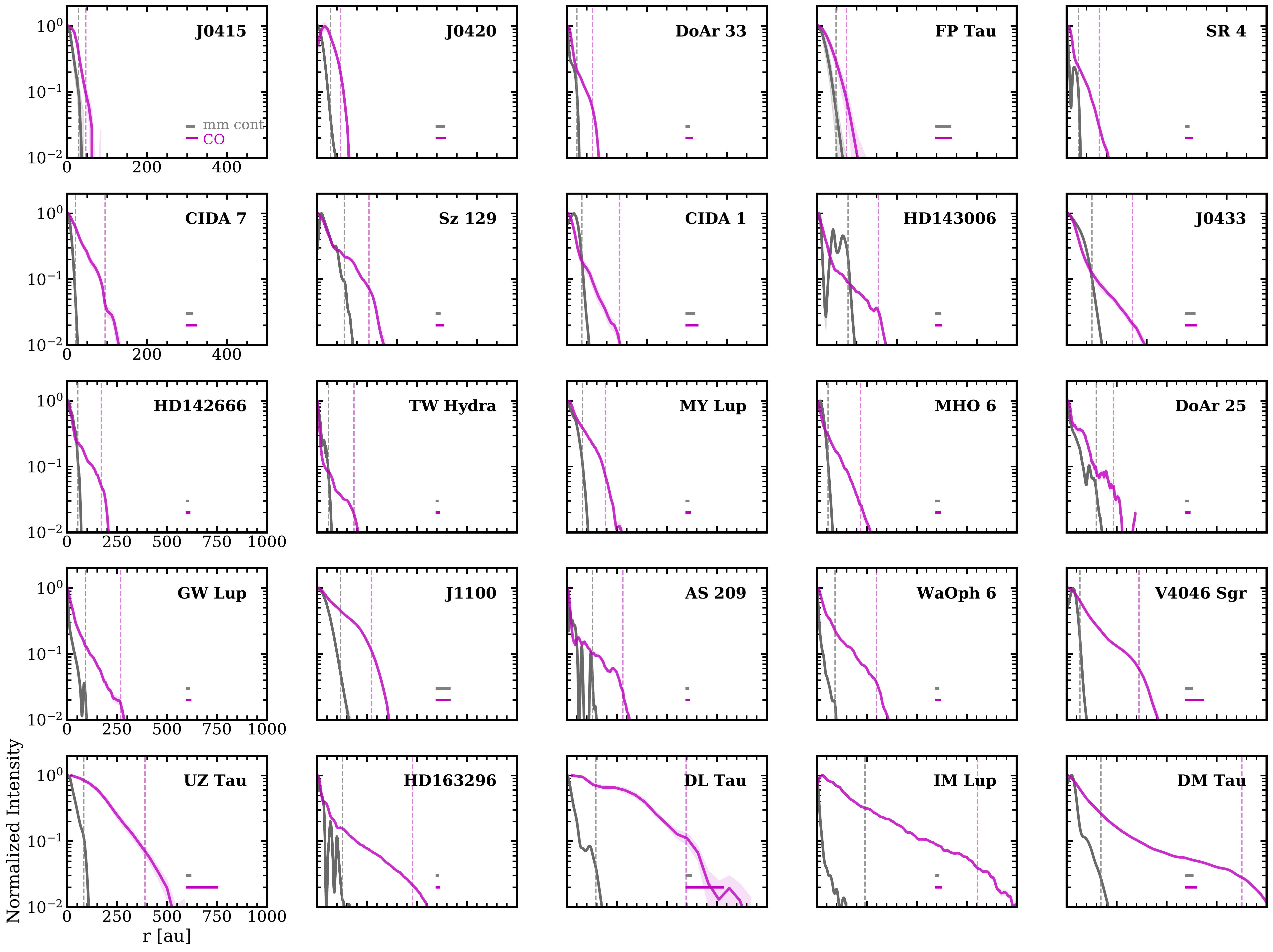}
\caption{The azimuthally averaged radial profiles for both $^{12}$CO line (in magenta) and mm continuum emission (in grey), in order of increasing CO size (excluding the 17 Lupus disks from \citet{Ansdell2018} and the most extended GO Tau disk shown in Figure~\ref{fig:gotau}). Each profile is normalized to its peak emission value. Each row has the same radial extent as labeled out in the x-axis labels of the left panels. The synthesized beam sizes are indicated by the horizontal bars. $\rm R_{mm}$ and $\rm R_{CO}$ are marked as vertical dashed lines.  \label{fig:rps}}
\end{figure*}

For the mm continuum size, we first deproject the continuum image using the known disk inclination and position angles from previous studies, and then extract the azimuthally averaged radial intensity profile. The size ($R_{\rm mm}$) is then measured from the cumulative intensity profile, which is constructed by integrating the radial intensity profile. 
The statistical uncertainty of the size is estimated via bootstrapping over 1000 random sample profiles. Each sample profile is drawn by perturbing the intensity at individual radial bins assuming a Gaussian noise distribution, where the noise level at each radial bin is calculated as the 1$\sigma$ scatter divided by the square root of the number of beams spanning the azimuthal angles over which the intensity is measured\footnote{The final adopted uncertainty of the size also adds the beam radius in quadrature to represent possible systematic errors in the measurements.}.
The adopted disk geometry parameters, continuum sizes, as well as some continuum observation details are summarized in Table~\ref{tab:data_prop}.

The CO emission size is measured in the same way, using the radial emission profile extracted from the CO moment-zero map, obtained by integrating the data cube along the velocity axis. To boost SNR for the faint line emission, especially in the outer disk, we apply a Keplerian mask customized for each disk when creating the moment-zero map. This ensures that only regions associated with disk emission are included. This method has been used in previous work to enable weak line detection and increase SNR for faint emission \citep[e.g.,][]{Salinas2017, Matra2017, Yen2018, Teague2018_CS, Loomis2018}. 
The basic emission morphology at each velocity channel can be well predicted given the following parameters: stellar mass and source distance, disk inclination and position angle, phase center offsets in R.A. and Dec, the systemic velocity, and a mask size\footnote{The code to make Keplerian masks can be found at \url{https://github.com/richteague/keplerian_mask}; see also a detailed technical procedure for Keplarian mask making in the Appendix of \citet{Trapman2020_Lupus}. For each disk, we create a series of masks with different sizes and determine the optimal mask as the one with which the measured gas radii no longer increase significantly (well within 1$\sigma$ statistical uncertainty).}.
In some cases where both the disk front and back surfaces are well detected (e.g., HD 163296, IM Lup), we employ the aspect ratio of the emission surface ($z/r$) to describe the 3D geometry of the disk in the generation of the Keplerian mask. The parameters used for creating the mask for individual disks are summarized in Appendix Table~\ref{tab:mask_rp_apx}.


Figure~\ref{fig:gotau} shows an example of the Keplerian mask applied to the GO Tau disk. With the extracted radial profile from the moment-zero map generated with the Keplerian mask, we then calculate the CO disk size ($R_{\rm CO}$) and its uncertainty as described above for the dust disk size. 
For sources like GO Tau where cloud contamination dominates in some velocity channels, an additional mask (as shown in the bottom left panel of Figure~\ref{fig:gotau}) is applied to avoid using that portion of azimuthal angles in extracting the radial profile\footnote{This procedure is implemented in the \texttt{radial\_profile} function of the \texttt{gofish} package \citep{GoFish}. }. We usually use the half-side of the disk associated with less cloud contamination. For disks with high inclinations, we use instead an azimuthal wedge centered around the disk major axis that encompasses the disk emitting area (see details for individual disks as listed in Appendix Table~\ref{tab:mask_rp_apx}). 
The flexibility in choosing optimal azimuthal angles for radial profile generation results in smaller statistical uncertainties in size estimates, compared to the increasing aperture method as applied in \citet{Ansdell2018}. 
In some cases (about 20\% of our sample, mostly in Taurus), the cloud absorption is located around the systemic velocity, for which the derived $R_{\rm CO}$ would be underestimated. Our tests show that, depending on the line width of the cloud emission, missing flux around central channels could lead to 10--30\% under-estimation in sizes (see Figure~\ref{fig:cloud-test} in the Appendix where we manually exclude emission from central channels to estimate disk size for this demonstration).

The extraction of a radial emission profile from the moment-zero map usually assumes a flat disk geometry, since the exact flaring structures for most disks are unknown. For the well-studied HD 163296 disk, we have performed radial profile extraction considering the $^{12}$CO emission surface \citep{Teague2019Natur}\footnote{The $^{12}$CO emission surface could be expressed as $Z(r)=0.265(r/1\farcs)^{1.29}-0.006(r/1\farcs)^{3.8}$, which falls off towards large radial distances.  Assuming a constant $z/r$ would stretch out the emission along minor axis further than a tapered vertical structure, resulting in a larger gas disk size.}, and no significant differences in the derived CO sizes are observed when compared with the result based on a flat geometry.  For disks with low to intermediate inclination angles, the flat disk geometry is a reasonable assumption.  We have also tested with mock CO data with varying inclination angles. The comparison of high (70$\degr$) and low (10$\degr$) inclination models reveals an uncertainty of $\sim$10$\%$. 
The derived radial profiles for both $^{12}$CO and continuum emission for individual disks are shown in Figure~\ref{fig:rps}, with the calculated disk sizes summarized in Table~\ref{tab:data_prop}. 

This effective radius definition depends on the distribution of emission within the disk. If the continuum emission is optically thin, $R_{\rm mm}$ traces the outer boundary of the dust density distribution in the disk midplane (but see \citealt{Rosotti2019}). The $^{12}$CO emission is usually optically thick across the disk and probes the gas temperature distribution in the elevated disk layers assuming local thermodynamic equilibrium.
The moment-zero maps could not directly reflect the mass distribution due to the high optical depth of CO line emission and the incorporation of the velocity information, for which line emission in the inner disk is associated with a broader velocity range thus higher weights.
Nevertheless, the adopted $R_{\rm mm}$ and $R_{\rm CO}$ are reasonable metrics for the spatial extent of the disk dust and gas distributions, though we should keep in mind that the two disk sizes have different underlying physical meanings.

\begin{deluxetable*}{cccccrcccccl}
\tabletypesize{\scriptsize}
\tablecaption{Disk Properties\label{tab:data_prop}}
\tablewidth{0pt}
\tablehead{
\colhead{Idx} & \colhead{Name} & \colhead{Incl} & \colhead{PA} & \multicolumn{3}{c}{Continuum Properties} & \multicolumn{3}{c}{CO Line Properties} & \colhead{$R_{\rm CO}$/$R_{\rm mm}$} &  \colhead{Refs} \\ 
\cmidrule(lr){5-7} \cmidrule(lr){8-10}
\colhead{} & \colhead{} & \colhead{} 		  & \colhead{}   & \colhead{Freq} & \colhead{$F_{\rm mm}$} & \colhead{$R_{\rm mm}$} 	&  \colhead{Line} & \colhead{$F_{\rm CO}$}  & \colhead{$R_{\rm CO}$} &  \colhead{} & \colhead{} \\ 
\colhead{} & \colhead{} & \colhead{($\degr$)} & \colhead{($\degr$)} & \colhead{(GHz)}  & \colhead{(mJy)} & \colhead{(au)} &  \colhead{}  & \colhead{(Jy km/s)}  & \colhead{(au)} &   \colhead{} & \colhead{} \\
}
\colnumbers
\startdata
   1 &    CX Tau & 55.1 & 246.2 & 227.7 &      9.8 &     29$\pm$4 &  2-1 &   1.94 &  115$\pm$13 &     3.9$\pm$0.7 & 1 \\
   2 &    DL Tau & 45.0 & 232.1 & 225.5 &    170.7 &    144$\pm$10 &  2-1 &   7.05 &  597$\pm$91 &     4.1$\pm$0.7 & 2,3 \\
   3 &    DM Tau & 36.0 & 338.0 & 283.9 &     99.1 &    178$\pm$14 &  2-1 &  15.21 &  876$\pm$23 &     4.9$\pm$0.4 & 4,5 \\
   4 &    GO Tau & 53.9 &  20.9 & 225.5 &     64.8 &    134$\pm$9 &  2-1 &  17.66 & 1014$\pm$83 &    7.6$\pm$0.8 & 2,3 \\
   5 &    UZ Tau & 56.1 & 270.4 & 225.5 &    129.5 &     83$\pm$8 &  2-1 &   7.32 &  389$\pm$75 &     4.7$\pm$1.0 & 2,3 \\
   6 &    FP Tau & 42.9 & 237.8 & 230.5 &      8.4 &     47$\pm$16 &  2-1 &   0.81 &   74$\pm$17 &    1.6$\pm$0.7 & 6 \\
   7 &    CIDA 1 & 38.5 &  11.5 & 338.6 &     35.9 &     38$\pm$9 &  3-2 &   1.25 &  132$\pm$14 &     3.5$\pm$0.9 & 7 \\
   8 &    CIDA 7 & 31.4 & 274.0 & 338.2 &     25.5 &     21$\pm$6 &  3-2 &   1.19 &   95$\pm$11 &     4.6$\pm$1.7 & 7 \\
   9 &     MHO 6 & 65.2 & 293.6 & 338.5 &     49.0 &     56$\pm$6 &  3-2 &   3.56 &  218$\pm$7 &    3.9$\pm$0.5 & 7 \\
  10 &     J0415 & 37.3 & 122.8 & 338.2 &      1.0 &     16$\pm$8 &  3-2 &   0.23 &   47$\pm$13 &    2.9$\pm$1.8 & 7 \\
  11 &     J0420 & 38.4 & 105.1 & 338.5 &     16.5 &     34$\pm$8 &  3-2 &    0.4 &   59$\pm$10 &    1.7$\pm$0.5 & 7 \\
  12 &     J0433 & 65.2 & 165.3 & 338.5 &     37.5 &     63$\pm$9 &  3-2 &    1.3 &  165$\pm$12 &     2.6$\pm$0.4 & 7 \\
  13 &     Sz 65 & 61.5 & 108.6 & 225.0 &     29.9 &     66$\pm$2 &  2-1 &    2.08 &  178$\pm$25 &     2.7$\pm$0.4 & 8 \\
  14 &     Sz 73 & 49.8 &  94.7 & 225.0 &     10.8 &     58$\pm$3 &  2-1 &    1.63 &  107$\pm$9 &      1.8$\pm$0.2 & 8 \\
  15 &     Sz 75 & 60.2 & 169.0 & 225.0 &     34.1 &     56$\pm$2 &  2-1 &    3.01 &  195$\pm$21 &      3.5$\pm$0.4 & 8 \\
  16 &     Sz 76 & 38.9 & 113.0 & 225.0 &      4.7 &     80$\pm$11 &  2-1 &   1.60 &  174$\pm$6 &      2.2$\pm$0.3 & 8 \\
  17 &     J1556 & 53.5 &  55.6 & 225.0 &     23.5 &     59$\pm$2 &  2-1 &    1.16 &  116$\pm$3 &      2.0$\pm$0.1 & 8 \\
  18 &     Sz 84 & 65.0 & 168.0 & 225.0 &     12.6 &     81$\pm$3 &  2-1 &    1.41 &  148$\pm$18 &      1.8$\pm$0.2 & 8 \\
  19 &    RY Lup & 68.0 & 109.0 & 225.0 &     86.1 &    141$\pm$3 &  2-1 &    3.66 &  263$\pm$66 &     1.9$\pm$0.5 & 8 \\
  20 &     J1600 & 65.7 & 160.5 & 225.0 &     50.0 &    123$\pm$3 &  2-1 &    2.45 &  291$\pm$49 &     2.4$\pm$0.4 & 8 \\
  21 &    EX Lup & 30.5 &  70.0 & 225.0 &     19.5 &     49$\pm$2 &  2-1 &    3.14 &  140$\pm$9 &     2.9$\pm$0.2 & 8 \\
  22 &    Sz 133 & 78.5 & 126.3 & 225.0 &     27.0 &    145$\pm$6 &  2-1 &    1.71 &  243$\pm$67 &   1.7$\pm$0.5 & 8 \\
  23 &     Sz 91 & 51.7 &  17.4 & 225.0 &      9.5 &    123$\pm$3 &  2-1 &    2.34 &  358$\pm$64 &      2.9$\pm$0.5 & 8 \\
  24 &     Sz 98 & 47.1 & 111.6 & 225.0 &    103.4 &    148$\pm$3 &  2-1 &    .. &  279$\pm$41 &     1.9$\pm$0.3 & 8 \\
  25 &    Sz 100 & 45.1 &  60.2 & 225.0 &     21.7 &     56$\pm$2 &  2-1 &    1.08 &  121$\pm$8 &      2.2$\pm$0.2 & 8 \\
  26 &     J1608 & 74.0 & 107.0 & 225.0 &     38.8 &    141$\pm$3 &  2-1 &    5.90 &  305$\pm$77 &     2.2$\pm$0.6 & 8 \\
  27 & V1094 Sco & 55.4 & 110.0 & 225.0 &    180.0 &    256$\pm$15 &  2-1 &   4.88 &  335$\pm$86 &     1.3$\pm$0.4 & 8 \\
  28 &    Sz 111 & 53.0 &  40.0 & 225.0 &     60.3 &    105$\pm$2 &  2-1 &    2.56 &  363$\pm$75 &     3.4$\pm$0.7 & 8 \\
  29 &   Sz 123A & 43.0 & 145.0 & 225.0 &     16.1 &     60$\pm$2 &  2-1 &    .. &  118$\pm$10 &     2.0$\pm$0.2 & 8 \\
  30 &    GW Lup & 38.7 &  37.6 & 240.0 &     89 &     92$\pm$3 &  2-1 &    3.1 &  267$\pm$8 &     2.9$\pm$0.1 & 9 \\
  31 &    IM Lup & 47.5 & 144.5 & 240.0 &    253 &    244$\pm$4 &  2-1 &  30.68 &  803$\pm$9 &      3.3$\pm$0.1 & 9 \\
  32 &    MY Lup & 73.2 & 238.8 & 240.0 &     79 &     77$\pm$3 &  2-1 &   2.86 &  192$\pm$7 &     2.5$\pm$0.1 & 9 \\
  33 &    Sz 129 & 34.1 & 151.0 & 240.0 &     86 &     68$\pm$3 &  2-1 &   1.56 &  130$\pm$8 &     1.9$\pm$0.1 & 9 \\
  34 &    AS 209 & 35.0 &  85.8 & 240.0 &    288 &    127$\pm$2 &  2-1 &  10.62 &  280$\pm$5 &      2.2$\pm$0.1 & 9 \\
  35 &      SR 4 & 22.0 &  18.0 & 240.0 &     69 &     29$\pm$2 &  2-1 &    1.5 &   82$\pm$7 &     2.8$\pm$0.3 & 9 \\
  36 &   DoAr 25 & 67.4 & 290.6 & 240.0 &    246 &    147$\pm$2 &  2-1 &   2.78 &  233$\pm$6 &    1.6$\pm$0.1 & 9 \\
  37 &   DoAr 33 & 41.8 &  81.1 & 240.0 &     35 &     25$\pm$2 &  2-1 &   1.12 &   64$\pm$6 &     2.6$\pm$0.3 & 9 \\
  38 &   WaOph 6 & 47.3 & 174.2 & 240.0 &    161 &     91$\pm$4 &  2-1 &  10.91 &  297$\pm$7 &    3.3$\pm$0.1 & 9 \\
  39 &  HD142666 & 62.2 & 162.1 & 240.0 &    120 &     53$\pm$2 &  2-1 &   4.24 &  171$\pm$5 &    3.2$\pm$0.2 & 9 \\
  40 &  HD143006 & 18.6 & 169.0 & 240.0 &     59 &     78$\pm$4 &  2-1 &   2.94 &  154$\pm$5 &      2.0$\pm$0.1 & 9 \\
  41 &  HD163296 & 46.7 & 313.3 & 240.0 &    715 &    137$\pm$2 &  2-1 &  60.56 &  478$\pm$5 &      3.5$\pm$0.1 & 9 \\
  42 &     J1100 & 19.1 & 340.3 & 225.0 &     25 &    118$\pm$31 &  2-1 &    1.3 &  273$\pm$31 &    2.3$\pm$0.7 & 6 \\
  43 &    TW Hya &  5.0 & 152.0 & 290.5 &    580 &     59$\pm$1 &  3-2 &   41.8 &  184$\pm$4 &    3.1$\pm$0.1 & 10 \\
  44 & V4046 Sgr & 33.5 & 256.0 & 283.9 &     49.6 &     66$\pm$12 &  2-1 &  35.55 &  362$\pm$39 &   5.5$\pm$1.2 & 4,11 \\
\enddata
\tablecomments{The CO line fluxes for the 17 Lupus targets in \citet{Ansdell2018} are adopted from \citet{Sanchis2021}. Position angle is counted from N to E for the redshifted disk side, except for the 17 Lupus disks.}

\tablerefs{1=\citet{Facchini2019}, 2=\citet{Long2018}, 3=This Work, 4=\citet{Qi2019}, 5=\citet{Flaherty2020}, 6=\citet{Pegues2021}, 7=\citet{Kurtovic2021}, 8=\citet{Ansdell2018}, 9=\citet{Andrews2018_dsharp}, 10=\citet{Huang2018_CO}, 11=\citet{Rosenfeld2012}}
\end{deluxetable*}

\begin{figure}[!t]
\centering
    \includegraphics[width=0.5\textwidth]{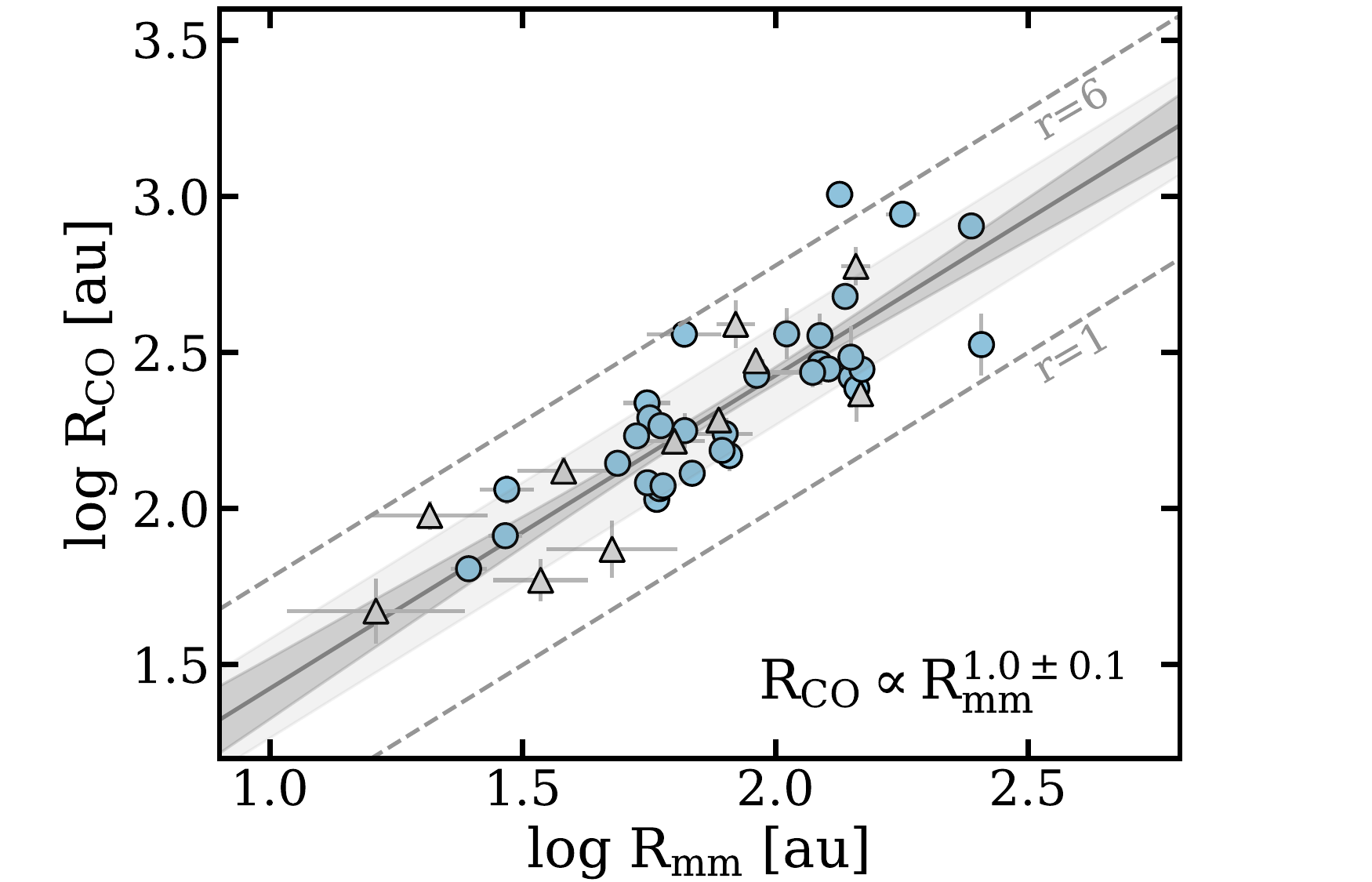} 
\caption{The comparison of disk mm continuum size ($R_{\rm mm}$) and CO emission size ($R_{\rm CO}$). Triangles represent sources with severe cloud absorption around the central velocities that $R_{\rm CO}$ would be underestimated. $R_{\rm CO}$ is generally more extended than $R_{\rm mm}$ by a factor of 2--3, along with large scatter. Two dashed lines denote the range of $R_{\rm CO}$/$R_{\rm mm}$ (1--6). The linear fit is marked as a grey line and indicated in the lower corner.  \label{fig:rg-rd}}
\end{figure}

\section{Results} \label{sec:results}

\subsection{Overview of Disk Sizes} 
Following the procedure described in Section~\ref{sec:measure}, we calculated $R_{\rm CO}$ and $R_{\rm mm}$ for 26 disks. Combining the results of 17 Lupus disks\footnote{Sizes of the 17 Lupus disks are updated with the Gaia DR2 distances individually, where \citet{Ansdell2018} adopted 200 and 150\,pc uniformly for disks located in Lupus III and other Lupus clouds, respectively.} from \citet{Ansdell2018} and the CX Tau disk sizes from \citet{Facchini2019}, we build up a sample of 44 disks with both $R_{\rm CO}$ and $R_{\rm mm}$ measured in a similar manner from the images. The measurements are summarized in Table~\ref{tab:data_prop} and Figure~\ref{fig:rg-rd}.

Recently, \citet{Sanchis2021} revisited the Lupus sample with new disk size measurements. They measured $R_{\rm CO}$ from elliptical Gaussian or Nuker function fitting of the CO moment-zero maps, and $R_{\rm mm}$ from modeling the continuum visibilities.
Their $R_{\rm CO}$ values are broadly consistent with those presented in \citet{Ansdell2018} (and adopted in this study), but the $R_{\rm mm}$ derived from visibility modeling is generally 20--30$\%$ smaller than those derived in the image plane due to limited spatial resolution (see more detailed comparison in Appendix~\ref{sec:lupus-comp}). This is similar to the six very low mass Taurus stars, for which \citet{Kurtovic2021} measured $R_{\rm mm}$ from \textit{uv-}plane modeling profiles. 
For the other four Lupus disks where observations from DSHARP are available and adopted in this analysis, we find that our size measurements generally agree with \citet{Ansdell2018}, except for the $R_{\rm CO}$ of the IM Lup (Sz 82) disk. The clear detection of the diffuse outer gas emission in the deep DSHARP data leads to a larger $R_{\rm CO}$ that is twice the size measured in \citet{Ansdell2018}.

\begin{figure*}[t]
\centering
    \includegraphics[width=0.329\linewidth]{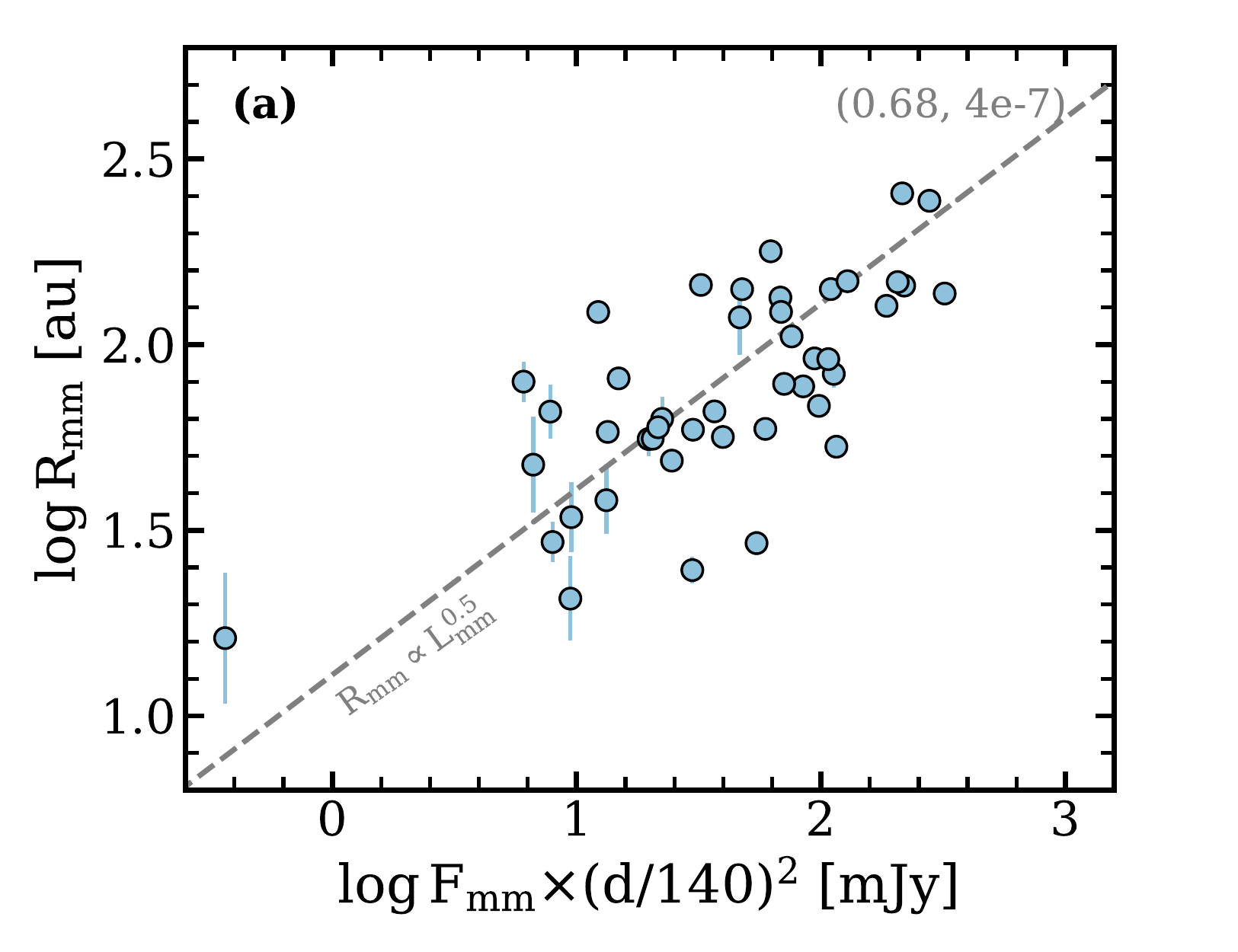} 
    \includegraphics[width=0.329\linewidth]{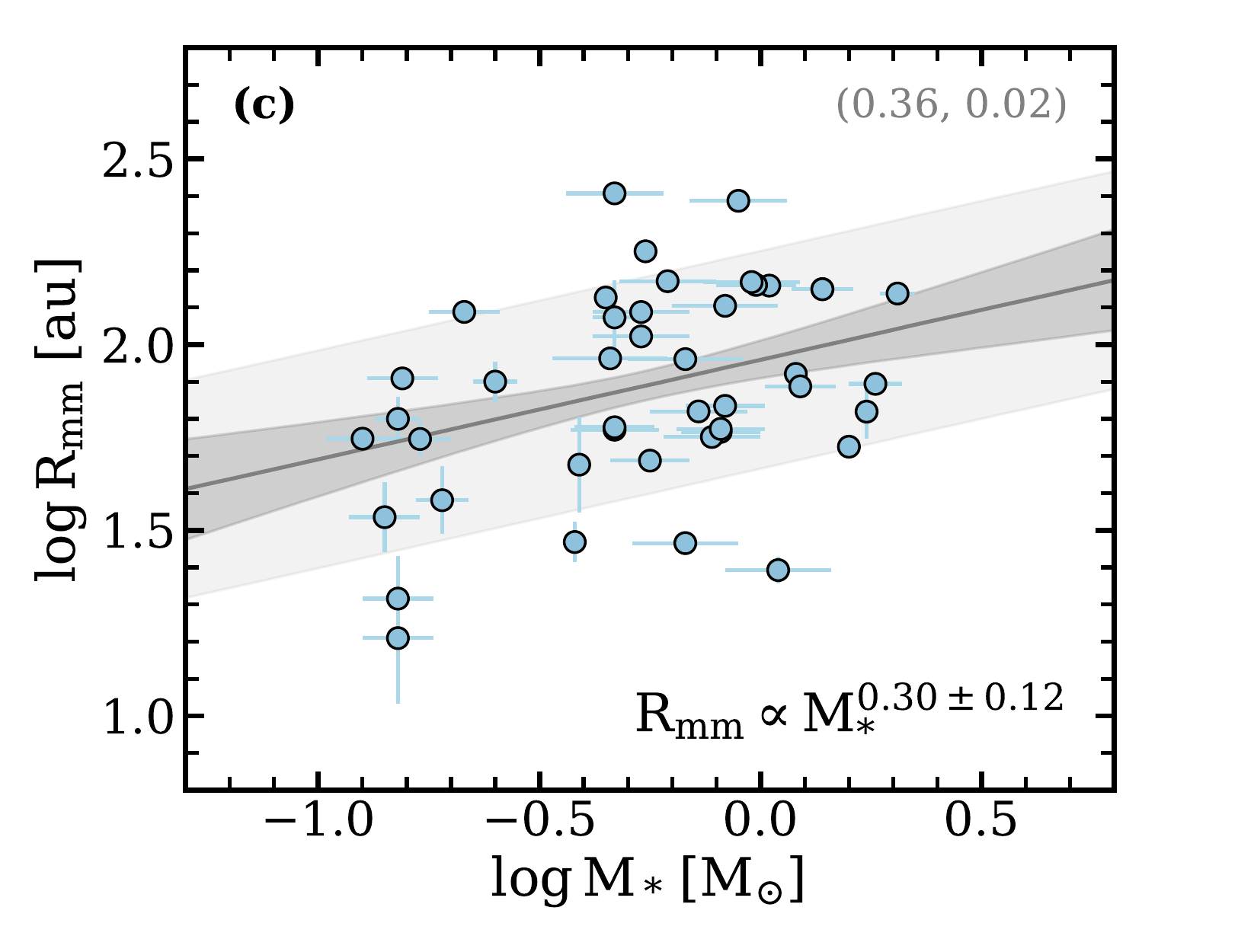}
    \includegraphics[width=0.329\linewidth]{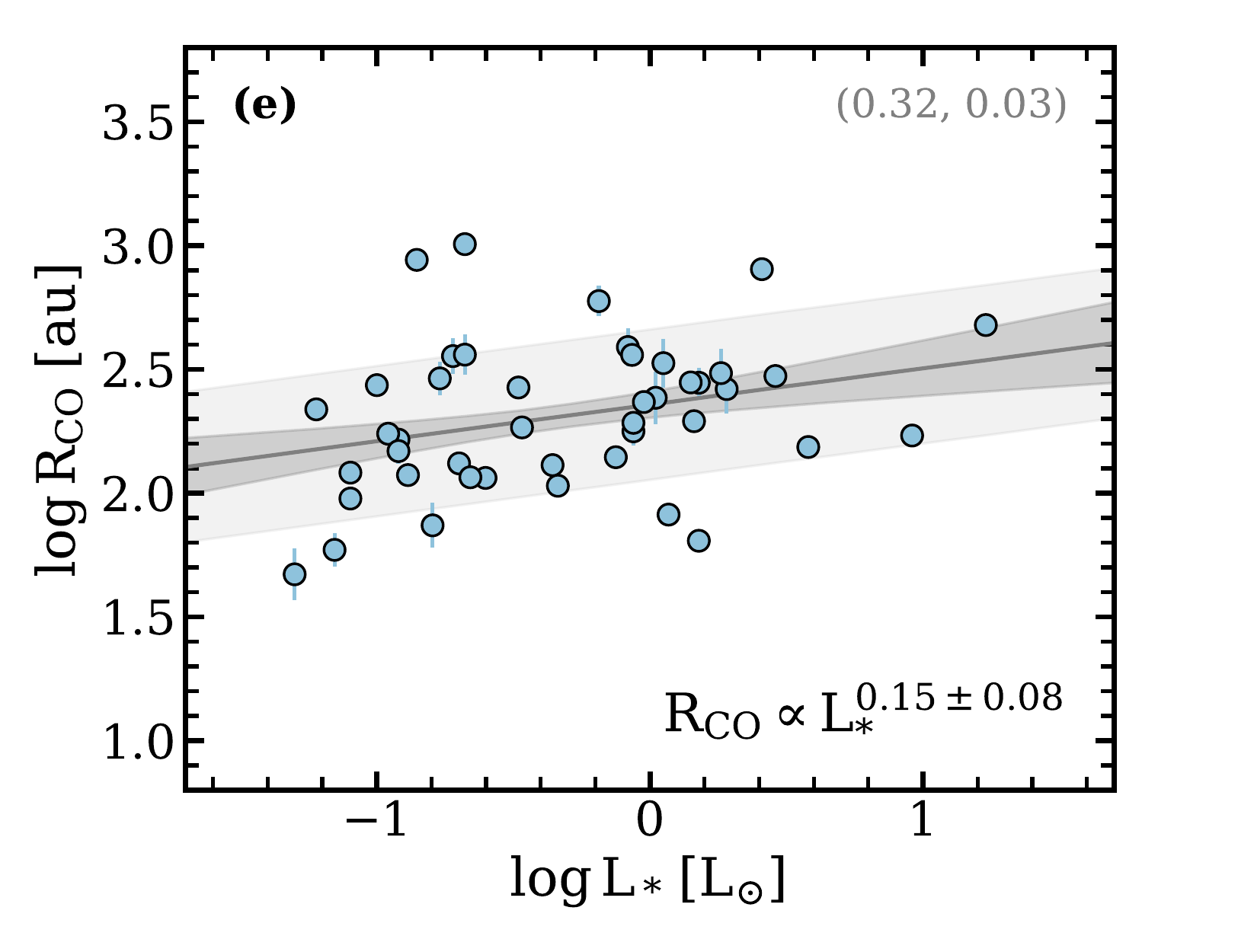} \\
    \includegraphics[width=0.329\linewidth]{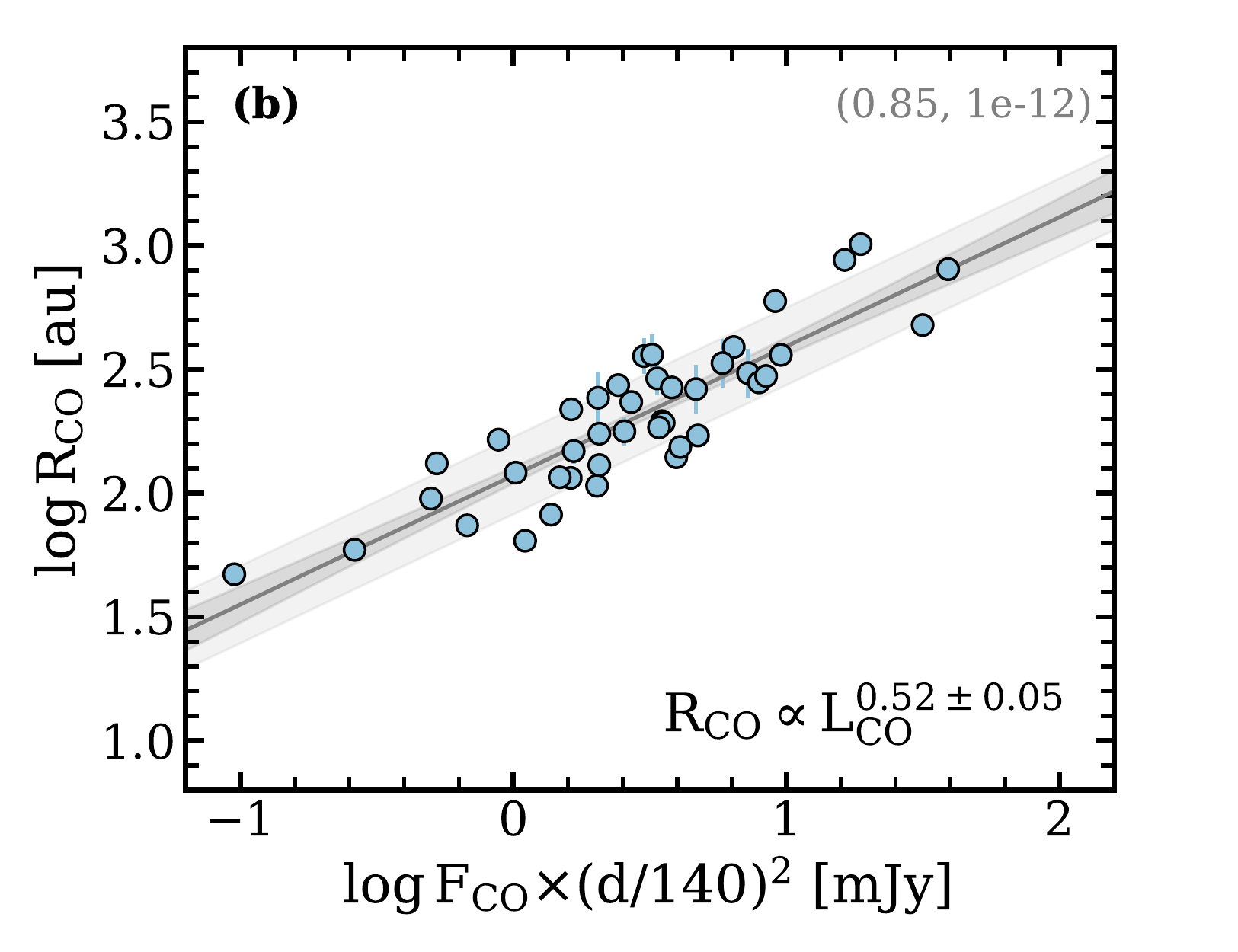} 
    \includegraphics[width=0.329\linewidth]{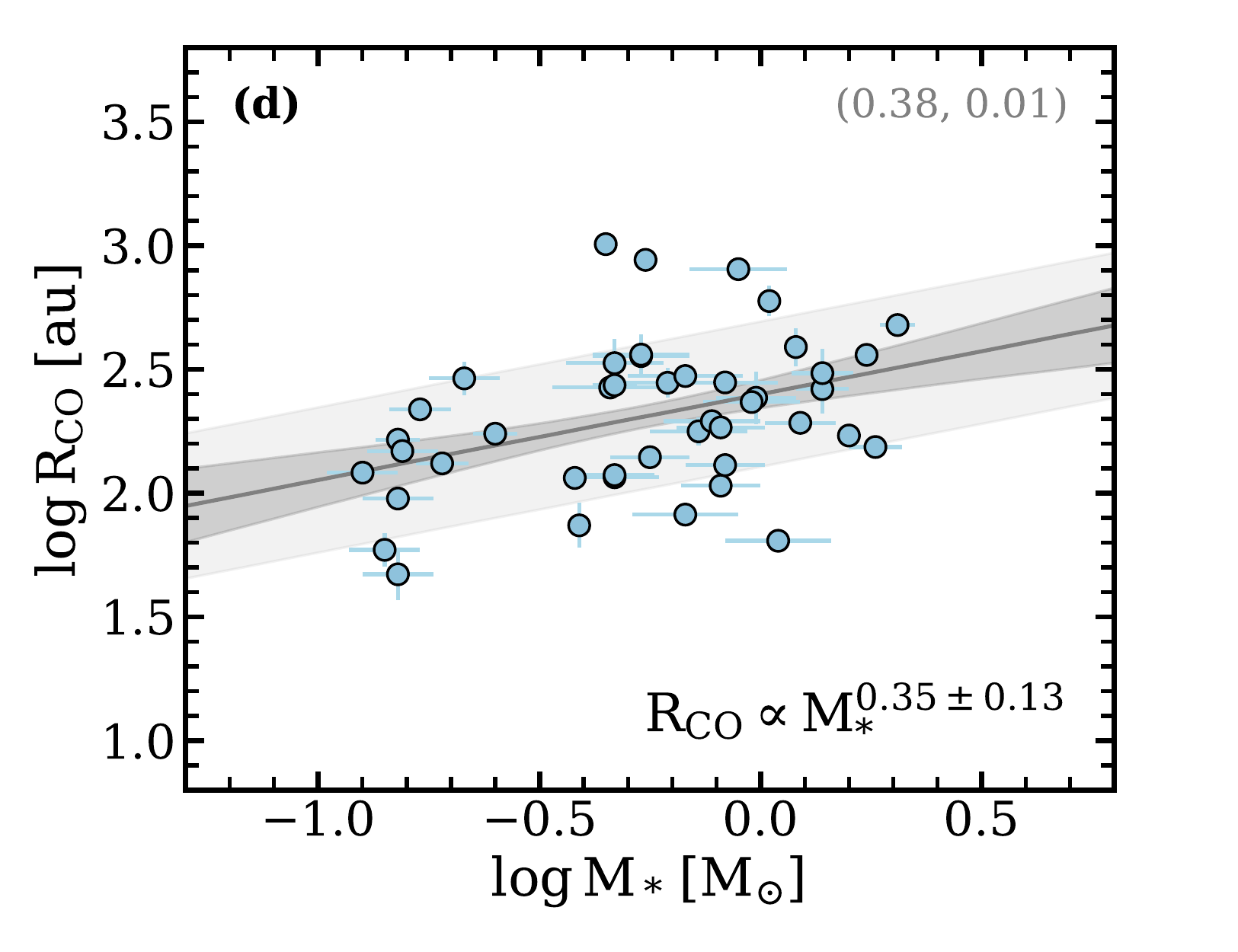} 
    \includegraphics[width=0.329\linewidth]{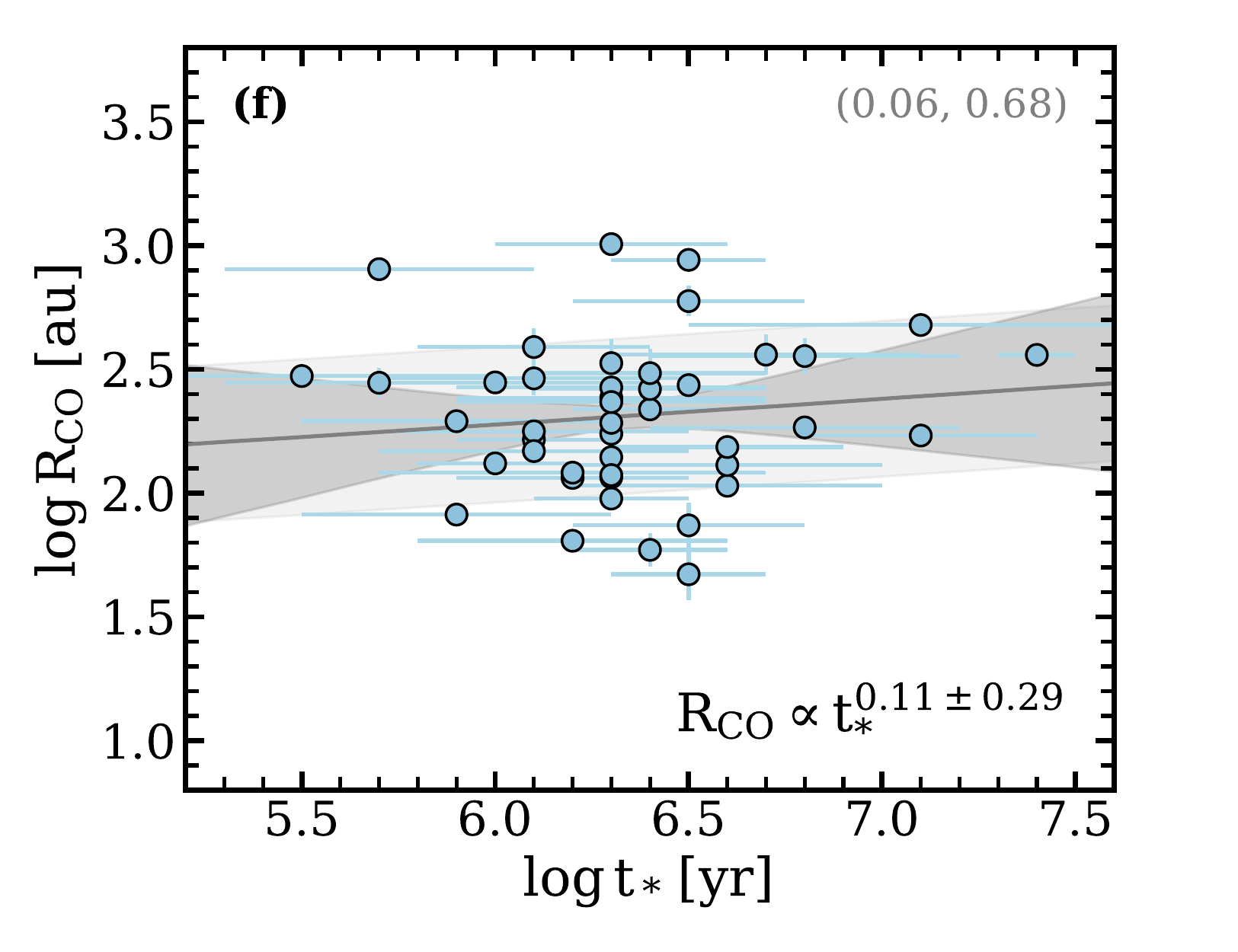}   \\
\caption{The comparison of disk sizes with stellar and disk properties. \textbf{(a)} $L_{\rm mm}$ vs. $R_{\rm mm}$. The scaled correlation from \citet{Andrews2018_Lmm} is marked as dashed line to demonstrate the generic properties of the sample.  \textbf{(b)} $L_{\rm CO 2-1}$ vs. $R_{\rm CO}$.
\textbf{(c)} $M_{\rm *}$ vs. $R_{\rm mm}$. \textbf{(d)} $M_{\rm *}$ vs. $R_{\rm CO}$. \textbf{(e)} $L_{\rm *}$ vs. $R_{\rm CO}$. \textbf{(f)} system age vs. $R_{\rm CO}$.  The derived scaling relation is shown in grey solid line with the 68\% confidence interval marked in grey shaded region. The light shaded region denotes the scatter around the mean relation. The Pearson correlation coefficient and the corresponding $p-$value are noted in the upper right corner in each panel.  
\label{fig:LR}}
\end{figure*}

Figure~\ref{fig:rg-rd} compares $R_{\rm CO}$ and $R_{\rm mm}$ of this sample, which span ranges from $\sim$50--1000\,au and $\sim$15--250\,au, respectively. Disks with severe visible cloud absorption around the central velocity channels are marked as triangles, for which the $R_{\rm CO}$ measurements are lower limits. The CO disks are all spatially resolved, and the majority of them are resolved in at least three independent resolution elements (see Table~\ref{tab:sizes_apx} for the comparison of $R_{\rm CO}$ and beam size). Though all the mm disks are also resolved, 9 of them are resolved in less than two independent resolution elements.
$R_{\rm CO}$ and $R_{\rm mm}$ are positively correlated. Employing the Bayesian linear regression method of \citet{Kelly2007} with its python implementation \textit{Linmix}\footnote{\url{ https://github.com/jmeyers314/linmix}}, we find a best-fit relation of log\,$R_{\rm CO}$ = (1.0$\pm$0.1)log\,$R_{\rm mm}$ +  (0.4$\pm$0.2), with 0.2\,dex scatter of the correlation (as the standard deviation $\sigma$ of an assumed Gaussian distribution around the mean relation).
We also derive a best-fit relation in the linear space as $R_{\rm CO}$ = (2.9$\pm$0.4)\,$R_{\rm mm}$ -- (7.5$\pm$39.5), demonstrating the average ratio of the two sizes.
Though the two disk sizes are correlated, the derived scaling relations also suggest substantial dispersion. The CO disk of GO Tau is the most extended case, followed by the DM Tau disk. Both systems host large mm continuum disks, but are significantly offset upward from the best-fit scaling relation. In contrast, V1094 Sco has a CO disk (measured from the shallow Lupus survey, but the CO emission is spatially resolved), that is much smaller than suggested from its mm disk following the size relation.

\subsection{Disk Size -- Luminosity Plane} \label{sec:Lplane}
The disk size -- luminosity scaling relation of $R_{\rm mm}\propto L_{\rm mm}^{0.5}$ has been established in nearby star-forming regions \citep{Tripathi2017,Andrews2018_Lmm}. 
This correlation is also suggested to slightly flatten with evolution \citep{Hendler2020}. Though the origins of this relation are unclear, it provides a straightforward view of the typical disk characteristics. Figure~\ref{fig:LR}a shows this sample in the $L_{\rm mm}-R_{\rm mm}$ plane, which is well consistent with the established scaling relation\footnote{We take the scaling relation from \citet{Andrews2018_Lmm} (slope=0.5, intercept=2.4 for $R_{\rm 95\%}$), and crudely shift the mm luminosity to 225\,GHz by applying a spectral index of 2.3.}. The mm luminosities are normalized to a common distance of 140\,pc and frequency of 225\,GHz (1.3\,mm), assuming a uniform disk-integrated spectral index of 2.3 \citep{ Andrews2020_review, Tazzari2021}. Data for a few disks were taken at 345\,GHz and we assume $R_{\rm mm}$ are similar at close frequencies of 225--345\,GHz. The sample disks with $R_{\rm mm}$ of $\sim$15--250\,au almost cover the full range of dust disk sizes revealed from previous surveys \citep{Andrews2020_review}. However, $L_{\rm mm}$ of this sample spans from a few mJy to 300\,mJy, mostly sampling the brighter end of the whole disk population, where more than half of known disks are fainter than 10\,mJy at this wavelength.

The CO sizes and line luminosities are tightly correlated (Figure~\ref{fig:LR}b). The line flux is calculated by integrating over the deprojected radial profile from the moment-zero map. The line fluxes at $J=3-2$ transition in a few disks are converted into $J=2-1$ fluxes using the ratio of the square of the line rest frequencies (assuming optically thick line emission),
and assuming similar emitting radii for both transitions. We derive a best-fit relation of log\,$R_{\rm CO}$ = (0.52$\pm$0.05)\,log\,$L_{\rm CO 2-1}$ + (2.07$\pm$0.03), with a scatter of 0.15\,dex.
The slope is consistent with the fact that $^{12}$CO emission is optically thick in disks and serves as a good proxy for the gas temperature of the emitting layer (see also \citealt{Sanchis2021}).

\subsection{Disk Size -- Host Star Properties} \label{sec:host-prop}
We consider here a search for any relationships between the disk sizes and host star properties, including mass ($M_{*}$), luminosity ($L_{*}$), and age ($t_*$). 
We find marginal relationships between disk sizes and $M_{*}$. Regression analyses suggest best-fit relations of log\,$R_{\rm mm}$ = (0.30$\pm$0.12)\,log\,$M_{*}$ + (1.96$\pm$0.05) with a scatter of 0.26\,dex,
and log\,$R_{\rm CO}$ = (0.35$\pm$0.13)\,log\,$M_{*}$ + (2.40$\pm$0.06) with a scatter of 0.29\,dex. 
These relationships are plotted in Figure~\ref{fig:LR}c and Figure~\ref{fig:LR}d. The shallow slope and the large dispersion are also consistent with no correlation between disk size and host stellar mass. Based on the well-established $L_{\rm mm}-M_{*}$ relation ($L_{\rm mm}\propto M_{*}^{1.7\pm0.3}$, \citealt{Andrews2013, Ansdell2016, Pascucci2016}) and $L_{\rm mm}-R_{\rm mm}$ relation \citep{Tripathi2017, Andrews2018_Lmm}, we expect $R_{\rm mm}\propto M_{*}^{0.9}$ (as also demonstrated in \citealt{Andrews2020_review}), which is much steeper than what has been demonstrated by this sample. As we focus on bright disks, this inconsistency can be explained by the exclusion of a large number of faint and small disks around lower mass stars. 
Given the strong correlation between $R_{\rm CO}$ and $R_{\rm mm}$, such bias in sample selection could also account for the derived relation between $R_{\rm CO}$ and $M_{*}$, and a steeper slope is expected when a more complete sample is considered.

Our linear regression analysis also suggests at most a weak relation between $R_{\rm CO}$ and $L_{*}$ ($R_{\rm CO}\propto L_{*}^{0.15\pm0.08}$, with a dispersion of 0.3\,dex, see Figure~\ref{fig:LR}e). This implies that the gas temperature in the disk surface layer may only weakly depend on stellar properties. 

Although disks in this sample have stellar ages spanning  0.5 -- 20\,Myr, no correlation between the age and the CO or continuum disk sizes is found (Figure~\ref{fig:LR}f). 
This is probably because most of these disks are located in Taurus and Lupus and have ages of 1--3\,Myr, accompanied with large uncertainties in individual age measurements.
As the young disks ($t<$1\,Myr) are all around K-type stars (see Figure~\ref{fig:age-mass}), we also compare the CO disk sizes in different age ranges considering only earlier type stars ($M_{*}>0.6\,M_{\odot}$). By splitting the sample into $t_*<1$\,Myr, $1\,\rm Myr<t_*<3\,Myr$, and $t_*>3$\,Myr,  We find comparable average CO sizes (also similar 1$\sigma$ scatter of $\sim$0.2\,dex) within the three bins.

\begin{figure}[!t]
\centering
    \includegraphics[width=\linewidth]{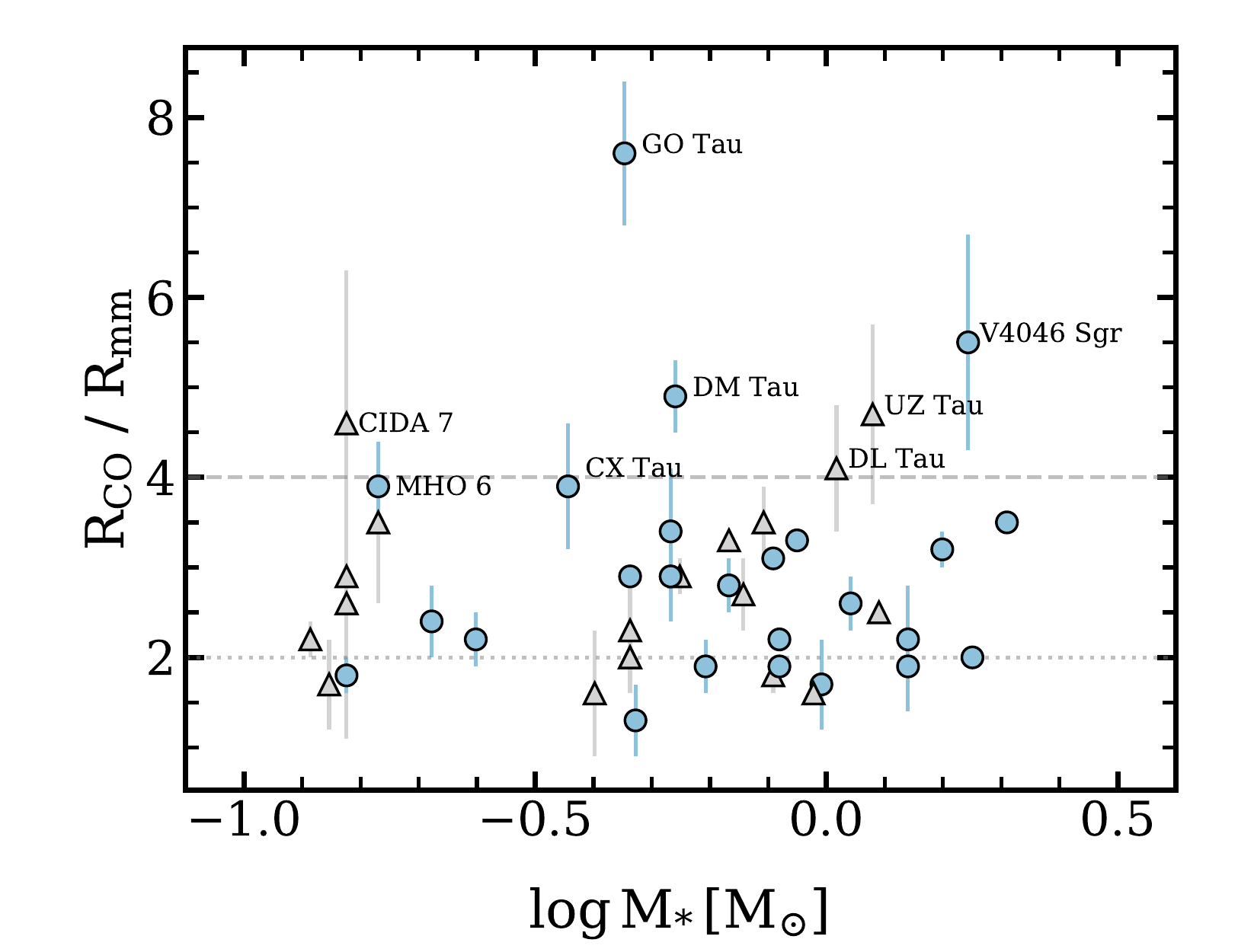}  \\
    \includegraphics[width=\linewidth]{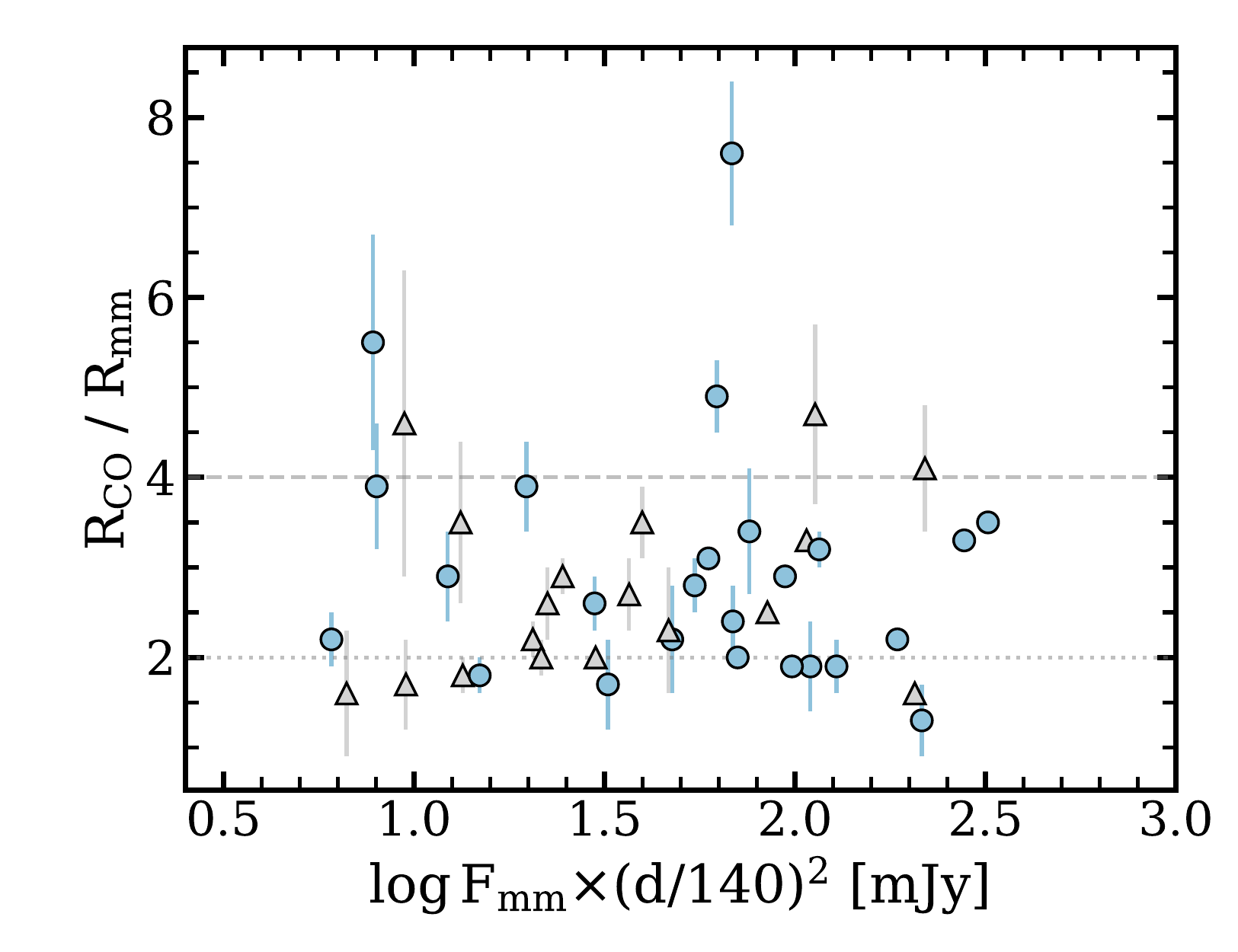} \\
\caption{The distribution of $R_{\rm CO}$/$R_{\rm mm}$ with stellar mass (upper) and disk mm luminosity (bottom). Grey triangles represent disks that either have cloud contamination around systemic velocities or $R_{\rm mm}$ is less than twice of the beam size, where in both cases $R_{\rm CO}$/$R_{\rm mm}$ is likely underestimated. \label{fig:ratio}}
\end{figure}

\begin{figure*}[!th]
\centering
    \includegraphics[width=0.32\linewidth]{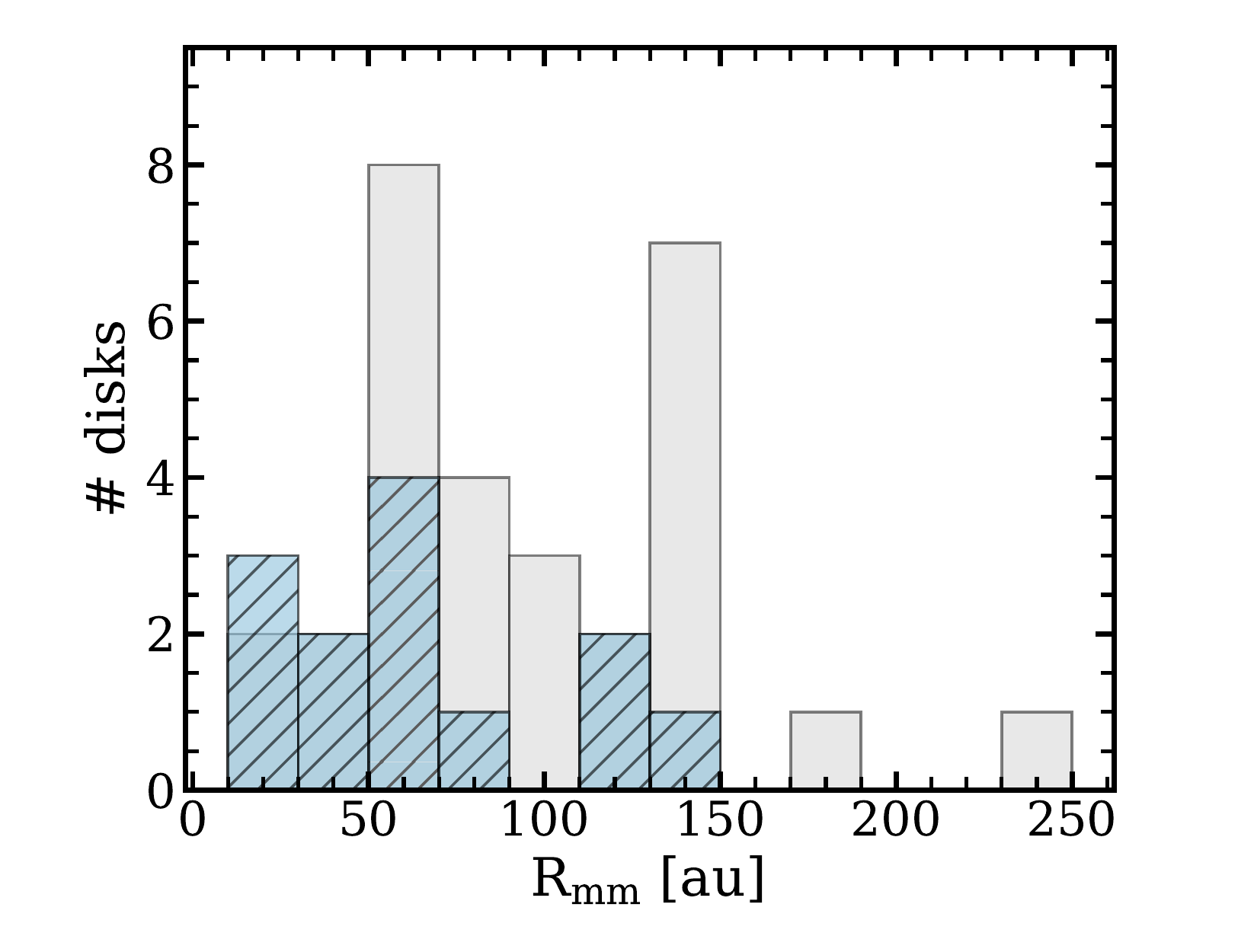}
    \includegraphics[width=0.32\linewidth]{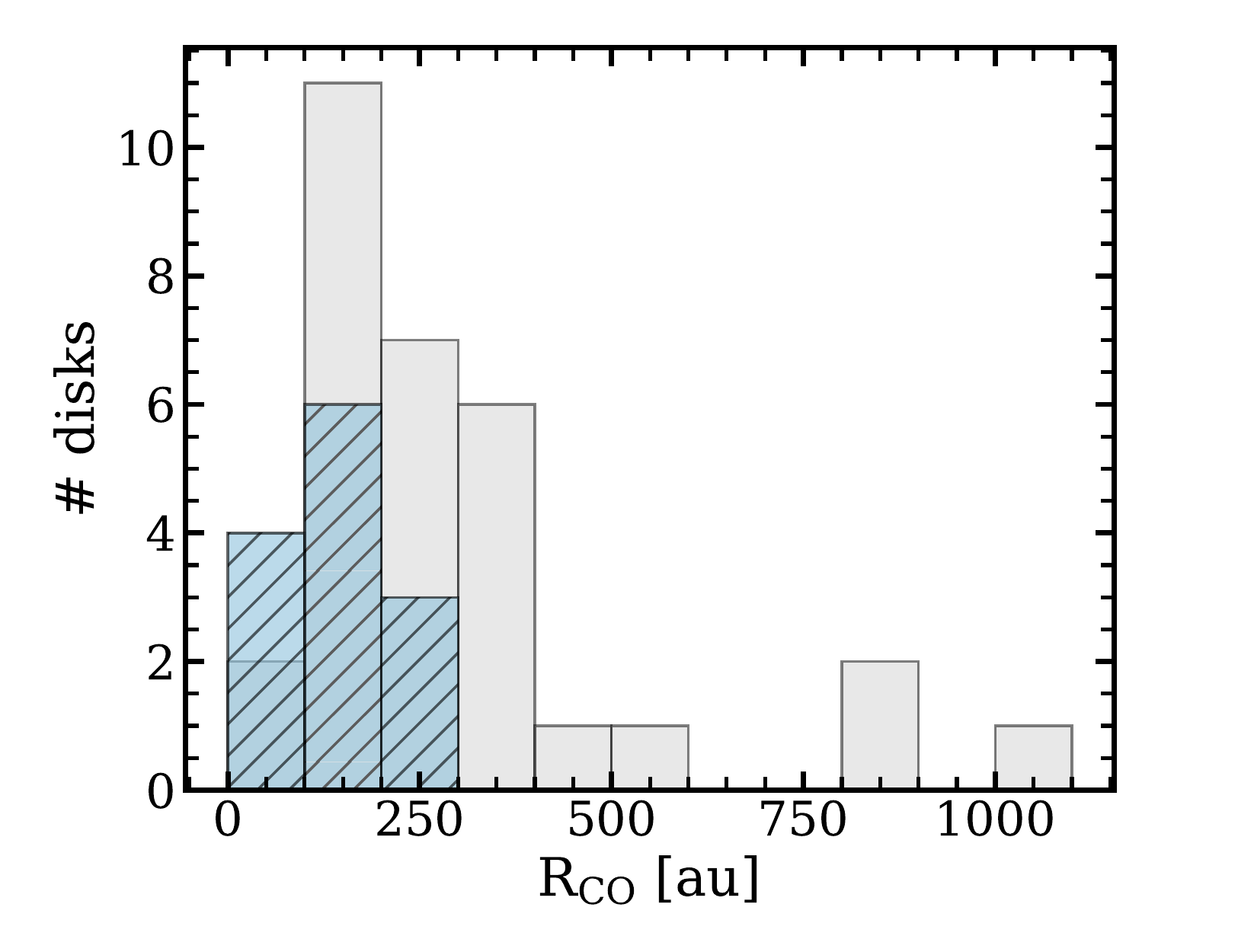} 
    \includegraphics[width=0.32\linewidth]{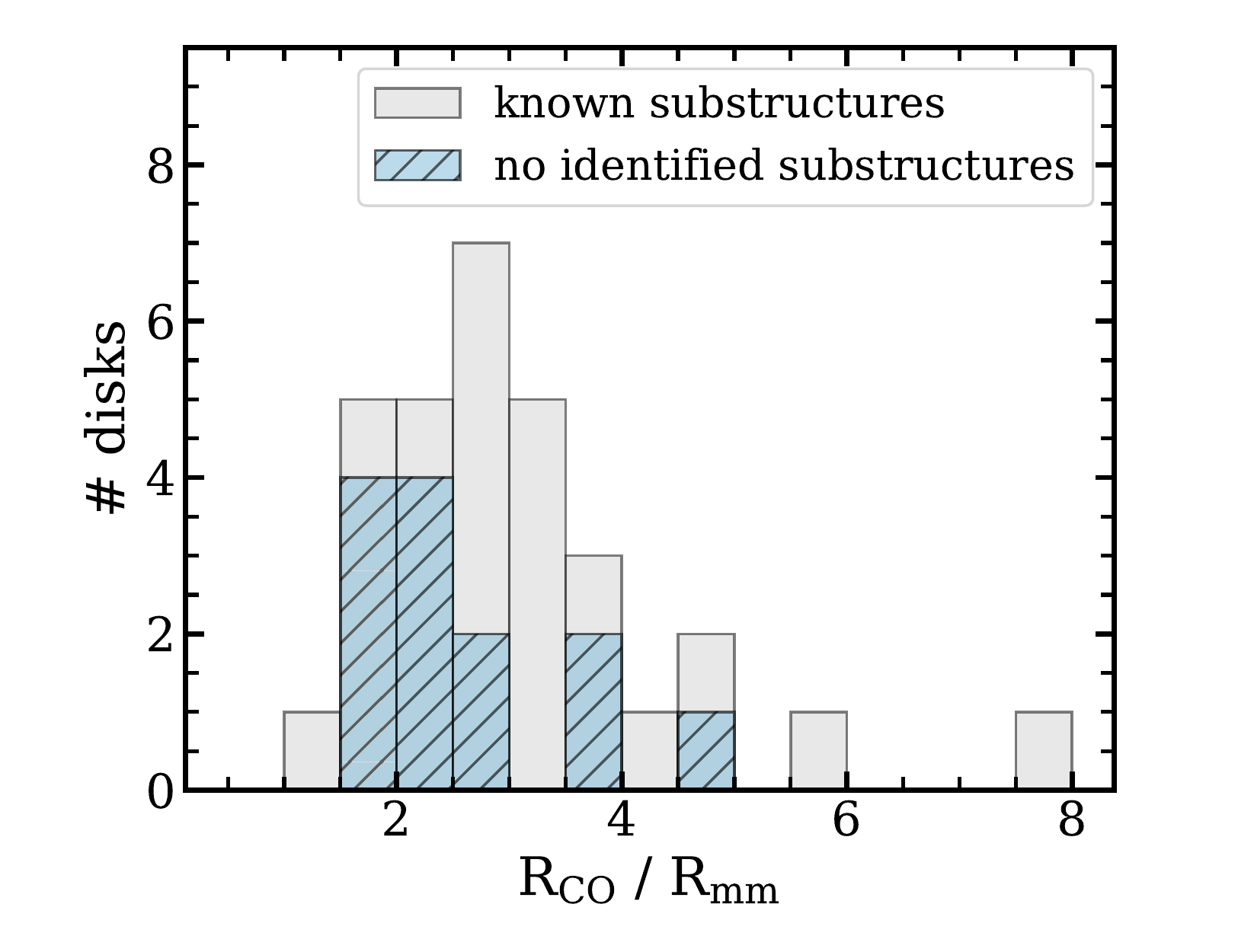} 
\caption{The distribution of $R_{\rm mm}$, $R_{\rm CO}$, $R_{\rm CO}$/$R_{\rm mm}$ for disks with observed dust substructures (grey) and disks that are featureless at the resolution of currently available data (blue). The latter group of disks are more compact, but their CO-to-continuum size ratios are comparable to disks with substructures. \label{fig:ratio_hist}}
\end{figure*}

\subsection{CO-to-Continuum Disk Size Ratio} \label{sec:ratio}
The sample presented in this work has universally larger $R_{\rm CO}$ than $R_{\rm mm}$, with the $R_{\rm CO}/R_{\rm mm}$ value spanning from 1.3 to the extreme case of 7.6. Most disks have $R_{\rm CO}/R_{\rm mm}$ of 2--4, providing an average ratio of 2.9$\pm$1.2 (Figure~\ref{fig:ratio}). Note that the measurements for 40\% of the sample only provide a lower limit on $R_{\rm CO}/R_{\rm mm}$, because $R_{\rm CO}$ is underestimated when cloud absorption is observed around the systemic velocity and/or $R_{\rm mm}$ could be overestimated when the continuum emission is not well resolved (those disks are marked as grey triangles in Figure~\ref{fig:ratio}). 
Previous observations of individual bright disks and the ALMA-Lupus survey have demonstrated a general feature that the CO sizes are 2--3 time more extended than the mm continuum sizes \citep{Isella2007, Panic2009, Andrews2012, Ansdell2018}. Our analysis confirms and extends this finding in a sample with wider ranges of stellar and disk properties.
The CX Tau disk, with $R_{\rm CO}/R_{\rm mm}$ of 3.9, was considered as one of the most extreme cases \citep{Facchini2019}. In the sample presented here, we have six additional disks with higher $R_{\rm CO}/R_{\rm mm}$ (another disk has ratio comparable to CX Tau). These disks with high $R_{\rm CO}/R_{\rm mm}$ cover the full range of stellar mass and disk mm luminosity (see Figure~\ref{fig:ratio}).

We found no relationship between $R_{\rm CO}/R_{\rm mm}$ with either stellar mass or disk mm luminosity. 
Theoretical calculation of dust evolution predicts higher drift velocity in disks around lower mass stars \citep{Pinilla2013}, thus larger $R_{\rm CO}/R_{\rm mm}$ is expected for these disks. The lack of correlation suggests that pressure bumps may form in individual disks at certain radii or time that regulate the inward drift and break any potential relationships. 


We explore here any relationship between the size ratio and the presence of substructures in disks. 
Within this sample, dust substructures have been identified in 31 disks, in the forms of inner cavities, gaps and rings, and/or spiral patterns (see corresponding references in Table~\ref{tab:data_prop}). In the remaining 13 disks\footnote{The so-far featureless disks include CX Tau, FP Tau, CIDA 7, J0415, J0420, Sz 65, Sz 73, Sz 75, Sz 76, J1556, J1600, Sz 133, and J1000, in which Sz 65, Sz 75, J1600 and Sz 133 are highly inclined systems.}, any substructure would be difficult to identify with the current data; four disks have large inclination angles ($>60\degr$) and all but CX Tau in this subsample have $R_{\rm mm}$ that is less than twice the data resolution. Figure~\ref{fig:ratio_hist} shows the histograms of $R_{\rm mm}$, $R_{\rm CO}$, and $R_{\rm CO}/R_{\rm mm}$ by dividing the sample based on the observed presence of dust substructures. 
In the 31 disks of this sample where dust substructures have been identified, we found an average $R_{\rm CO}/R_{\rm mm}$ of 3.0$\pm$1.3. Though the remaining 13 disks are in general more compact from both the dust and gas components, their average $R_{\rm CO}/R_{\rm mm}$ (2.6$\pm$0.9) is comparable within uncertainties to that of the substructure group (see Figure~\ref{fig:ratio_hist}).
Those with high $R_{\rm CO}/R_{\rm mm}$ include both extended substructure disks (GO Tau, DM Tau, V4046 Sgr, UZ Tau, DL Tau, and MHO 6) and compact disks without large-scale substructures (CX Tau and CIDA 7). More specifically, the two largest CO disks, GO Tau and DM Tau, both surround M3 stars and host extended continuum disks with multiple gaps and rings detected \citep{Long2018, Hashimoto2021}. Around the same type of host star, the continuum disk of CX Tau only extends to 30\,au and remains featureless with a high resolution of 5\,au \citep{Facchini2019}.  
Our findings suggest that $R_{\rm CO}/R_{\rm mm}$ may not strongly depend on the detailed dust distribution in disks (or every disk is highly structured and $R_{\rm mm}$ is largely determined by where the last pressure bump can be built in the disk, so does $R_{\rm CO}/R_{\rm mm}$). 
The evolution of $R_{\rm CO}$ may also differ in different systems so that $R_{\rm CO}/R_{\rm mm}$ is not a simple reflection of the dust radial drift efficiency.


\section{Discussion} \label{sec:diss}
In this section, we first discuss if CO disk sizes can reveal the dominant disk evolution mechanism, assuming $R_{\rm CO}$ well traces the disk gas component. We then explore how CO-to-continuum size ratios probe the evolution and distribution of disk solids. Finally, we employ the GO Tau disk to demonstrate the influence of a large gas disk on dust evolution.

\subsection{Gas Disk Evolution}
In the classical theory of disk viscous evolution, turbulence transports angular momentum outward and enables disk material to be accreted onto the star \citep[e.g.,][]{Hartmann2016}. This re-distribution of angular momentum leads to the growth of the disk size. The expansion rate largely depends on the turbulent viscosity. \citet{Trapman2020_viscous} recently explored how the disk sizes (using the same definition as 90\% fractional radius) measured from $^{12}$CO emission vary with time in viscous evolution models using a simplified $\alpha-$prescription for the viscosity \citep{Shakura1973, Lynden-Bell1974}. This model involves three key parameters: viscosity coefficient $\alpha$, initial disk mass $M_{\rm d,0}$, and initial characteristic disk radius $R_{\rm c,0}$. For a viscous disk, $M_{\rm d,0}$ is related to the stellar accretion rate \citep{Hartmann1998}, which scales with the central stellar mass. Therefore, we considered two sets of models with stellar masses of 0.32 and 1\,$M_{\odot}$, and adopted the average stellar accretion rate among the corresponding host stars.

\begin{figure*}[!t]
\centering
    \includegraphics[width=0.45\linewidth]{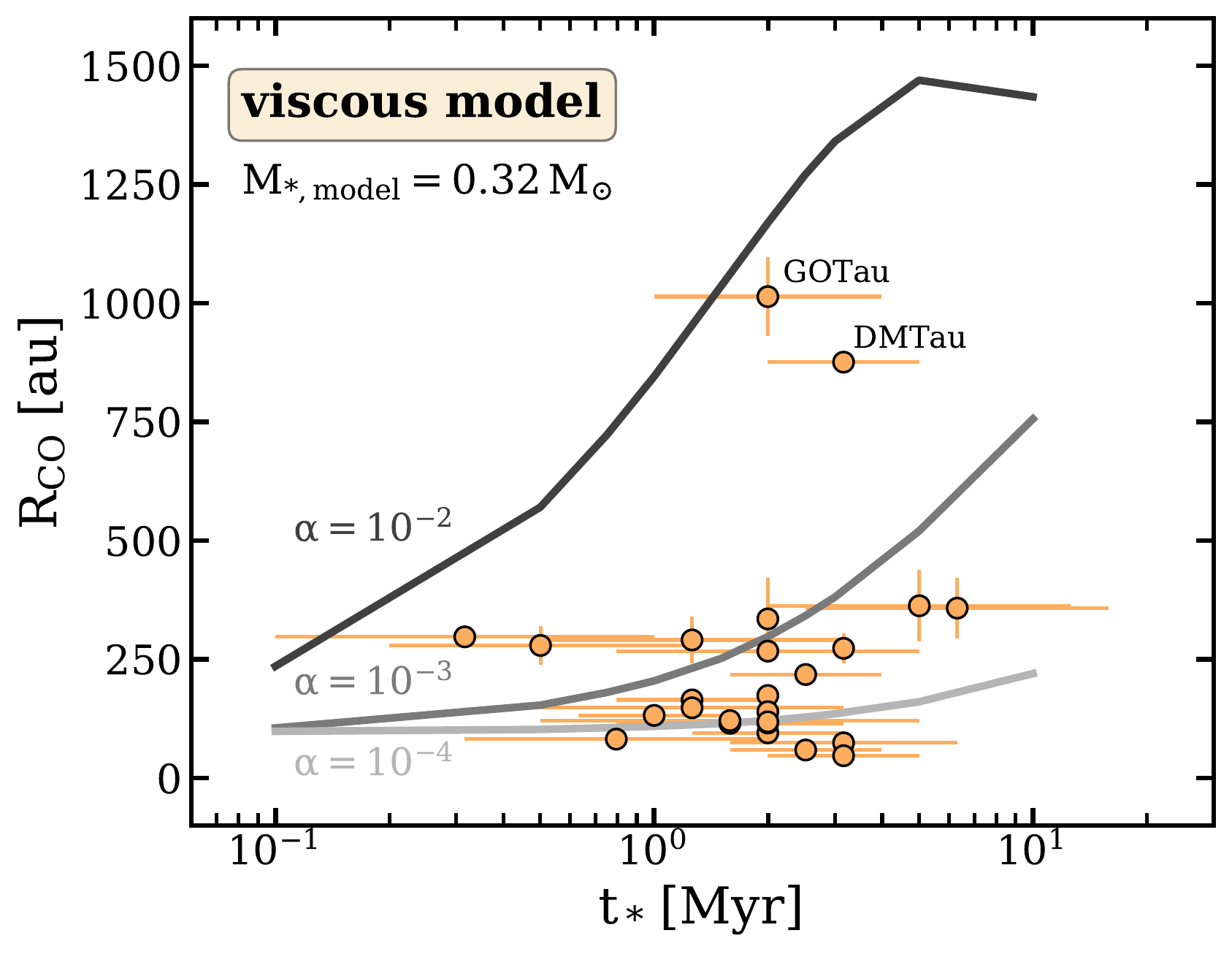}
    \includegraphics[width=0.45\linewidth]{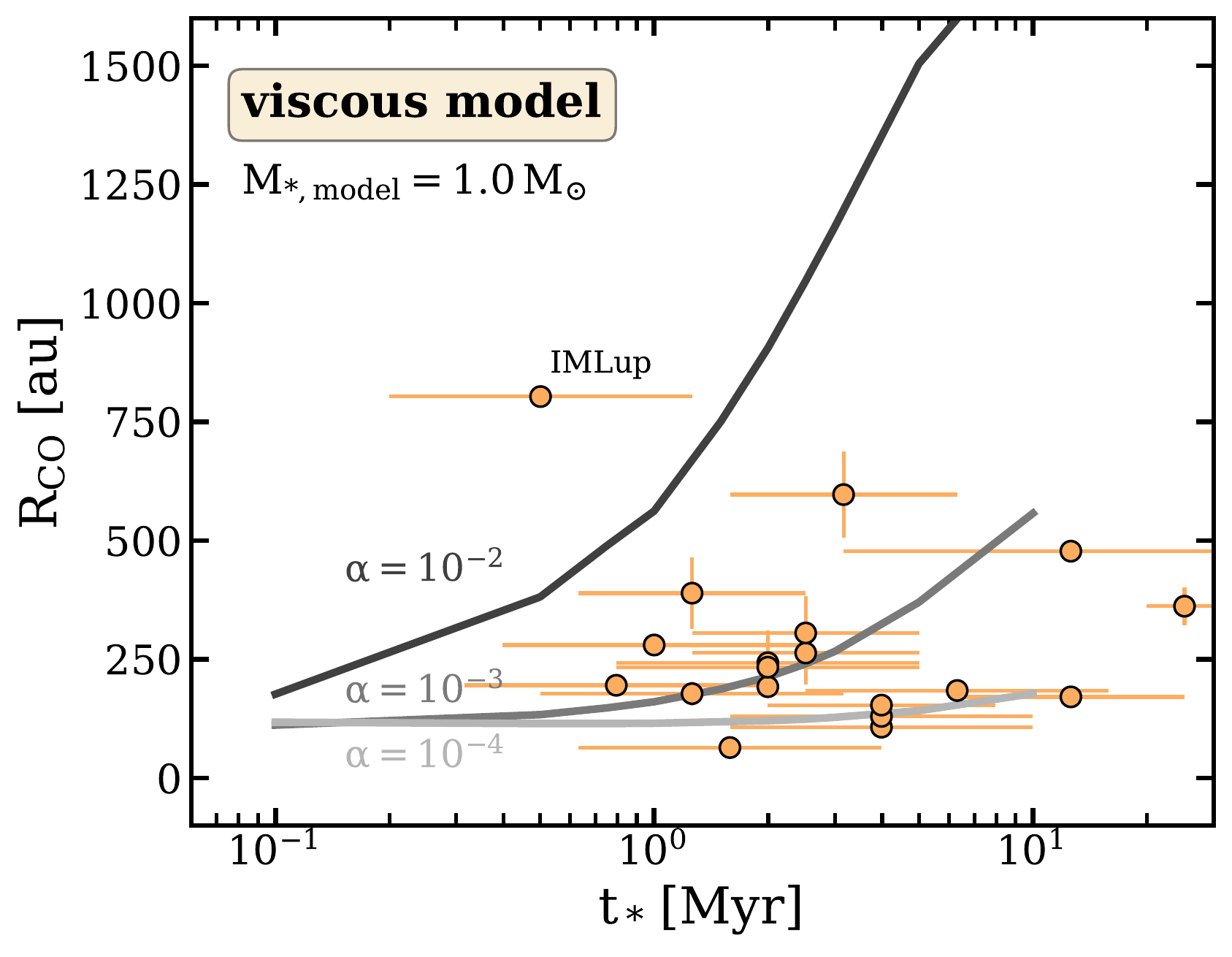} \\
\caption{The comparison of observed CO disk sizes (dots) with disk viscous evolution models from \citet{Trapman2020_viscous} that adopt the same size definition (lines). Models with three values of $\alpha$ viscosity parameter and two values of stellar mass are presented. The initial characteristic radius in the model is $R_{\rm c,0}$=10\,au. Systems with $M_{*}<0.7M_{\odot}$ are compared to models with $M_{*}=0.32\,M_{\odot}$ (\textbf{left}), and the remaining systems are compared to models with $M_{*}=1.0\,M_{\odot}$ (\textbf{right}). \label{fig:alpha-model}}
\end{figure*}

A comparison of the CO disk sizes derived here with model grids from \citet{Trapman2020_viscous} suggests that most cases can be explained by viscous evolution models with low $\alpha$ in the range of 10$^{-4}$--10$^{-3}$ and small $R_{\rm c,0}$=10\,au (Figure~\ref{fig:alpha-model}). 
The three most extended gas disks in this sample (GO Tau, DM Tau, and IM Lup) are better described by models with high $\alpha$ ($\sim$0.01), though the large $R_{\rm CO}$ of IM Lup was suggested to result from external photoevaporation in a weak radiation field \citep{Haworth2017}.
Measuring the spectral line broadening provides observational constraints on disk turbulence, which suggests that weak turbulence ($\alpha<10^{-3}$) might be common in disks \citep{Guilloteau2012, Flaherty2015, Flaherty2018, Teague2018_CS}. DM Tau is the only case with measurable turbulence in the outer disk \citep{Flaherty2020}; such measurements are not yet available for GO Tau and IM Lup.
An alternative solution is to start with larger initial sizes. Disk models with $R_{\rm c,0}=$ 50\,au can spread to $R_{\rm CO}=$ 500\,au with a lower $\alpha$ of 10$^{-3}$ by 1\,Myr \citep{Trapman2020_viscous}. Future constraints on turbulence from nonthermal line broadening in these extended disks will potentially distinguish the two scenarios. 

\begin{figure*}[!t]
\centering
     \includegraphics[width=0.45\linewidth]{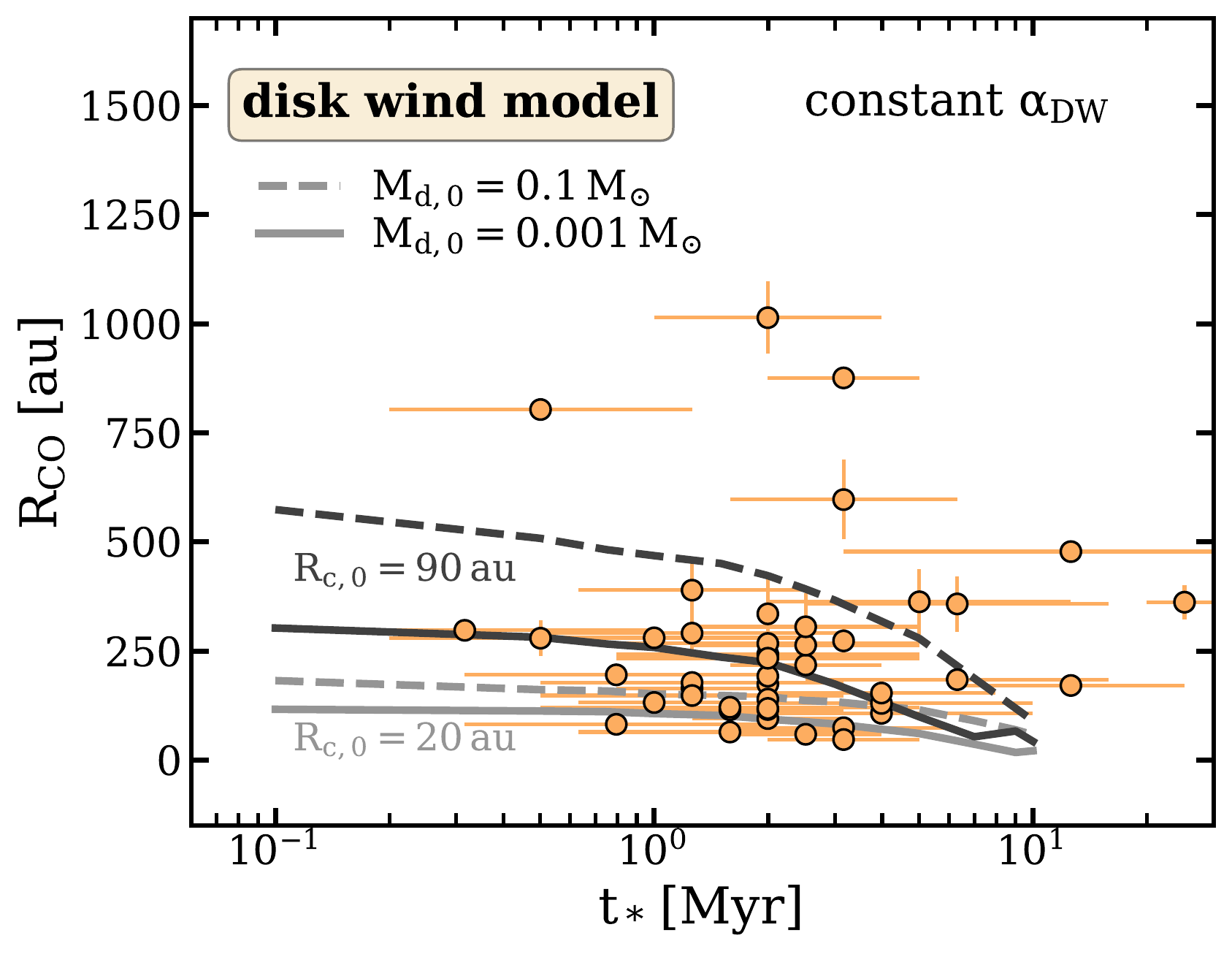}
    \includegraphics[width=0.45\linewidth]{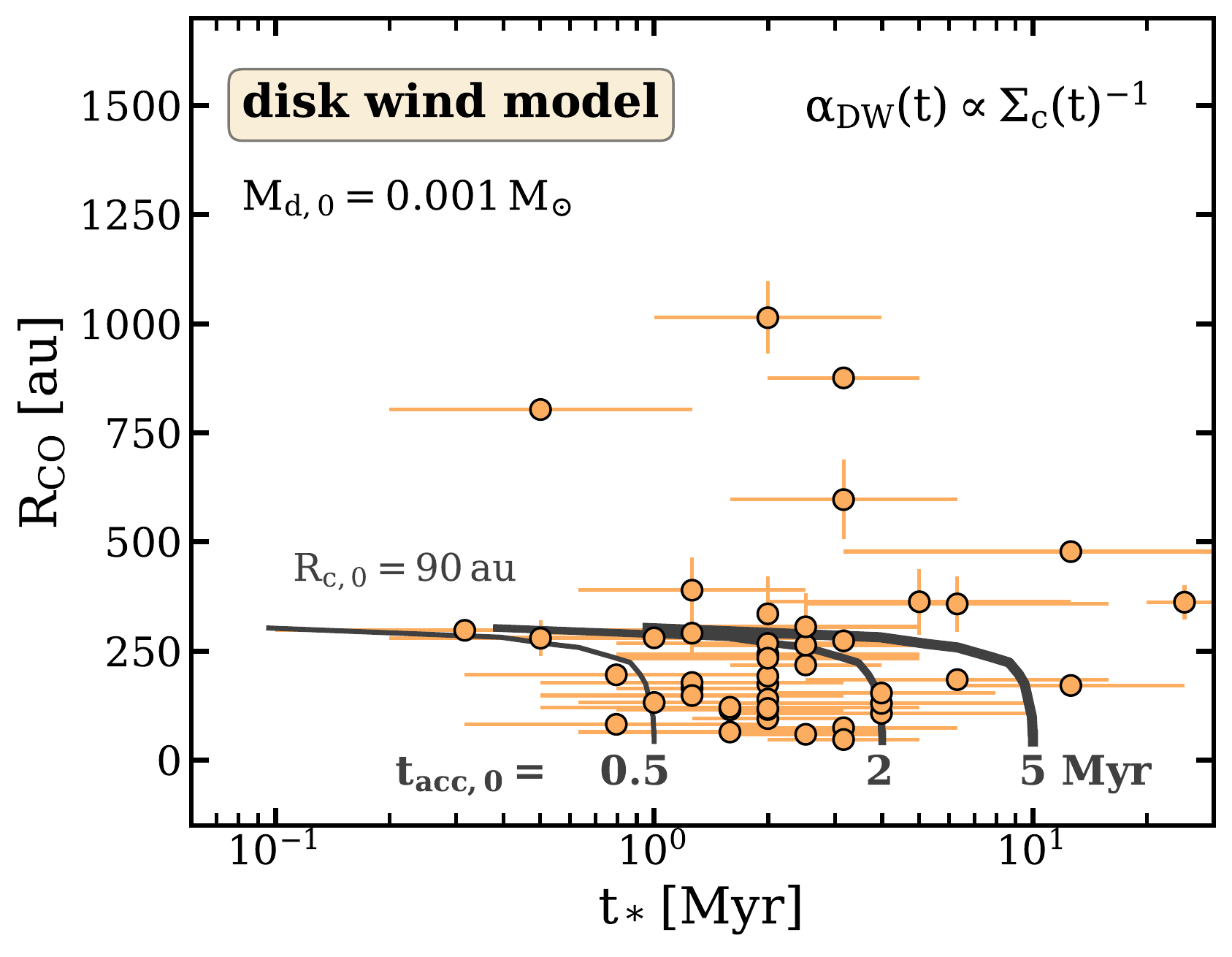} \\
\caption{The comparison of observed CO disk sizes (dots) with MHD disk wind models from \citet{Trapman2021arXiv} that adopt the same size definition (lines). In contrast to viscous models, $R_{\rm CO}$ decreases with stellar ages in a wind-driven scenario (constant $\alpha_{DW}$, \textbf{left}), while the decrease of $R_{\rm CO}$ slows down at late times when considering a time-dependent $\alpha_{DW}$ before disk disperses (\textbf{right}). Models with different $R_{\rm c,0}$ are indicated by the gradients of the color scale and different initial disk masses $M_{\rm d,0}$ are indicated by the line styles. In the right panel, three models with different initial accretion timescale $t_{\rm acc,0}$ are shown with different line width. \label{fig:wind-model}}
\end{figure*}

The magneto-rotational instability (MRI, \citealt{Balbus1998}) is taken as the leading mechanism in generating the needed turbulence for the viscous transport of disk angular momentum. However, the full operation of MRI throughout the disk is often questioned due to the weak ionization conditions in large disk areas (e.g., \citealt{Turner2014}). Numerical simulations with proper treatments of non-ideal MHD effects demonstrate that MRI turbulence is largely suppressed in the cold, low-ionization disk regions (e.g., \citealt{Bai2013, Gressel2015}). The magnetized disk wind concept \citep{Blandford1982} has thus re-emerged as a promising alternative (e.g., \citealt{Bai2016, Lesur2021}). In this wind-driven accretion scenario, angular momentum is not transported within the disk but directly removed through MHD disk winds. Therefore the sizes of gas disks need not expand to enable mass accretion. 

Following the $\alpha$-framework for viscous evolution, a simple parameterized description of disk evolution for a MHD disk wind was recently provided by \citet{Tabone2021a_arXiv}.
This introduced a similar dimensionless parameter $\alpha_{DW}$ that describes the wind torque and relates to the local accretion rate driven by the wind. This framework therefore controls disk evolution with three key parameters: $\alpha_{DW}$, initial disk mass $M_{\rm d,0}$, and initial characteristic disk radii $R_{\rm c,0}$ (the accretion timescale $t_{\rm acc,0}$ is related to $\alpha_{DW}$ and $R_{\rm c,0}$, which will be used in the following discussion). 
Based on the analytical solution from \citet{Tabone2021a_arXiv}, \citet{Trapman2021arXiv} then examined how $R_{\rm CO}$ evolves in the pure disk wind scenario (ignoring the viscous term).
When considering a constant $\alpha_{DW}$, $R_{\rm CO}$ gradually declines with time (between 0.1 to 10\,Myr, as shown in the left panel of Figure~\ref{fig:wind-model}), which is in direct contrast to the case of viscous evolution. 
The declining rate of $R_{\rm CO}$ depends on the choice of $M_{d,0}$ and $R_{c,0}$, but in general models with higher $M_{d,0}$ and larger $R_{c,0}$ result in larger $R_{\rm CO}$ at specific time steps. Therefore, disk wind models with different $M_{d,0}$ (from 0.0001 to 0.1\,$M_{\odot}$) and $R_{c,0}$ (from 20 to 90\,au) can explain most of our disk sample. As $R_{c,0}$=90\,au is the most extreme case considered in the models of \citet{Trapman2021arXiv} (guided by the Lupus survey, \citealt{Ansdell2018}), the few very extended disks and the disks around older systems in the sample are difficult to account for without even larger initial disk sizes.

The time evolution of $R_{\rm CO}$ in this disk wind scenario largely depends on the rate at which the disk surface density decreases. If we consider a time-dependent $\alpha_{DW}$ that scales with $\Sigma_{c}(t)^{-1}$, the decrease of $R_{\rm CO}$ becomes much shallower until a sharp drop occurs when the disk starts to dissipate (at 2$t_{\rm acc,0}$, \citealt{Tabone2021a_arXiv}). This difference is demonstrated in the two panels of Figure~\ref{fig:wind-model}. In this time-dependent case, the evolution of $R_{\rm CO}$ is mostly determined by the ratio between $t_{\rm acc,0}$ and $t_{*}$. Therefore, old disks with large $R_{\rm CO}$ of 200--300\,au may be initialized with large $t_{\rm acc,0}$ and evolve with varying $\alpha_{DW}$.  Considering a distribution of $t_{\rm acc,0}$ in any individual cluster that follows the evolution of disk fraction with cluster age (e.g., \citealt{Fedele2010}), \citet{Tabone2021b_arXiv} show that the correlation between mass accretion rate and disk mass, as well as its large scatter, can be well reproduced by MHD disk wind models with time-dependent $\alpha_{DW}$. It is likely that these old and large disks represent the long-lived outliers (those with long initial accretion timescale) in individual star-forming regions (e.g, TW Hya).

The comparisons above suggest that the measured CO sizes for this disk sample can be explained by either viscous models or MHD disk wind models, assuming varying initial conditions for individual disks. As shown in Figure~\ref{fig:alpha-model} and \ref{fig:wind-model}, distinguishing between the two scenarios requires a more complete sample of older systems (see also Figure~\ref{fig:age-mass}). Considering the typical disk dispersal timescale, the characterization of such disks will need deep observations (the ongoing ALMA Cycle 8 Large Program AGE-PRO will hopefully provide such measurements). The inclusion of a large sample is also necessary to account for any possible biases in the distribution of initial conditions.

With uniformly determined CO sizes, the sample of Class II disks presented here do not show any trend of $R_{\rm CO}$ with evolution (with stellar ages from 0.5 to 20\,Myr, Figure~\ref{fig:LR}f).
Our small number statistics (especially for older disks) limit the evidence for disk viscous evolution during the Class II phase. 
However, in a comparison of gas disk sizes between the younger Class 0/I and more evolved Class II disks, \citet{Najita2018} found that gas disk sizes of Class II disks are systematically larger and explained this as a result of viscous spreading. Given the lack of large disks with $R_{\rm CO}>800$\,au at $t_{*}<0.5$\,Myr ($R_{\rm CO, Class\,0/I}\sim$ 50--300\,au, e.g., \citealt{Harsono2014})\footnote{Extended gas emission has been reported for a few Class I systems (e.g., DG Tau B, \citealt{deValon2020}), while the lack of proper modeling that takes into account the envelope emission makes the inference of a Keplerian disk size challenging.}, MHD disk wind models alone would be challenging to explain the extended gas distribution in some Class II objects. We speculate that disks evolve in a hybrid mode, where substantial growth of sizes may occur early on for some systems to account for the overall size differences at different evolutionary stages. For individual disks, different mechanisms may dominate the evolution at different times, depending on the physical conditions in the disk and of the large-scale environment. It also remains unclear how the mass and angular momentum transfer from the envelope impacts the disk size evolution, as the envelope is not treated in the current model framework.
One important caveat of the comparison in \citet{Najita2018} lies in the usage of various gas tracers and methods in calculating the disk size with literature results directly adopted. In addition, the difficulties in disentangling the disk and envelop emission result in only a small number of Class I sources with well-determined (Keplerian-rotating) disk radii. High spatial and spectral resolution ALMA observations towards these younger systems in the coming years will certainly enlarge the sample. A more robust test of viscous spreading would then be possible with sufficient lever arm in stellar age as well as comparable stellar mass distributions across the age spectrum.

\subsection{$R_{\rm CO}$/$R_{\rm mm}$ Tracing Dust Evolution}

Our analysis shows that $R_{\rm CO}$/$R_{\rm mm}$ can vary from 1.3 to 7.6, with an average ratio of 2.9$\pm$1.2 (see Sec~\ref{sec:ratio}). Such size differences have been widely attributed to two effects: emission optical depth and dust evolution. The $^{12}$CO line optical depth is always much higher than the nearby continuum emission \citep[e.g.,][]{BeckwithSargent1993}. The observed size discrepancy can simply be a manifestation of this optical depth difference, since the thicker line emission is easier to detect at low densities in the outer disk \citep{Facchini2017}.

From the dust evolution perspective, dust particles in disks will grow to larger sizes, and the shorter grain growth timescale at smaller radial distances results in a concentration of larger grains in the inner disk \citep[e.g.,][]{Testi2014}.  
Meanwhile, once grains in the outer disk reach some critical sizes, the gas drag imposed by gas-dust rotational velocity difference will push the large grains to move inward toward higher pressure regions \citep[][]{Weidenschilling1977}. The combination of the two processes means that large grains preferentially accumulate in the inner disk. 
The millimeter continuum emission tracing these particles will therefore appear much more compact than the gas disk. In addition, \citet{Rosotti2019} have suggested that small grains (with sizes smaller than 100\,$\mu$m), which are still well-coupled with the gas and present at large radial distances, could not produce sufficient mm continuum emission due to a sharp dust opacity drop.

The two effects of optical depth and dust evolution act in concert. However, their relative importance can not be easily retrieved without detailed knowledge of the gas surface density and other disk conditions that affect the grain growth and drift efficiencies (e.g., viscosity). Thermochemical models for the HD163296 disk have demonstrated that its $R_{\rm CO}$ and $R_{\rm mm}$ difference can be largely explained by the optical depth effect \citep{Facchini2017}. Using a grid of thermochemical models with varying disk conditions, \citet{Trapman2019} suggest that $R_{\rm CO}$/$R_{\rm mm} > 4$ is a clear sign of dust evolution in action, regardless of the detailed stellar/disk properties. About 18\% (8/44) of the disks in our sample have size ratio around or above 4. Presumably these disks have experienced substantial grain growth and radial drift. We recall that these high ratio systems do not occupy any preferred parameter space of stellar mass and disk millimeter brightness. 
Therefore, it remains unclear how $R_{\rm CO}$/$R_{\rm mm}$ can be better employed as a generic diagnostic of dust evolution. 
Through quantitative comparisons to thermochemical models without dust radial drift for a few individual systems, \citet{Trapman2020_Lupus} also revealed that dust evolution is required in five Lupus disks to account for the observed disk size differences with $R_{\rm CO}$/$R_{\rm mm}$ of 1.8--3; all of them are included in this sample.

Based on theoretical models, radial drift is expected to proceed in a short timescale (e.g., \citealt{TakeuchiLin2002}, \citealt{Brauer2008}), which will lead to a very rapid shrinking of $R_{\rm mm}$. \citet{Toci2021} recently explored the secular evolution of dust and gas disk sizes with a set of model grids, considering grain growth, radial drift, and disk viscous evolution under smooth gas distributions. 
In model disks with $R_{\rm c}$=10\,au surrounding solar-type stars, $R_{\rm CO}$ reaches 50--300\,au after 1--2\,Myr, while $R_{\rm mm}$ shrinks to 10--30\,au, leading to universally high $R_{\rm CO}$/$R_{\rm mm}$ ($>5$). Though $R_{\rm CO}$ in these models match with the peak distribution of the disk sample presented in this study (see Figure~\ref{fig:ratio_hist}), $R_{\rm mm}$ are too small compared to observations.

The presence of local gas pressure maxima, due to either the influence of planets or fluid instabilities, will stop or slow down the dust inward drift (e.g., \citealt{Pinilla2012_bump}).
About two thirds of the disks in this sample show visible dust substructures, which may trace pressure bumps that help sustain large grains at large radial distances. The remaining sample, though with highly comparable $R_{\rm CO}$ to models, show a wide distribution of $R_{\rm mm}$ ($15\sim150$\,au) and low size ratios mainly clustering around 2 (Figure~\ref{fig:ratio_hist}).
None of the observations toward these disks are capable of detecting substructures with sizes comparable to the gas pressure scale height in their outer disk. Substructures may hide in current data for these disks given their $R_{\rm mm}$ and $R_{\rm CO}$/$R_{\rm mm}$. 

\begin{figure}[!t]
\centering
    \includegraphics[width=0.95\linewidth]{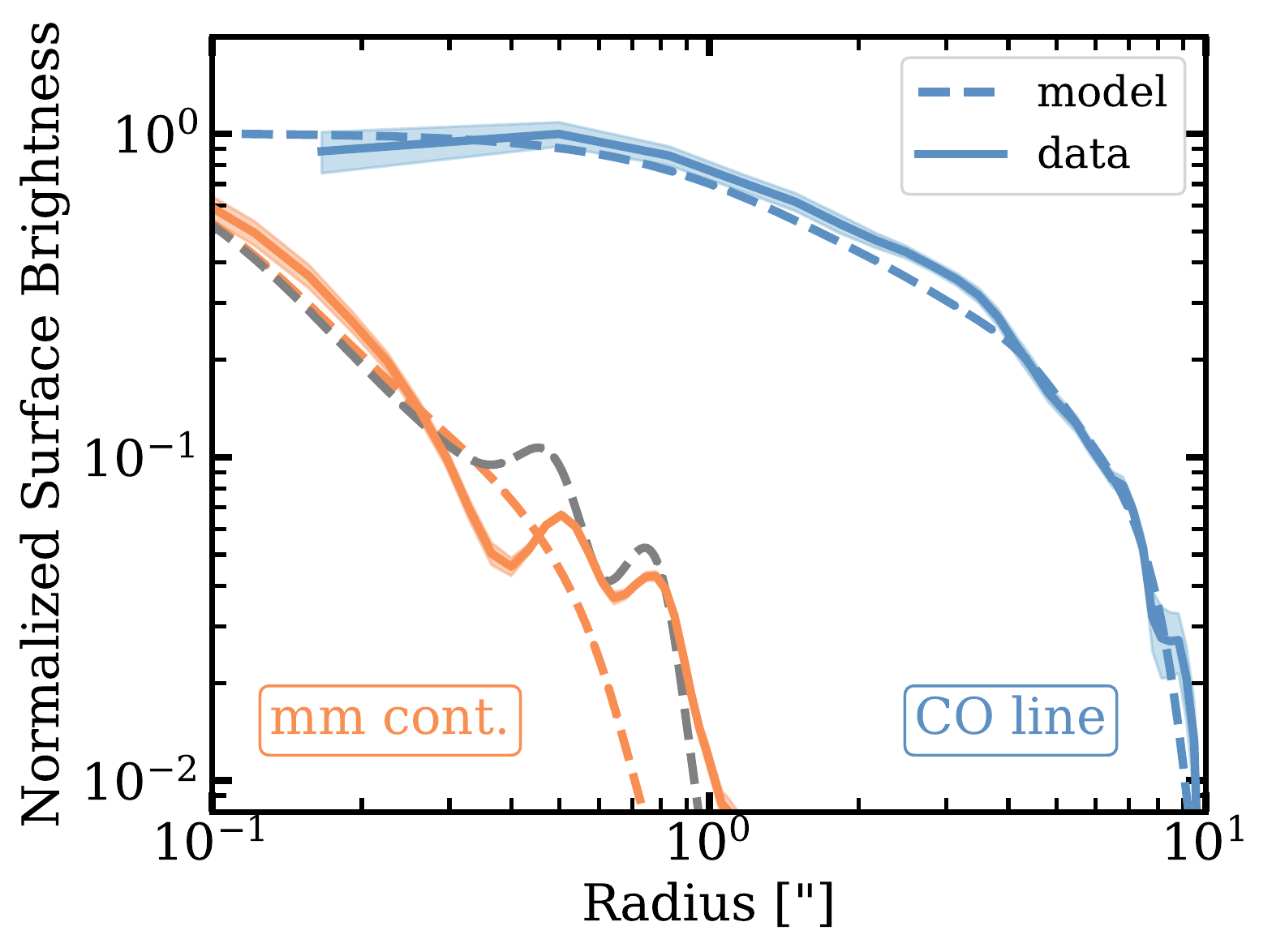}
\caption{The comparison of GO Tau data radial profiles with model profiles convolved with our data resolution (blue for the CO emission and orange for the mm continuum emission). This fiducial model with a smooth gas distribution is directly adopted from \citet{Toci2021}, selected to roughly match the large CO emission size of GO Tau. The grey dashed line shows the revised model with two pressure bumps added on top of the fiducial model aiming at reproducing the two observed dust rings (the corresponding CO profile is almost identical to the fiducial model). \label{fig:gotau-model}}
\end{figure}

\subsection{The Most Extended CO disk of GO Tau}


The GO Tau disk has the largest $R_{\rm CO}$ within this sample, approaching $\sim$1000\,au. Viscous spreading likely plays a leading role in the past evolution for such extended disks (including e.g., DM Tau). Future measurements of the line broadening will help assess the disk turbulence level and therefore constrain some initial disk conditions. We also note that both GO Tau and DM Tau have comparably faint optical \oi{} 6300\,$\mathring{\rm A}$ line emission and without high velocity line component, indicating the absence of strong disk winds \citep{Simon2016}. 

Such a large gas disk will affect the dust evolution in GO Tau. As a point of comparison, we selected a model from the \citet{Toci2021} grids that has a similar $R_{\rm CO}$ to the GO Tau disk. We also re-ran the simulation with a perturbed version of this model, with two pressure bumps added to mimic the morphology of the mm continuum emission rings at 73 and 110\,au, respectively. As seen in Figure~\ref{fig:gotau-model}, the radial profile from the pressure bump model matches well with the data. However, even without gas pressure bumps, the simulated mm continuum emission extends beyond the first prominent dust ring, preserving a large $R_{\rm mm}$. 
Therefore, a large $R_{\rm mm}$ is not necessarily set by the presence of pressure bumps. It could also result from the extended gas distribution (or a large initial disk size), where grain growth takes longer and the drifting particles have just reached the current mm disk outer edge. The models presented here are not fine-tuned to match the details of the GO Tau disk, but are used for a qualitative demonstration. While a low $R_{\rm CO}$/$R_{\rm mm}$ ratio suggests the presence of pressure bumps within the dust disk, a high ratio, like the case of GO Tau, only signifies prominent dust evolution outside the dust disk.  This latter scenario could also be applied to the disks of CX Tau and CIDA 7, both of which have high $R_{\rm CO}$/$R_{\rm mm}$ ratio. Their lack of dust substructures can be explained if the growth and subsequent radial migration of dust grains in these systems take quite a long time, though substructures may be present but harder to identify in these compact disks.

\section{Summary} \label{sec:sum}

We presented a joint analysis of disk sizes for a sample of 44 protoplanetary disks around central stars with masses of 0.15--2\,M$_{\odot}$. The gas and dust disk sizes were calculated as the 90\% flux fractional radii from the $^{12}$CO line and millimeter continuum radial profiles. This sample covers a wide range in $R_{\rm mm}$ of 15--250\,au and $R_{\rm CO}$ of 50--1000\,au. Based on the distribution of the sample in the $R_{\rm mm}$--$L_{\rm mm}$ plane, it is more representative of the brighter, larger disk population. 
We then explored how $R_{\rm CO}$ and $R_{\rm CO}$/$R_{\rm mm}$ are related to various stellar and disk properties, and considered their behavior in the context of disk evolution mechanisms. Our main results are summarized as follows.

\begin{enumerate}

\item $R_{\rm CO}$ in this sample shows no evolution with stellar age within 0.5--20\,Myr. Both viscous and MHD disk wind models can explain the sizes of individual disks in this sample, when considering varying initial conditions (e.g., initial disk characteristic radius, disk mass, accretion timescale). Disks with very large $R_{\rm CO}$ are more readily explained in the context of the viscous models. 
Though the lack of apparent $R_{\rm CO}$ evolution in this sample could be more consistent with the wind-driven accretion scenario, a more robust test would require a larger population study, especially towards both younger (Class I) objects and more evolved systems.

\item The disks in this sample have universally larger $R_{\rm CO}$ than $R_{\rm mm}$. The measured $R_{\rm CO}$/$R_{\rm mm}$ vary from 1.3 to 7.6, with an average value of 2.9$\pm$1.2. In 8/44 disks, high $R_{\rm CO}$/$R_{\rm mm}$ values 
($>\sim$4) suggest that substantial grain growth and radial drift have already occurred. However, this sub-sample does not show any preference in stellar and disk properties. 
A significant fraction of the sample disks (31/44, 70\%) exhibit dust substructures, which could mitigate dust radial drift and naturally explain their low $R_{\rm CO}$/$R_{\rm mm}$. The low disk size ratios in many currently featureless disks may then imply the presence of yet-unseen disk substructures.

\item Among this sample, the GO Tau disk stands out as an extreme outlier, with $R_{\rm CO}$ extending to 1000\,au and $R_{\rm CO}$/$R_{\rm mm}$ of 7.6. There are two possible explanations for such an extended gas distribution: 1) a disk with an initial characteristic radius $R_{\rm c,0}$=10\,au experienced significant viscous spreading with $\alpha\sim0.01$; 2) the disk was born large ($R_{\rm c,0}\gtrsim$50\,au). Measuring the non-thermal spectral line broadening that constrains the disk turbulence would be a promising approach to discern the two possibilities. 
Though GO Tau has two dust rings near the outer continuum edge, these substructures are not the prerequisite to sustain its large dust disk ($\sim$130\,au) due to the extended gas emission. 
From the dust evolution perspective, smooth disks with a high $R_{\rm CO}$/$R_{\rm mm}$ are not prohibited if the gas disks were initially large enough. 

\end{enumerate}

We are still at the early stage in exploring the gas content of planet-forming disks and their roles in tracing disk evolution, especially considering the small number of disks that have high quality molecular line data. Improved investments in ALMA gas observations are fundamental in addressing many key questions of disk evolution and planet formation.


\paragraph{Acknowledgments}
We thank the referee for careful reading of this manuscript and the useful comments.
F.L. thanks Kevin Flaherty, Charlie Qi, Jamila Pegues, and Nico Kurtovic for their kindness in sharing some of the data. F.L. thanks Claudia Toci in providing their model results, and Enrique Sanchis for sharing the CO fluxes of Lupus disks.  F.L. is grateful to Xuening Bai for insightful discussions. 
F.L. and R.T. acknowledge support from the Smithsonian Institution as the Submillimeter Array (SMA) Fellow. F.L. and S.A. acknowledge funding support from the National Aeronautics and Space Administration under grant No.17-XRP17$\_$2-0012 issued through the Exoplanets Research Program. G.R. acknowledges support from the Netherlands Organisation for Scientific Research (NWO, program number 016.Veni.192.233) and from an STFC Ernest Rutherford Fellowship (grant number ST/T003855/1). D.H. is supported by CICA through a grant and grant number 110J0353I9 from the Ministry of Education of Taiwan. P.P. acknowledges support provided by the Alexander von Humboldt Foundation in the framework of the Sofja Kovalevskaja Award endowed by the Federal Ministry of Education and Research.

The Submillimeter Array is a joint project between the Smithsonian Astrophysical Observatory and the Academia Sinica Institute of Astronomy and Astrophysics and is funded by the Smithsonian Institution and the Academia Sinica. This paper makes use of the following ALMA data: 2015.1.00222.S, 2015.1.00678.S, 2016.1.00484.L, 2016.1.00715.S, 2016.1.00724.S, 2016.1.01164.S, 2017.1.01107.S, 2018.1.00310.S, and 2018.1.00614.S. ALMA is a partnership of ESO (representing its member states), NSF (USA), and NINS (Japan), together with NRC (Canada), MOST and ASIAA (Taiwan), and KASI (Republic of Korea), in cooperation with the Republic of Chile. The Joint ALMA Observatory is operated by ESO, AUI/NRAO, and NAOJ.

\facilities{SMA, ALMA}
 

\software{analysisUtils~(\url{https://casaguides.nrao.edu/index.php/Analysis_Utilities}), AstroPy~\citep{Astropy2013}, CASA~\citep{McMullin2007}, matplotlib~\citep{matplotlib2007}, Scipy~(\url{http://www.scipy.org)}  }

\bibliography{ms}{}

\begin{thebibliography}{}
\expandafter\ifx\csname natexlab\endcsname\relax\def\natexlab#1{#1}\fi
\providecommand{\url}[1]{\href{#1}{#1}}
\providecommand{\dodoi}[1]{doi:~\href{http://doi.org/#1}{\nolinkurl{#1}}}
\providecommand{\doeprint}[1]{\href{http://ascl.net/#1}{\nolinkurl{http://ascl.net/#1}}}
\providecommand{\doarXiv}[1]{\href{https://arxiv.org/abs/#1}{\nolinkurl{https://arxiv.org/abs/#1}}}

\bibitem[{{Andrews}(2020)}]{Andrews2020_review}
{Andrews}, S.~M. 2020, \araa, 58, 483,
  \dodoi{10.1146/annurev-astro-031220-010302}

\bibitem[{{Andrews} {et~al.}(2013){Andrews}, {Rosenfeld}, {Kraus}, \&
  {Wilner}}]{Andrews2013}
{Andrews}, S.~M., {Rosenfeld}, K.~A., {Kraus}, A.~L., \& {Wilner}, D.~J. 2013,
  \apj, 771, 129, \dodoi{10.1088/0004-637X/771/2/129}

\bibitem[{{Andrews} {et~al.}(2018{\natexlab{a}}){Andrews}, {Terrell},
  {Tripathi}, {Ansdell}, {Williams}, \& {Wilner}}]{Andrews2018_Lmm}
{Andrews}, S.~M., {Terrell}, M., {Tripathi}, A., {et~al.} 2018{\natexlab{a}},
  \apj, 865, 157, \dodoi{10.3847/1538-4357/aadd9f}

\bibitem[{{Andrews} {et~al.}(2012){Andrews}, {Wilner}, {Hughes}, {Qi},
  {Rosenfeld}, {{\"O}berg}, {Birnstiel}, {Espaillat}, {Cieza}, {Williams},
  {Lin}, \& {Ho}}]{Andrews2012}
{Andrews}, S.~M., {Wilner}, D.~J., {Hughes}, A.~M., {et~al.} 2012, \apj, 744,
  162, \dodoi{10.1088/0004-637X/744/2/162}

\bibitem[{{Andrews} {et~al.}(2018{\natexlab{b}}){Andrews}, {Huang},
  {P{\'e}rez}, {Isella}, {Dullemond}, {Kurtovic}, {Guzm{\'a}n}, {Carpenter},
  {Wilner}, {Zhang}, {Zhu}, {Birnstiel}, {Bai}, {Benisty}, {Hughes},
  {{\"O}berg}, \& {Ricci}}]{Andrews2018_dsharp}
{Andrews}, S.~M., {Huang}, J., {P{\'e}rez}, L.~M., {et~al.} 2018{\natexlab{b}},
  \apjl, 869, L41, \dodoi{10.3847/2041-8213/aaf741}

\bibitem[{{Ansdell} {et~al.}(2016){Ansdell}, {Williams}, {van der Marel},
  {Carpenter}, {Guidi}, {Hogerheijde}, {Mathews}, {Manara}, {Miotello},
  {Natta}, {Oliveira}, {Tazzari}, {Testi}, {van Dishoeck}, \& {van
  Terwisga}}]{Ansdell2016}
{Ansdell}, M., {Williams}, J.~P., {van der Marel}, N., {et~al.} 2016, \apj,
  828, 46, \dodoi{10.3847/0004-637X/828/1/46}

\bibitem[{{Ansdell} {et~al.}(2018){Ansdell}, {Williams}, {Trapman}, {van
  Terwisga}, {Facchini}, {Manara}, {van der Marel}, {Miotello}, {Tazzari},
  {Hogerheijde}, {Guidi}, {Testi}, \& {van Dishoeck}}]{Ansdell2018}
{Ansdell}, M., {Williams}, J.~P., {Trapman}, L., {et~al.} 2018, \apj, 859, 21,
  \dodoi{10.3847/1538-4357/aab890}

\bibitem[{{Astropy Collaboration} {et~al.}(2013){Astropy Collaboration},
  {Robitaille}, {Tollerud}, {Greenfield}, {Droettboom}, {Bray}, {Aldcroft},
  {Davis}, {Ginsburg}, {Price-Whelan}, {Kerzendorf}, {Conley}, {Crighton},
  {Barbary}, {Muna}, {Ferguson}, {Grollier}, {Parikh}, {Nair}, {Unther},
  {Deil}, {Woillez}, {Conseil}, {Kramer}, {Turner}, {Singer}, {Fox}, {Weaver},
  {Zabalza}, {Edwards}, {Azalee Bostroem}, {Burke}, {Casey}, {Crawford},
  {Dencheva}, {Ely}, {Jenness}, {Labrie}, {Lim}, {Pierfederici}, {Pontzen},
  {Ptak}, {Refsdal}, {Servillat}, \& {Streicher}}]{Astropy2013}
{Astropy Collaboration}, {Robitaille}, T.~P., {Tollerud}, E.~J., {et~al.} 2013,
  \aap, 558, A33, \dodoi{10.1051/0004-6361/201322068}

\bibitem[{{Bai} \& {Stone}(2013)}]{Bai2013}
{Bai}, X.-N., \& {Stone}, J.~M. 2013, \apj, 769, 76,
  \dodoi{10.1088/0004-637X/769/1/76}

\bibitem[{{Bai} {et~al.}(2016){Bai}, {Ye}, {Goodman}, \& {Yuan}}]{Bai2016}
{Bai}, X.-N., {Ye}, J., {Goodman}, J., \& {Yuan}, F. 2016, \apj, 818, 152,
  \dodoi{10.3847/0004-637X/818/2/152}

\bibitem[{{Balbus} \& {Hawley}(1998)}]{Balbus1998}
{Balbus}, S.~A., \& {Hawley}, J.~F. 1998, Reviews of Modern Physics, 70, 1,
  \dodoi{10.1103/RevModPhys.70.1}

\bibitem[{{Baraffe} {et~al.}(2015){Baraffe}, {Homeier}, {Allard}, \&
  {Chabrier}}]{baraffe2015}
{Baraffe}, I., {Homeier}, D., {Allard}, F., \& {Chabrier}, G. 2015, \aap, 577,
  A42, \dodoi{10.1051/0004-6361/201425481}

\bibitem[{{Beckwith} \& {Sargent}(1993)}]{BeckwithSargent1993}
{Beckwith}, S. V.~W., \& {Sargent}, A.~I. 1993, \apj, 402, 280,
  \dodoi{10.1086/172131}

\bibitem[{{Benz} {et~al.}(2014){Benz}, {Ida}, {Alibert}, {Lin}, \&
  {Mordasini}}]{Benz2014prpl}
{Benz}, W., {Ida}, S., {Alibert}, Y., {Lin}, D., \& {Mordasini}, C. 2014, in
  Protostars and Planets VI, ed. H.~{Beuther}, R.~S. {Klessen}, C.~P.
  {Dullemond}, \& T.~{Henning}, 691,
  \dodoi{10.2458/azu\_uapress\_9780816531240-ch030}

\bibitem[{{Birnstiel} \& {Andrews}(2014)}]{Birnstiel2014}
{Birnstiel}, T., \& {Andrews}, S.~M. 2014, \apj, 780, 153,
  \dodoi{10.1088/0004-637X/780/2/153}

\bibitem[{{Blandford} \& {Payne}(1982)}]{Blandford1982}
{Blandford}, R.~D., \& {Payne}, D.~G. 1982, \mnras, 199, 883,
  \dodoi{10.1093/mnras/199.4.883}

\bibitem[{{Boyden} \& {Eisner}(2020)}]{Boyden2020}
{Boyden}, R.~D., \& {Eisner}, J.~A. 2020, \apj, 894, 74,
  \dodoi{10.3847/1538-4357/ab86b7}

\bibitem[{{Brauer} {et~al.}(2008){Brauer}, {Dullemond}, \&
  {Henning}}]{Brauer2008}
{Brauer}, F., {Dullemond}, C.~P., \& {Henning}, T. 2008, \aap, 480, 859,
  \dodoi{10.1051/0004-6361:20077759}

\bibitem[{{Czekala} {et~al.}(2019){Czekala}, {Chiang}, {Andrews}, {Jensen},
  {Torres}, {Wilner}, {Stassun}, \& {Macintosh}}]{Czekala2019}
{Czekala}, I., {Chiang}, E., {Andrews}, S.~M., {et~al.} 2019, \apj, 883, 22,
  \dodoi{10.3847/1538-4357/ab287b}

\bibitem[{{de Valon} {et~al.}(2020){de Valon}, {Dougados}, {Cabrit}, {Louvet},
  {Zapata}, \& {Mardones}}]{deValon2020}
{de Valon}, A., {Dougados}, C., {Cabrit}, S., {et~al.} 2020, \aap, 634, L12,
  \dodoi{10.1051/0004-6361/201936950}

\bibitem[{{Facchini} {et~al.}(2017){Facchini}, {Birnstiel}, {Bruderer}, \& {van
  Dishoeck}}]{Facchini2017}
{Facchini}, S., {Birnstiel}, T., {Bruderer}, S., \& {van Dishoeck}, E.~F. 2017,
  \aap, 605, A16, \dodoi{10.1051/0004-6361/201630329}

\bibitem[{{Facchini} {et~al.}(2019){Facchini}, {van Dishoeck}, {Manara},
  {Tazzari}, {Maud}, {Cazzoletti}, {Rosotti}, {van der Marel}, {Pinilla}, \&
  {Clarke}}]{Facchini2019}
{Facchini}, S., {van Dishoeck}, E.~F., {Manara}, C.~F., {et~al.} 2019, \aap,
  626, L2, \dodoi{10.1051/0004-6361/201935496}

\bibitem[{{Fedele} {et~al.}(2010){Fedele}, {van den Ancker}, {Henning},
  {Jayawardhana}, \& {Oliveira}}]{Fedele2010}
{Fedele}, D., {van den Ancker}, M.~E., {Henning}, T., {Jayawardhana}, R., \&
  {Oliveira}, J.~M. 2010, \aap, 510, A72, \dodoi{10.1051/0004-6361/200912810}

\bibitem[{{Feiden}(2016)}]{feiden2016}
{Feiden}, G.~A. 2016, \aap, 593, A99, \dodoi{10.1051/0004-6361/201527613}

\bibitem[{{Ferreira}(1997)}]{Ferreira1997}
{Ferreira}, J. 1997, \aap, 319, 340.
\newblock \doarXiv{astro-ph/9607057}

\bibitem[{{Flaherty} {et~al.}(2020){Flaherty}, {Hughes}, {Simon}, {Qi}, {Bai},
  {Bulatek}, {Andrews}, {Wilner}, \& {K{\'o}sp{\'a}l}}]{Flaherty2020}
{Flaherty}, K., {Hughes}, A.~M., {Simon}, J.~B., {et~al.} 2020, \apj, 895, 109,
  \dodoi{10.3847/1538-4357/ab8cc5}

\bibitem[{{Flaherty} {et~al.}(2015){Flaherty}, {Hughes}, {Rosenfeld},
  {Andrews}, {Chiang}, {Simon}, {Kerzner}, \& {Wilner}}]{Flaherty2015}
{Flaherty}, K.~M., {Hughes}, A.~M., {Rosenfeld}, K.~A., {et~al.} 2015, \apj,
  813, 99, \dodoi{10.1088/0004-637X/813/2/99}

\bibitem[{{Flaherty} {et~al.}(2018){Flaherty}, {Hughes}, {Teague}, {Simon},
  {Andrews}, \& {Wilner}}]{Flaherty2018}
{Flaherty}, K.~M., {Hughes}, A.~M., {Teague}, R., {et~al.} 2018, \apj, 856,
  117, \dodoi{10.3847/1538-4357/aab615}

\bibitem[{{Gaia Collaboration} {et~al.}(2018){Gaia Collaboration}, {Brown},
  {Vallenari}, {Prusti}, {de Bruijne}, {Babusiaux}, {Bailer-Jones}, {Biermann},
  {Evans}, {Eyer}, \& et~al.}]{Gaia2018}
{Gaia Collaboration}, {Brown}, A.~G.~A., {Vallenari}, A., {et~al.} 2018, \aap,
  616, A1, \dodoi{10.1051/0004-6361/201833051}

\bibitem[{{Ghez} {et~al.}(1997){Ghez}, {McCarthy}, {Patience}, \&
  {Beck}}]{Ghez1997}
{Ghez}, A.~M., {McCarthy}, D.~W., {Patience}, J.~L., \& {Beck}, T.~L. 1997,
  \apj, 481, 378, \dodoi{10.1086/304031}

\bibitem[{{Gressel} {et~al.}(2015){Gressel}, {Turner}, {Nelson}, \&
  {McNally}}]{Gressel2015}
{Gressel}, O., {Turner}, N.~J., {Nelson}, R.~P., \& {McNally}, C.~P. 2015,
  \apj, 801, 84, \dodoi{10.1088/0004-637X/801/2/84}

\bibitem[{{Guilloteau} \& {Dutrey}(1998)}]{Guilloteau1998}
{Guilloteau}, S., \& {Dutrey}, A. 1998, \aap, 339, 467

\bibitem[{{Guilloteau} {et~al.}(2012){Guilloteau}, {Dutrey}, {Wakelam},
  {Hersant}, {Semenov}, {Chapillon}, {Henning}, \&
  {Pi{\'e}tu}}]{Guilloteau2012}
{Guilloteau}, S., {Dutrey}, A., {Wakelam}, V., {et~al.} 2012, \aap, 548, A70,
  \dodoi{10.1051/0004-6361/201220331}

\bibitem[{{Harsono} {et~al.}(2014){Harsono}, {J{\o}rgensen}, {van Dishoeck},
  {Hogerheijde}, {Bruderer}, {Persson}, \& {Mottram}}]{Harsono2014}
{Harsono}, D., {J{\o}rgensen}, J.~K., {van Dishoeck}, E.~F., {et~al.} 2014,
  \aap, 562, A77, \dodoi{10.1051/0004-6361/201322646}

\bibitem[{{Hartmann} {et~al.}(1998){Hartmann}, {Calvet}, {Gullbring}, \&
  {D'Alessio}}]{Hartmann1998}
{Hartmann}, L., {Calvet}, N., {Gullbring}, E., \& {D'Alessio}, P. 1998, \apj,
  495, 385, \dodoi{10.1086/305277}

\bibitem[{{Hartmann} {et~al.}(2016){Hartmann}, {Herczeg}, \&
  {Calvet}}]{Hartmann2016}
{Hartmann}, L., {Herczeg}, G., \& {Calvet}, N. 2016, \araa, 54, 135,
  \dodoi{10.1146/annurev-astro-081915-023347}

\bibitem[{{Hashimoto} {et~al.}(2021){Hashimoto}, {Muto}, {Dong}, {Liu}, {van
  der Marel}, {Francis}, {Hasegawa}, \& {Tsukagoshi}}]{Hashimoto2021}
{Hashimoto}, J., {Muto}, T., {Dong}, R., {et~al.} 2021, \apj, 911, 5,
  \dodoi{10.3847/1538-4357/abe59f}

\bibitem[{{Haworth} {et~al.}(2017){Haworth}, {Facchini}, {Clarke}, \&
  {Cleeves}}]{Haworth2017}
{Haworth}, T.~J., {Facchini}, S., {Clarke}, C.~J., \& {Cleeves}, L.~I. 2017,
  \mnras, 468, L108, \dodoi{10.1093/mnrasl/slx037}

\bibitem[{{Hendler} {et~al.}(2020){Hendler}, {Pascucci}, {Pinilla}, {Tazzari},
  {Carpenter}, {Malhotra}, \& {Testi}}]{Hendler2020}
{Hendler}, N., {Pascucci}, I., {Pinilla}, P., {et~al.} 2020, \apj, 895, 126,
  \dodoi{10.3847/1538-4357/ab70ba}

\bibitem[{{Herczeg} \& {Hillenbrand}(2014)}]{HH2014}
{Herczeg}, G.~J., \& {Hillenbrand}, L.~A. 2014, \apj, 786, 97,
  \dodoi{10.1088/0004-637X/786/2/97}

\bibitem[{{Herczeg} \& {Hillenbrand}(2015)}]{HH2015}
---. 2015, \apj, 808, 23, \dodoi{10.1088/0004-637X/808/1/23}

\bibitem[{{Ho} {et~al.}(2004){Ho}, {Moran}, \& {Lo}}]{Ho2004}
{Ho}, P. T.~P., {Moran}, J.~M., \& {Lo}, K.~Y. 2004, \apjl, 616, L1,
  \dodoi{10.1086/423245}

\bibitem[{{Huang} {et~al.}(2018{\natexlab{a}}){Huang}, {Andrews}, {Cleeves},
  {{\"O}berg}, {Wilner}, {Bai}, {Birnstiel}, {Carpenter}, {Hughes}, {Isella},
  {P{\'e}rez}, {Ricci}, \& {Zhu}}]{Huang2018_CO}
{Huang}, J., {Andrews}, S.~M., {Cleeves}, L.~I., {et~al.} 2018{\natexlab{a}},
  \apj, 852, 122, \dodoi{10.3847/1538-4357/aaa1e7}

\bibitem[{{Huang} {et~al.}(2018{\natexlab{b}}){Huang}, {Andrews}, {Dullemond},
  {Isella}, {P{\'e}rez}, {Guzm{\'a}n}, {{\"O}berg}, {Zhu}, {Zhang}, {Bai},
  {Benisty}, {Birnstiel}, {Carpenter}, {Hughes}, {Ricci}, {Weaver}, \&
  {Wilner}}]{Huang2018_ring}
{Huang}, J., {Andrews}, S.~M., {Dullemond}, C.~P., {et~al.} 2018{\natexlab{b}},
  \apjl, 869, L42, \dodoi{10.3847/2041-8213/aaf740}

\bibitem[{{Hughes} {et~al.}(2008){Hughes}, {Wilner}, {Qi}, \&
  {Hogerheijde}}]{Hughes2008}
{Hughes}, A.~M., {Wilner}, D.~J., {Qi}, C., \& {Hogerheijde}, M.~R. 2008, \apj,
  678, 1119, \dodoi{10.1086/586730}

\bibitem[{{Hunter}(2007)}]{matplotlib2007}
{Hunter}, J.~D. 2007, Computing in Science and Engineering, 9, 90,
  \dodoi{10.1109/MCSE.2007.55}

\bibitem[{{Isella} {et~al.}(2007){Isella}, {Testi}, {Natta}, {Neri}, {Wilner},
  \& {Qi}}]{Isella2007}
{Isella}, A., {Testi}, L., {Natta}, A., {et~al.} 2007, \aap, 469, 213,
  \dodoi{10.1051/0004-6361:20077385}

\bibitem[{{Kelly}(2007)}]{Kelly2007}
{Kelly}, B.~C. 2007, \apj, 665, 1489, \dodoi{10.1086/519947}

\bibitem[{{Kraus} \& {Hillenbrand}(2009)}]{Kraus2009}
{Kraus}, A.~L., \& {Hillenbrand}, L.~A. 2009, \apj, 704, 531,
  \dodoi{10.1088/0004-637X/704/1/531}

\bibitem[{{Kraus} {et~al.}(2012){Kraus}, {Ireland}, {Hillenbrand}, \&
  {Martinache}}]{Kraus2012}
{Kraus}, A.~L., {Ireland}, M.~J., {Hillenbrand}, L.~A., \& {Martinache}, F.
  2012, \apj, 745, 19, \dodoi{10.1088/0004-637X/745/1/19}

\bibitem[{{Kurtovic} {et~al.}(2021){Kurtovic}, {Pinilla}, {Long}, {Benisty},
  {Manara}, {Natta}, {Pascucci}, {Ricci}, {Scholz}, \& {Testi}}]{Kurtovic2021}
{Kurtovic}, N.~T., {Pinilla}, P., {Long}, F., {et~al.} 2021, \aap, 645, A139,
  \dodoi{10.1051/0004-6361/202038983}

\bibitem[{{Lesur}(2021)}]{Lesur2021}
{Lesur}, G. R.~J. 2021, \aap, 650, A35, \dodoi{10.1051/0004-6361/202040109}

\bibitem[{{Long} {et~al.}(2017){Long}, {Herczeg}, {Pascucci}, {Drabek-Maunder},
  {Mohanty}, {Testi}, {Apai}, {Hendler}, {Henning}, {Manara}, \&
  {Mulders}}]{Long2017}
{Long}, F., {Herczeg}, G.~J., {Pascucci}, I., {et~al.} 2017, \apj, 844, 99,
  \dodoi{10.3847/1538-4357/aa78fc}

\bibitem[{{Long} {et~al.}(2018){Long}, {Pinilla}, {Herczeg}, {Harsono},
  {Dipierro}, {Pascucci}, {Hendler}, {Tazzari}, {Ragusa}, {Salyk}, {Edwards},
  {Lodato}, {van de Plas}, {Johnstone}, {Liu}, {Boehler}, {Cabrit}, {Manara},
  {Menard}, {Mulders}, {Nisini}, {Fischer}, {Rigliaco}, {Banzatti}, {Avenhaus},
  \& {Gully-Santiago}}]{Long2018}
{Long}, F., {Pinilla}, P., {Herczeg}, G.~J., {et~al.} 2018, \apj, 869, 17,
  \dodoi{10.3847/1538-4357/aae8e1}

\bibitem[{{Long} {et~al.}(2019){Long}, {Herczeg}, {Harsono}, {Pinilla},
  {Tazzari}, {Manara}, {Pascucci}, {Cabrit}, {Nisini}, {Johnstone}, {Edwards},
  {Salyk}, {Menard}, {Lodato}, {Boehler}, {Mace}, {Liu}, {Mulders}, {Hendler},
  {Ragusa}, {Fischer}, {Banzatti}, {Rigliaco}, {van de Plas}, {Dipierro},
  {Gully-Santiago}, \& {Lopez-Valdivia}}]{Long2019}
{Long}, F., {Herczeg}, G.~J., {Harsono}, D., {et~al.} 2019, \apj, 882, 49,
  \dodoi{10.3847/1538-4357/ab2d2d}

\bibitem[{{Loomis} {et~al.}(2018){Loomis}, {{\"O}berg}, {Andrews}, {Walsh},
  {Czekala}, {Huang}, \& {Rosenfeld}}]{Loomis2018}
{Loomis}, R.~A., {{\"O}berg}, K.~I., {Andrews}, S.~M., {et~al.} 2018, \aj, 155,
  182, \dodoi{10.3847/1538-3881/aab604}

\bibitem[{{Lynden-Bell} \& {Pringle}(1974)}]{Lynden-Bell1974}
{Lynden-Bell}, D., \& {Pringle}, J.~E. 1974, \mnras, 168, 603,
  \dodoi{10.1093/mnras/168.3.603}

\bibitem[{{Matr{\`a}} {et~al.}(2017){Matr{\`a}}, {MacGregor}, {Kalas}, {Wyatt},
  {Kennedy}, {Wilner}, {Duchene}, {Hughes}, {Pan}, {Shannon}, {Clampin},
  {Fitzgerald}, {Graham}, {Holland }, {Pani{\'c}}, \& {Su}}]{Matra2017}
{Matr{\`a}}, L., {MacGregor}, M.~A., {Kalas}, P., {et~al.} 2017, \apj, 842, 9,
  \dodoi{10.3847/1538-4357/aa71b4}

\bibitem[{{McMullin} {et~al.}(2007){McMullin}, {Waters}, {Schiebel}, {Young},
  \& {Golap}}]{McMullin2007}
{McMullin}, J.~P., {Waters}, B., {Schiebel}, D., {Young}, W., \& {Golap}, K.
  2007, in Astronomical Society of the Pacific Conference Series, Vol. 376,
  Astronomical Data Analysis Software and Systems XVI, ed. R.~A. {Shaw},
  F.~{Hill}, \& D.~J. {Bell}, 127

\bibitem[{{Morbidelli} \& {Raymond}(2016)}]{MorbidelliRaymond2016}
{Morbidelli}, A., \& {Raymond}, S.~N. 2016, Journal of Geophysical Research
  (Planets), 121, 1962, \dodoi{10.1002/2016JE005088}

\bibitem[{{Najita} \& {Bergin}(2018)}]{Najita2018}
{Najita}, J.~R., \& {Bergin}, E.~A. 2018, \apj, 864, 168,
  \dodoi{10.3847/1538-4357/aad80c}

\bibitem[{{Pani{\'c}} {et~al.}(2009){Pani{\'c}}, {Hogerheijde}, {Wilner}, \&
  {Qi}}]{Panic2009}
{Pani{\'c}}, O., {Hogerheijde}, M.~R., {Wilner}, D., \& {Qi}, C. 2009, \aap,
  501, 269, \dodoi{10.1051/0004-6361/200911883}

\bibitem[{{Pascucci} {et~al.}(2016){Pascucci}, {Testi}, {Herczeg}, {Long},
  {Manara}, {Hendler}, {Mulders}, {Krijt}, {Ciesla}, {Henning}, {Mohanty},
  {Drabek-Maunder}, {Apai}, {Sz{\H{u}}cs}, {Sacco}, \&
  {Olofsson}}]{Pascucci2016}
{Pascucci}, I., {Testi}, L., {Herczeg}, G.~J., {et~al.} 2016, \apj, 831, 125,
  \dodoi{10.3847/0004-637X/831/2/125}

\bibitem[{{Pegues} {et~al.}(2021){Pegues}, {Czekala}, {Andrews}, {{\"O}berg},
  {Herczeg}, {Bergner}, {Ilsedore Cleeves}, {Guzm{\'a}n}, {Huang}, {Long},
  {Teague}, \& {Wilner}}]{Pegues2021}
{Pegues}, J., {Czekala}, I., {Andrews}, S.~M., {et~al.} 2021, \apj, 908, 42,
  \dodoi{10.3847/1538-4357/abd4eb}

\bibitem[{{Pi{\'e}tu} {et~al.}(2007){Pi{\'e}tu}, {Dutrey}, \&
  {Guilloteau}}]{Pietu2007}
{Pi{\'e}tu}, V., {Dutrey}, A., \& {Guilloteau}, S. 2007, \aap, 467, 163,
  \dodoi{10.1051/0004-6361:20066537}

\bibitem[{{Pinilla} {et~al.}(2013){Pinilla}, {Birnstiel}, {Benisty}, {Ricci},
  {Natta}, {Dullemond}, {Dominik}, \& {Testi}}]{Pinilla2013}
{Pinilla}, P., {Birnstiel}, T., {Benisty}, M., {et~al.} 2013, \aap, 554, A95,
  \dodoi{10.1051/0004-6361/201220875}

\bibitem[{{Pinilla} {et~al.}(2012){Pinilla}, {Birnstiel}, {Ricci}, {Dullemond},
  {Uribe}, {Testi}, \& {Natta}}]{Pinilla2012_bump}
{Pinilla}, P., {Birnstiel}, T., {Ricci}, L., {et~al.} 2012, \aap, 538, A114,
  \dodoi{10.1051/0004-6361/201118204}

\bibitem[{{Qi} {et~al.}(2019){Qi}, {{\"O}berg}, {Espaillat}, {Robinson},
  {Andrews}, {Wilner}, {Blake}, {Bergin}, \& {Cleeves}}]{Qi2019}
{Qi}, C., {{\"O}berg}, K.~I., {Espaillat}, C.~C., {et~al.} 2019, \apj, 882,
  160, \dodoi{10.3847/1538-4357/ab35d3}

\bibitem[{{Rosenfeld} {et~al.}(2012){Rosenfeld}, {Andrews}, {Wilner}, \&
  {Stempels}}]{Rosenfeld2012}
{Rosenfeld}, K.~A., {Andrews}, S.~M., {Wilner}, D.~J., \& {Stempels}, H.~C.
  2012, \apj, 759, 119, \dodoi{10.1088/0004-637X/759/2/119}

\bibitem[{{Rosotti} {et~al.}(2019){Rosotti}, {Tazzari}, {Booth}, {Testi},
  {Lodato}, \& {Clarke}}]{Rosotti2019}
{Rosotti}, G.~P., {Tazzari}, M., {Booth}, R.~A., {et~al.} 2019, \mnras, 486,
  4829, \dodoi{10.1093/mnras/stz1190}

\bibitem[{{Salinas} {et~al.}(2017){Salinas}, {Hogerheijde}, {Mathews},
  {{\"O}berg}, {Qi}, {Williams}, \& {Wilner}}]{Salinas2017}
{Salinas}, V.~N., {Hogerheijde}, M.~R., {Mathews}, G.~S., {et~al.} 2017, \aap,
  606, A125, \dodoi{10.1051/0004-6361/201731223}

\bibitem[{{Sanchis} {et~al.}(2021){Sanchis}, {Testi}, {Natta}, {Facchini},
  {Manara}, {Miotello}, {Ercolano}, {Henning}, {Preibisch}, {Carpenter}, {de
  Gregorio-Monsalvo}, {Jayawardhana}, {Lopez}, {Mu{\v{z}}i{\'c}}, {Pascucci},
  {Santamar{\'\i}a-Miranda}, {van Terwisga}, \& {Williams}}]{Sanchis2021}
{Sanchis}, E., {Testi}, L., {Natta}, A., {et~al.} 2021, \aap, 649, A19,
  \dodoi{10.1051/0004-6361/202039733}

\bibitem[{{Shakura} \& {Sunyaev}(1973)}]{Shakura1973}
{Shakura}, N.~I., \& {Sunyaev}, R.~A. 1973, \aap, 24, 337

\bibitem[{{Simon} {et~al.}(2019){Simon}, {Guilloteau}, {Beck}, {Chapillon}, {Di
  Folco}, {Dutrey}, {Feiden}, {Grosso}, {Pi{\'e}tu}, {Prato}, \&
  {Schaefer}}]{Simon2019}
{Simon}, M., {Guilloteau}, S., {Beck}, T.~L., {et~al.} 2019, \apj, 884, 42,
  \dodoi{10.3847/1538-4357/ab3e3b}

\bibitem[{{Simon} {et~al.}(2016){Simon}, {Pascucci}, {Edwards}, {Feng},
  {Gorti}, {Hollenbach}, {Rigliaco}, \& {Keane}}]{Simon2016}
{Simon}, M.~N., {Pascucci}, I., {Edwards}, S., {et~al.} 2016, \apj, 831, 169,
  \dodoi{10.3847/0004-637X/831/2/169}

\bibitem[{{Tabone} {et~al.}(2021{\natexlab{a}}){Tabone}, {Rosotti}, {Cridland},
  {Armitage}, \& {Lodato}}]{Tabone2021a_arXiv}
{Tabone}, B., {Rosotti}, G.~P., {Cridland}, A.~J., {Armitage}, P.~J., \&
  {Lodato}, G. 2021{\natexlab{a}}, arXiv e-prints, arXiv:2111.10145.
\newblock \doarXiv{2111.10145}

\bibitem[{{Tabone} {et~al.}(2021{\natexlab{b}}){Tabone}, {Rosotti}, {Lodato},
  {Armitage}, {Cridland}, \& {van Dishoeck}}]{Tabone2021b_arXiv}
{Tabone}, B., {Rosotti}, G.~P., {Lodato}, G., {et~al.} 2021{\natexlab{b}},
  arXiv e-prints, arXiv:2111.14473.
\newblock \doarXiv{2111.14473}

\bibitem[{{Takeuchi} \& {Lin}(2002)}]{TakeuchiLin2002}
{Takeuchi}, T., \& {Lin}, D.~N.~C. 2002, \apj, 581, 1344,
  \dodoi{10.1086/344437}

\bibitem[{{Tazzari} {et~al.}(2021){Tazzari}, {Testi}, {Natta}, {Williams},
  {Ansdell}, {Carpenter}, {Facchini}, {Guidi}, {Hogherheijde}, {Manara},
  {Miotello}, \& {van der Marel}}]{Tazzari2021}
{Tazzari}, M., {Testi}, L., {Natta}, A., {et~al.} 2021, \mnras,
  \dodoi{10.1093/mnras/stab1912}

\bibitem[{Teague(2019)}]{GoFish}
Teague, R. 2019, The Journal of Open Source Software, 4, 1632,
  \dodoi{10.21105/joss.01632}

\bibitem[{{Teague} {et~al.}(2019){Teague}, {Bae}, \&
  {Bergin}}]{Teague2019Natur}
{Teague}, R., {Bae}, J., \& {Bergin}, E.~A. 2019, \nat, 574, 378,
  \dodoi{10.1038/s41586-019-1642-0}

\bibitem[{{Teague} {et~al.}(2018){Teague}, {Henning}, {Guilloteau}, {Bergin},
  {Semenov}, {Dutrey}, {Flock}, {Gorti}, \& {Birnstiel}}]{Teague2018_CS}
{Teague}, R., {Henning}, T., {Guilloteau}, S., {et~al.} 2018, \apj, 864, 133,
  \dodoi{10.3847/1538-4357/aad80e}

\bibitem[{{Testi} {et~al.}(2014){Testi}, {Birnstiel}, {Ricci}, {Andrews},
  {Blum}, {Carpenter}, {Dominik}, {Isella}, {Natta}, {Williams}, \&
  {Wilner}}]{Testi2014}
{Testi}, L., {Birnstiel}, T., {Ricci}, L., {et~al.} 2014, Protostars and
  Planets VI, 339, \dodoi{10.2458/azu_uapress_9780816531240-ch015}

\bibitem[{{Toci} {et~al.}(2021){Toci}, {Rosotti}, {Lodato}, {Testi}, \&
  {Trapman}}]{Toci2021}
{Toci}, C., {Rosotti}, G., {Lodato}, G., {Testi}, L., \& {Trapman}, L. 2021,
  \mnras, 507, 818, \dodoi{10.1093/mnras/stab2112}

\bibitem[{{Trapman} {et~al.}(2020{\natexlab{a}}){Trapman}, {Ansdell},
  {Hogerheijde}, {Facchini}, {Manara}, {Miotello}, {Williams}, \&
  {Bruderer}}]{Trapman2020_Lupus}
{Trapman}, L., {Ansdell}, M., {Hogerheijde}, M.~R., {et~al.}
  2020{\natexlab{a}}, \aap, 638, A38, \dodoi{10.1051/0004-6361/201834537}

\bibitem[{{Trapman} {et~al.}(2019){Trapman}, {Facchini}, {Hogerheijde}, {van
  Dishoeck}, \& {Bruderer}}]{Trapman2019}
{Trapman}, L., {Facchini}, S., {Hogerheijde}, M.~R., {van Dishoeck}, E.~F., \&
  {Bruderer}, S. 2019, \aap, 629, A79, \dodoi{10.1051/0004-6361/201834723}

\bibitem[{{Trapman} {et~al.}(2020{\natexlab{b}}){Trapman}, {Rosotti}, {Bosman},
  {Hogerheijde}, \& {van Dishoeck}}]{Trapman2020_viscous}
{Trapman}, L., {Rosotti}, G., {Bosman}, A.~D., {Hogerheijde}, M.~R., \& {van
  Dishoeck}, E.~F. 2020{\natexlab{b}}, \aap, 640, A5,
  \dodoi{10.1051/0004-6361/202037673}

\bibitem[{{Trapman} {et~al.}(2021){Trapman}, {Tabone}, {Rosotti}, \&
  {Zhang}}]{Trapman2021arXiv}
{Trapman}, L., {Tabone}, B., {Rosotti}, G., \& {Zhang}, K. 2021, arXiv
  e-prints, arXiv:2112.00645.
\newblock \doarXiv{2112.00645}

\bibitem[{{Tripathi} {et~al.}(2017){Tripathi}, {Andrews}, {Birnstiel}, \&
  {Wilner}}]{Tripathi2017}
{Tripathi}, A., {Andrews}, S.~M., {Birnstiel}, T., \& {Wilner}, D.~J. 2017,
  \apj, 845, 44, \dodoi{10.3847/1538-4357/aa7c62}

\bibitem[{{Turner} {et~al.}(2014){Turner}, {Fromang}, {Gammie}, {Klahr},
  {Lesur}, {Wardle}, \& {Bai}}]{Turner2014}
{Turner}, N.~J., {Fromang}, S., {Gammie}, C., {et~al.} 2014, in Protostars and
  Planets VI, ed. H.~{Beuther}, R.~S. {Klessen}, C.~P. {Dullemond}, \&
  T.~{Henning}, 411, \dodoi{10.2458/azu\_uapress\_9780816531240-ch018}

\bibitem[{{Weidenschilling}(1977)}]{Weidenschilling1977}
{Weidenschilling}, S.~J. 1977, \mnras, 180, 57, \dodoi{10.1093/mnras/180.1.57}

\bibitem[{{Yang} \& {Bai}(2021)}]{YangBai2021arXiv}
{Yang}, H., \& {Bai}, X.-N. 2021, arXiv e-prints, arXiv:2108.10485.
\newblock \doarXiv{2108.10485}

\bibitem[{{Yen} {et~al.}(2018){Yen}, {Koch}, {Manara}, {Miotello}, \&
  {Testi}}]{Yen2018}
{Yen}, H.-W., {Koch}, P.~M., {Manara}, C.~F., {Miotello}, A., \& {Testi}, L.
  2018, \aap, 616, A100, \dodoi{10.1051/0004-6361/201732196}

\bibitem[{{Zurlo} {et~al.}(2020){Zurlo}, {Cieza}, {P{\'e}rez}, {Christiaens},
  {Williams}, {Guidi}, {C{\'a}novas}, {Casassus}, {Hales}, {Principe},
  {Ru{\'\i}z-Rodr{\'\i}guez}, \& {Fernand ez-Figueroa}}]{Zurlo2020}
{Zurlo}, A., {Cieza}, L.~A., {P{\'e}rez}, S., {et~al.} 2020, \mnras, 496, 5089,
  \dodoi{10.1093/mnras/staa1886}

\end{thebibliography}
\bibliographystyle{aasjournal}

\appendix
\restartappendixnumbering
\twocolumngrid

\section{Channel maps and line spectra of CO emission}

\begin{figure}[!th]
\centering
    \includegraphics[width=0.9\linewidth]{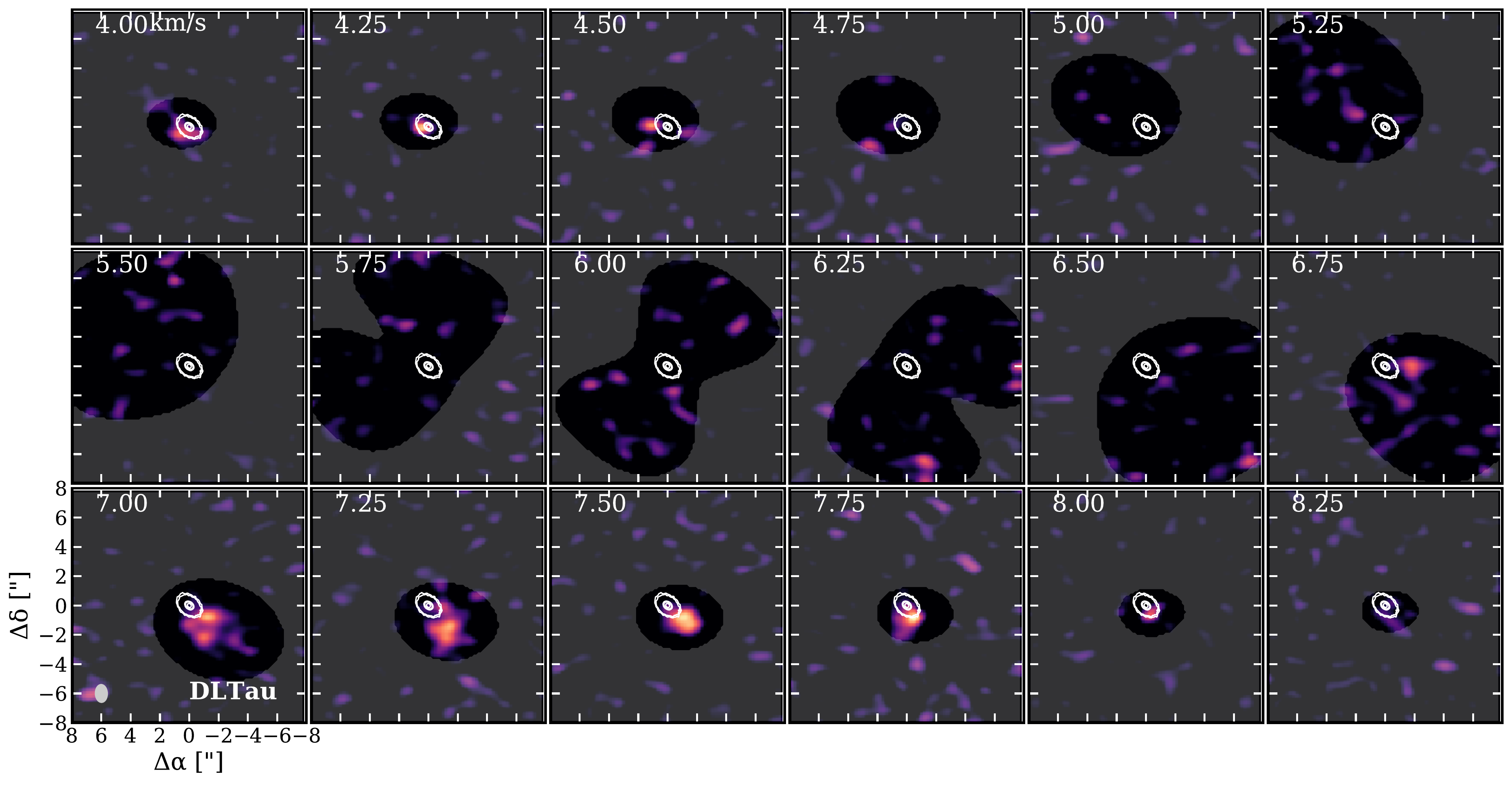} \\ 
\caption{Channel maps of $^{12}$CO emission for DL Tau, with Keplerian masks applied in each channel. Contours are starting from 5$\sigma$ for the 1.3\,mm continuum image. The blueshifted side of the disk, as well as central velocity channels, are heavily cloud contaminated. \label{fig:dltau}}
\end{figure}

\begin{figure}[!th]
\centering
    \includegraphics[width=0.9\linewidth]{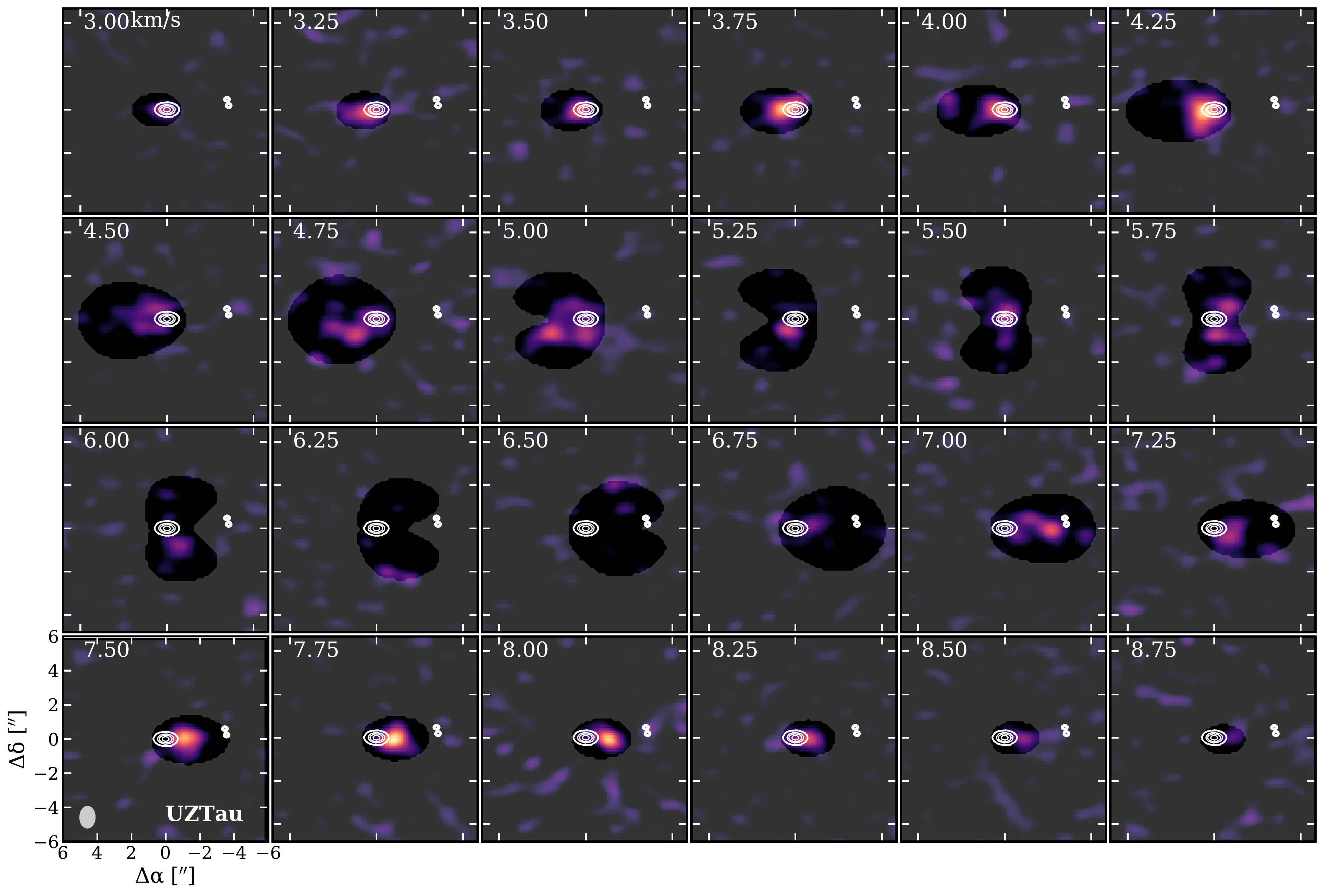} \\ 
\caption{Channel maps of $^{12}$CO emission for UZ Tau stellar system, with Keplerian masks applied in each channel. Contours are starting from 5$\sigma$ for the 1.3\,mm continuum image. The close binary to the west (UZ Tau Wab) are also shown in continuum contours in the field, for which the $^{12}$CO emission around this binary is not detected in our SMA observation. Cloud absorption is visible around central channels.  \label{fig:uztau}}
\end{figure}

\begin{figure*}[!th]
\centering
    \includegraphics[width=0.9\linewidth]{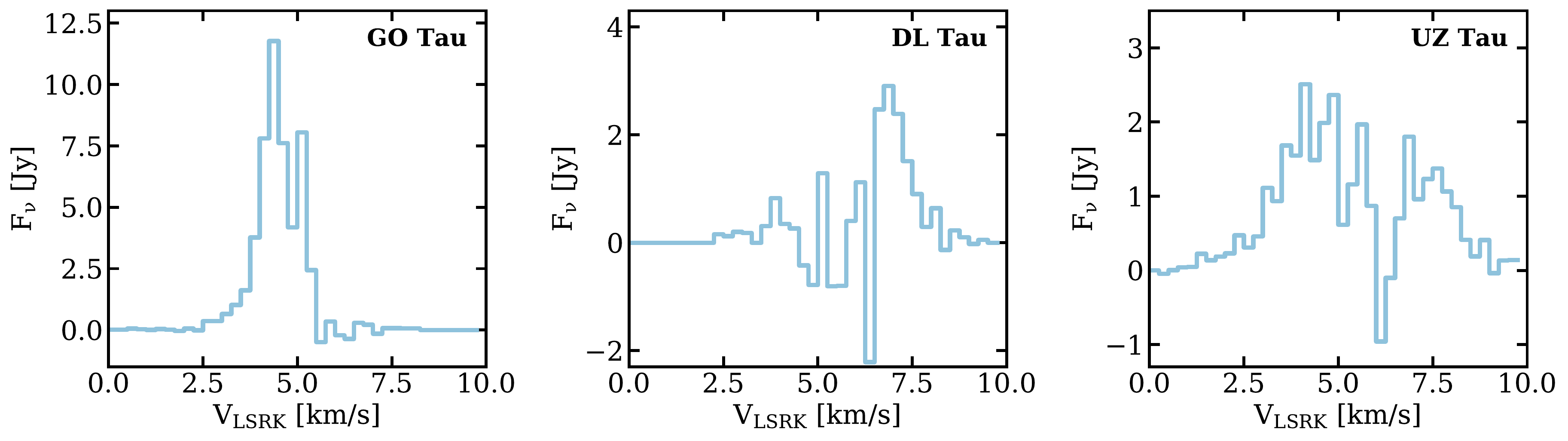} \\ 
\caption{Line spectra of the $^{12}$CO emission for the three Taurus disks with SMA observations, extracted within the Keplerian mask. Cloud contamination regions are visible around 5-6\,km\,s$^{-1}$.  \label{fig:sma-spectra}}
\end{figure*}

The channel maps of the $^{12}$CO $J=2-1$ emission for DL Tau and UZ Tau from the SMA observations are shown in Figure~\ref{fig:dltau} and Figure~\ref{fig:uztau}, respectively. Both disks suffer from severe cloud contamination near the systemic velocity (see Figure~\ref{fig:sma-spectra} for the extracted line profile).

\section{Notes on Keplerian masks and CO Size measurments for individual disks} \label{sec:notes}

Table~\ref{tab:mask_rp_apx} summarizes the mask size ($R_{\rm mask}$) and the CO emission height ($z/r$) for individual disks when creating the Keplerian mask, which will ensure the inclusion of the full CO emission region. We also specify in Table~\ref{tab:mask_rp_apx} the status of cloud contamination and the azimuthal angle adopted for radial profile extraction in each disk.

\begin{figure*}[!th]
\centering
    \includegraphics[width=0.9\linewidth]{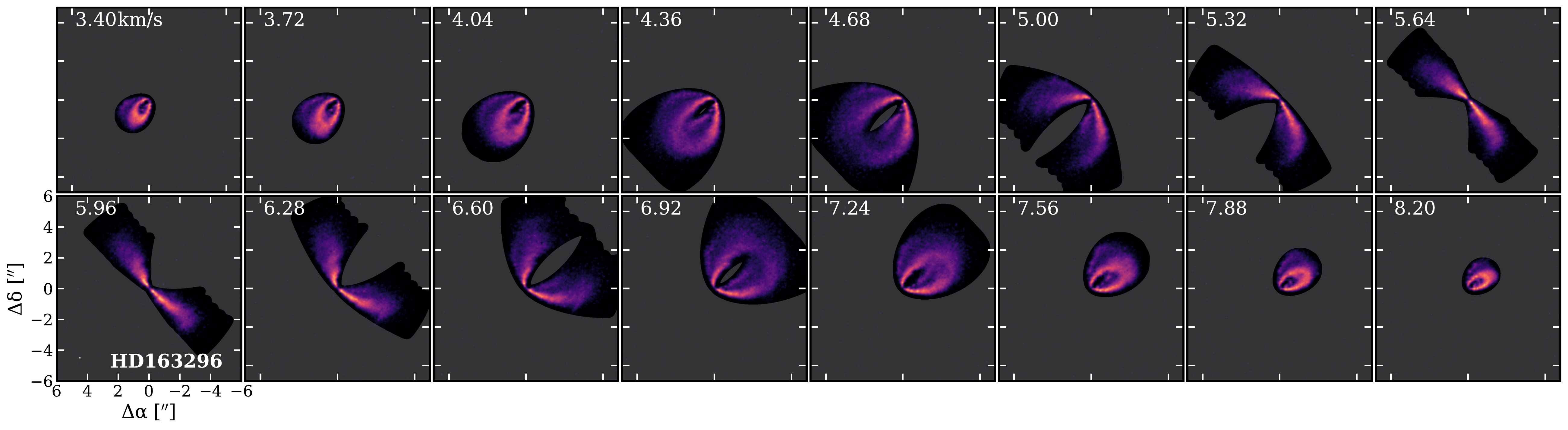} \\ 
    \includegraphics[width=0.9\linewidth]{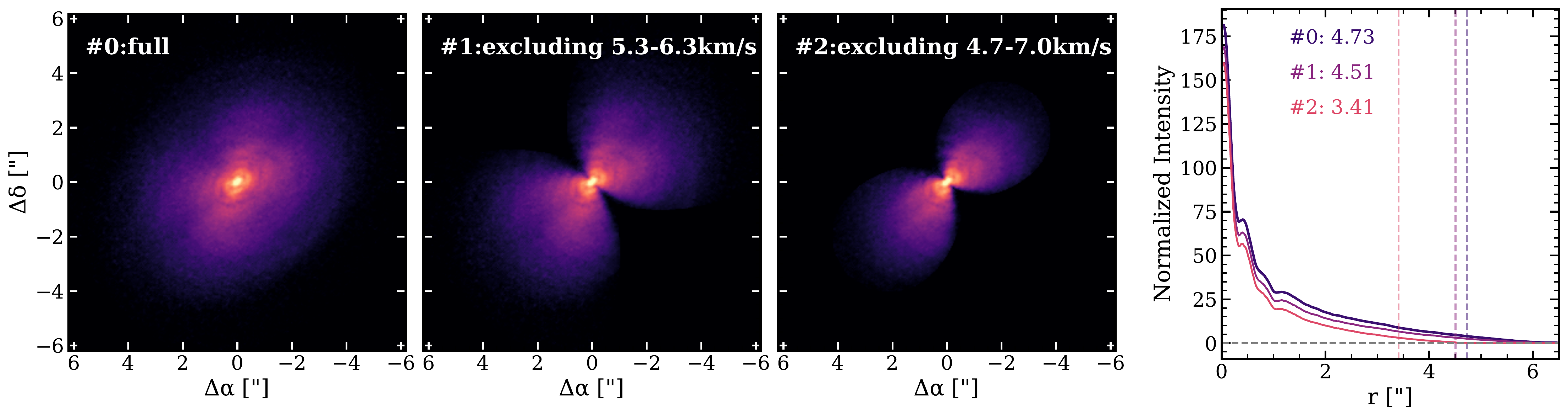} \\ 
\caption{\textbf{Top:} Channel maps of $^{12}$CO emission for HD 163296, with Keplerian masks applied in each channel. \textbf{Bottom:} Moment-zero maps created using the full velocity range ($\#$0), excluding the velocity range of 5.3--6.3\,km\,s$^{-1}$ ($\#$1), and excluding the velocity range of 4.0--7.0\,km\,s$^{-1}$ ($\#$2), and azimuthal averaged radial profiles from the three moment-zero maps. The $R_{90\%}$ for each map is marked as dashed line and noted.  \label{fig:cloud-test}}
\end{figure*}

In Figure~\ref{fig:cloud-test}, we use HD 163296 to demonstrate how cloud absorption in central velocity channels affect the gas disk size measurement. Considering a typical cloud emission line width of 1-2\,km\,s$^{-1}$ (see Table~\ref{tab:mask_rp_apx}), we calculate $R_{\rm CO}$ by excluding a similar velocity range in the central channels. Our experiment shows that the disk sizes can be underestimated by 10--30\%.

\begin{deluxetable*}{cccccll}
\tabletypesize{\scriptsize}
\tablecaption{Notes on Keplerian Mask and Radial Profile Extraction \label{tab:mask_rp_apx}}
\tablewidth{0pt}
\tablehead{
\colhead{Idx} & \colhead{Name} & \colhead{$V_{\rm sys}$} & \colhead{R$_{\rm mask}$} & \colhead{$z/r$} & \colhead{Cloud Absorption}  & \colhead{Radial Profile Extraction}  \\
\colhead{} & \colhead{} & \colhead{[$\rm km\,s^{-1}$]} & \colhead{[$''$]}   & \colhead{} & \colhead{} & \colhead{}\\
}
\colnumbers
\startdata
   2 &    DL Tau & 6.0 & 8.0 & ..  & strong around $V_{\rm sys}$ and weak towards blueshifted (4.7-6.5) & redshifted side with 60$\degr$ wedge \\
   3 &    DM Tau & 6.0 & 8.0 & 0.4 & no cloud absorption & full azimuthal   \\
   4 &    GO Tau & 4.9 & 8.0 & 0.3 & weak around $V_{\rm sys}$ and strong towards redshifted & blueshifted disk side  \\
   5 &    UZ Tau & 5.7 & 4.0 & ..  & strong around $V_{\rm sys}$ (5.2-6.8) & blueshifted/East disk side    \\
   6 &    FP Tau & 8.3 & 0.8 & ..  & strong around $V_{\rm sys}$ (7.8-8.4) & full azimuthal  \\
   7 &    CIDA 1 & 6.5 & 1.5 & ..  & strong around $V_{\rm sys}$ and weak towards redshifted (5.7-7.5) & blueshifted side  \\
   8 &    CIDA 7 & 6.0 & 1.0 & ..  & strong around $V_{\rm sys}$ and weak towards redshifted & blueshifted side  \\
   9 &     MHO 6 & 5.6 & 1.5 & 0.2 & weak around $V_{\rm sys}$ & full azimuthal  \\
  11 &     J0420 & 7.3 & 0.5 & ..  & strong around $V_{\rm sys}$ and towards blueshifted (2.0-7.5) & redshifted side  \\
  12 &     J0433 & 6.0 & 1.5 & 0.3 & strong around $V_{\rm sys}$ (5.2-7.0) & full azimuthal  \\
  13 &    GW Lup & 3.7 & 2.5 & ..  & weak around $V_{\rm sys}$ and towards redshifted (3.7-5.0) & full azimuthal  \\
  14 &    IM Lup & 4.5 & 7.0 & 0.5 & weak around $V_{\rm sys}$ & full azimuthal  \\
  15 &    MY Lup & 4.5 & 2.0 & 0.3 & weak around $V_{\rm sys}$ and towards blueshifted (3.7-5.0) & full azimuthal  \\
  16 &    Sz 129 & 4.0 & 1.5 & ..  & no cloud absorption & full azimuthal  \\
  34 &    AS 209 & 4.6 & 3.0 & ..  & weak towards blueshifted (3.7-4.4) & redshifted side  \\
  35 &      SR 4 & 5.1 & 1.0 & ..  & strong at blueshifted (1.0-4.6) & redshifted side  \\
  36 &   DoAr 25 & 3.2 & 3.0 & 0.3 & strong around $V_{\rm sys}$ and towards blueshifted (1.5-4.7) & redshifted side with 60$\degr$ wedge \\
  37 &   DoAr 33 & 2.7 & 1.0 & ..  & weak around $V_{\rm sys}$ &  full azimuthal  \\
  38 &   WaOph 6 & 4.1 & 3.0 & 0.3 & strong around $V_{\rm sys}$ and weak towards blueshifted (2.0-4.2) & redshifted side  \\
  39 & HD 142666 & 4.6 & 2.0 & ..  & no cloud absorption & full azimuthal   \\
  40 & HD 143006 & 7.8 & 2.0 & ..  & no cloud absorption & full azimuthal   \\
  41 & HD 163296 & 5.76& 7.0 & 0.3 & no cloud absorption & full azimuthal   \\
  42 &     J1100 & 4.72& 2.0 & ..  & no cloud absorption & full azimuthal  \\
  43 &    TW Hya & 2.84& 5.0 & ..  & no cloud absorption & full azimuthal  \\
  44 & V4046 Sgr & 2.9 & 7.0 & ..  & no cloud absorption & full azimuthal  \\
\enddata
\tablecomments{Column (3) is the systemic velocity of the CO emission. 
Column (4) and (5) are the mask radius and the assumed CO emission height, respectively, for the generation of Keplerian mask. In IM Lup, AS 209, and DoAr 25, the mask is made with a 20\% higher stellar mass, which matches better the line emission at high velocity channels. The parenthesis in column (6) denotes the cloud velocity range. }
\end{deluxetable*}

\begin{deluxetable*}{cccc|ccc|ccc|ccc}
\tabletypesize{\scriptsize}
\tablecaption{Disk Sizes\label{tab:sizes_apx}}
\tablewidth{0pt}
\tablehead{
\colhead{Idx} & \colhead{Name} & \colhead{mm beam size} & \colhead{CO beam size} & \colhead{R$_{\rm mm, 68\%}$} & \colhead{$R_{\rm CO, 68\%}$} & \colhead{Ratio$_{68\%}$} & \colhead{$R_{\rm mm, 90\%}$} & \colhead{$R_{\rm CO, 90\%}$} & \colhead{Ratio$_{90\%}$} & \colhead{$R_{\rm mm, 95\%}$} & \colhead{$R_{\rm CO, 95\%}$} & \colhead{Ratio$_{95\%}$} \\
\cmidrule(lr){5-7} \cmidrule(lr){8-10} \cmidrule(lr){11-13 }
\colhead{} & \colhead{} & \colhead{($''\times ''$)} & \colhead{($''\times ''$)}   & \colhead{$('')$} & \colhead{$('')$} & \colhead{$('')$} & \colhead{$('')$} & \colhead{$('')$} & \colhead{$('')$}  & \colhead{$('')$} & \colhead{$('')$} & \colhead{$('')$} \\
}
\colnumbers
\startdata
   1 &    CX Tau &      0.06$\times$0.03  & 0.17$\times$0.13 &     0.11 &   0.59 &     5.36 &     0.23 &   0.90 &    3.91 &     0.29 &  1.09 &    3.76 \\
   2 &    DL Tau &      0.14$\times$0.11  & 1.32$\times$0.94 &     0.69 &   2.65 &     3.87 &     0.91 &   3.75 &    4.14 &     0.99 &  4.20 &    4.22 \\
   3 &    DM Tau &      0.21$\times$0.18  & 0.36$\times$0.27 &     0.73 &   4.35 &     5.93 &     1.23 &   6.04 &    4.91 &     1.55 &  6.60 &    4.25 \\
   4 &    GO Tau &      0.14$\times$0.11  & 1.33$\times$0.94 &     0.66 &   4.87 &     7.34 &     0.93 &   7.04 &    7.60 &     1.06 &  7.83 &    7.36 \\
   5 &    UZ Tau &      0.13$\times$0.10  & 1.33$\times$0.95 &     0.46 &   1.99 &     4.33 &     0.64 &   2.97 &    4.67 &     0.69 &  3.39 &    4.90 \\
   6 &    FP Tau &      0.29$\times$0.23  & 0.30$\times$0.23 &     0.25 &   0.40 &     1.62 &     0.37 &   0.58 &    1.56 &     0.44 &  0.66 &    1.51 \\
   7 &    CIDA 1 &      0.15$\times$0.11  & 0.22$\times$0.16 &     0.21 &   0.56 &     2.64 &     0.28 &   0.97 &    3.49 &     0.31 &  1.25 &    4.01 \\
   8 &    CIDA 7 &      0.10$\times$0.08  & 0.17$\times$0.15 &     0.10 &   0.49 &     4.86 &     0.15 &   0.70 &    4.60 &     0.19 &  0.82 &    4.34 \\
   9 &     MHO 6 &      0.10$\times$0.07  & 0.12$\times$0.08 &     0.28 &   1.02 &     3.69 &     0.39 &   1.53 &    3.92 &     0.45 &  1.77 &    3.92 \\
  11 &    J0420 &      0.11$\times$0.08  & 0.13$\times$0.10 &     0.13 &   0.26 &     2.07 &     0.20 &   0.35 &    1.72 &     0.26 &  0.38 &    1.46 \\
  12 &    J0433 &      0.13$\times$0.09  & 0.17$\times$0.11 &     0.26 &   0.61 &     2.33 &     0.36 &   0.95 &    2.61 &     0.42 &  1.08 &    2.56 \\
  13 &    GW Lup &      0.04$\times$0.04  & 0.11$\times$0.08 &     0.38 &   1.07 &     2.81 &     0.59 &   1.72 &    2.91 &     0.68 &  2.04 &    2.99 \\
  14 &    IM Lup &      0.04$\times$0.04  & 0.12$\times$0.12 &     1.00 &   3.49 &     3.48 &     1.54 &   5.08 &    3.29 &     1.70 &  5.71 &    3.37 \\
  15 &    MY Lup &      0.04$\times$0.04  & 0.10$\times$0.08 &     0.36 &   0.90 &     2.46 &     0.49 &   1.23 &    2.49 &     0.55 &  1.38 &    2.50 \\
  16 &    Sz 129 &      0.04$\times$0.03  & 0.11$\times$0.08 &     0.30 &   0.59 &     1.97 &     0.42 &   0.81 &    1.90 &     0.48 &  0.87 &    1.82 \\
  34 &    AS 209 &      0.04$\times$0.04  & 0.10$\times$0.07 &     0.68 &   1.77 &     2.61 &     1.05 &   2.31 &    2.20 &     1.15 &  2.63 &    2.30 \\
  35 &      SR 4 &      0.03$\times$0.03  & 0.11$\times$0.09 &     0.18 &   0.40 &     2.22 &     0.22 &   0.61 &    2.80 &     0.23 &  0.72 &    3.11 \\
  36 &   DoAr 25 &      0.04$\times$0.02  & 0.10$\times$0.08 &     0.81 &   1.18 &     1.46 &     1.07 &   1.69 &    1.58 &     1.19 &  1.86 &    1.56 \\
  37 &   DoAr 33 &      0.04$\times$0.02  & 0.10$\times$0.08 &     0.14 &   0.33 &     2.37 &     0.18 &   0.46 &    2.60 &     0.19 &  0.50 &    2.62 \\
  38 &   WaOph 6 &      0.06$\times$0.05  & 0.13$\times$0.12 &     0.48 &   1.69 &     3.50 &     0.74 &   2.42 &    3.25 &     0.84 &  2.69 &    3.20 \\
  39 & HD142666 &      0.03$\times$0.02  & 0.08$\times$0.06 &     0.27 &   0.86 &     3.15 &     0.36 &   1.16 &    3.22 &     0.40 &  1.25 &    3.14 \\
  40 & HD143006 &      0.05$\times$0.04  & 0.07$\times$0.05 &     0.41 &   0.66 &     1.61 &     0.47 &   0.93 &    1.96 &     0.51 &  1.02 &    2.03 \\
  41 & HD163296 &      0.05$\times$0.04  & 0.10$\times$0.10 &     0.94 &   3.34 &     3.57 &     1.36 &   4.73 &    3.48 &     1.67 &  5.22 &    3.13 \\
  42 &    J1100 &      0.39$\times$0.25  & 0.40$\times$0.25 &     0.41 &   1.07 &     2.65 &     0.62 &   1.43 &    2.30 &     0.76 &  1.56 &    2.05 \\
  43 &    TW Hya &      0.04$\times$0.03  & 0.14$\times$0.13 &     0.74 &   2.32 &     3.14 &     0.99 &   3.07 &    3.11 &     1.12 &  3.31 &    2.94 \\
  44 & V4046 Sgr &      0.38$\times$0.29  & 1.21$\times$0.98 &     0.69 &   3.57 &     5.19 &     0.91 &   4.99 &    5.48 &     1.04 &  5.52 &    5.31 \\
\enddata
\tablecomments{Column (3) and (4) are the beam sizes for data images used for disk size calculation. The 17 Lupus disks and the marginally resolve disk J0415 are not included.}
\end{deluxetable*}

\section{Comparison of Lupus disks} \label{sec:lupus-comp}

\begin{figure*}[!t]
\centering
    \includegraphics[width=0.45\linewidth]{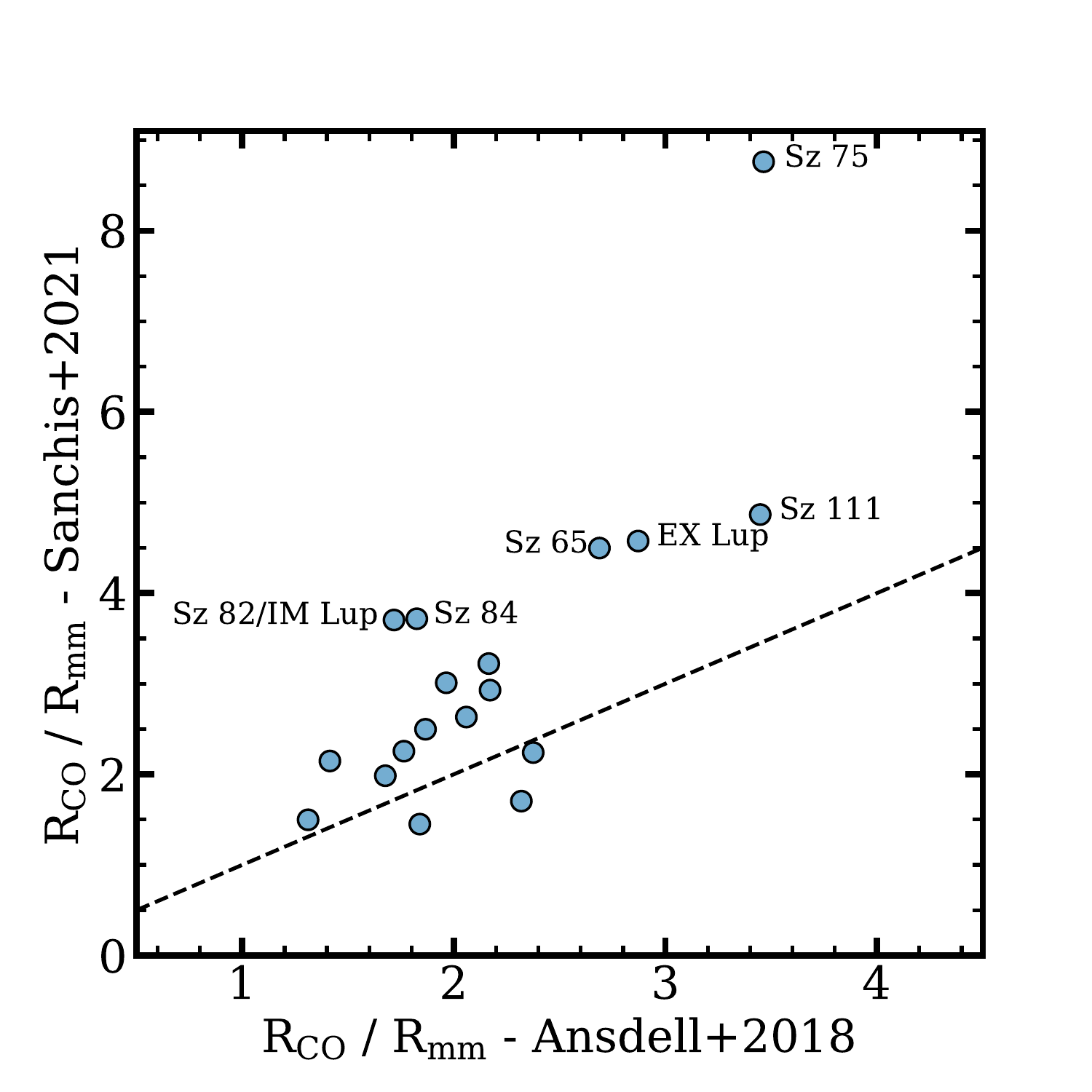} 
    \includegraphics[width=0.45\linewidth]{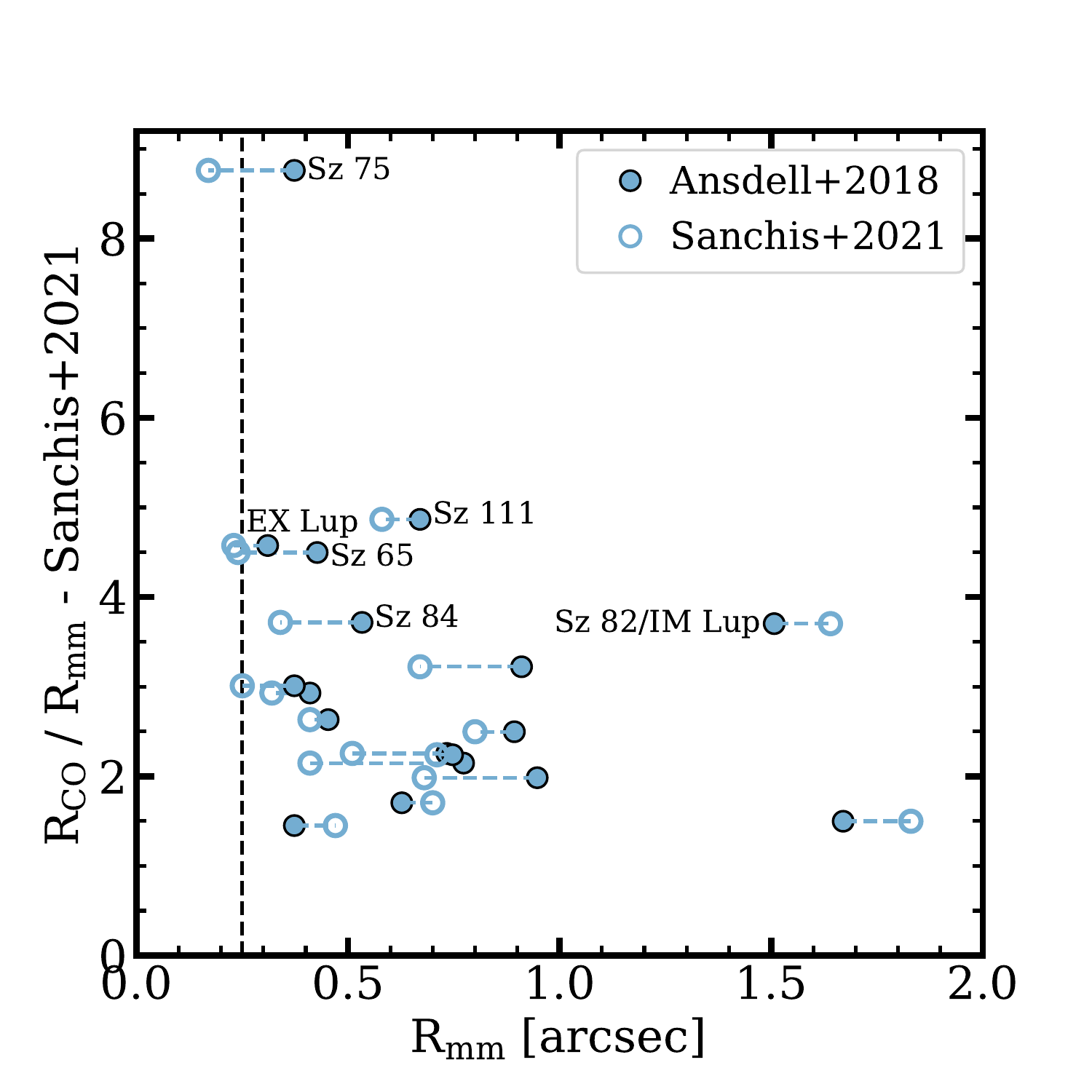} 
\caption{\textbf{Left:} The comparison of disk CO-to-mm continuum size ratio for Lupus disks from \citet{Ansdell2018} and \citet{Sanchis2021}.   \textbf{Right:} The comparison of disk CO-to-mm continuum size ratio for Lupus disks with the mm continuum size. The vertical dashed line marks the typical beam size of the ALMA Lupus survey data.  \label{fig:lupus_comp}}
\end{figure*}

We adopted the mm continuum and CO disk size measurements from \citet{Ansdell2018} for 17 Lupus disks. Recently, \citet{Sanchis2021} provided new size measurements for this sample, but employing different methods. They calculated $R_{\rm CO}$ through elliptical Gaussian or Nuker profile fitting on moment-zero maps, and $R_{\rm mm}$ through Nuker profile fitting in the $uv$-plane. Figure~\ref{fig:lupus_comp} shows the comparison of disk sizes with their ratios from both studies. As \citet{Sanchis2021} only reported $R_{\rm CO,68\%}$, we calculated the $R_{\rm CO,90\%}$ based on the model profile parameters provided in the paper to enable a fair comparison. The higher CO-to-mm size ratios in \citet{Sanchis2021} are largely attributed to the smaller $R_{\rm mm}$ as estimated by visibility fitting, especially for sources that are only resolved in 2--3 beams in the mm images.

\section{The Dependence of $R_{\rm CO}$/$R_{\rm mm}$ on stellar and disk properties} \label{sec:ratio-more}
\begin{figure*}[!t]
\centering
    \includegraphics[width=0.45\linewidth]{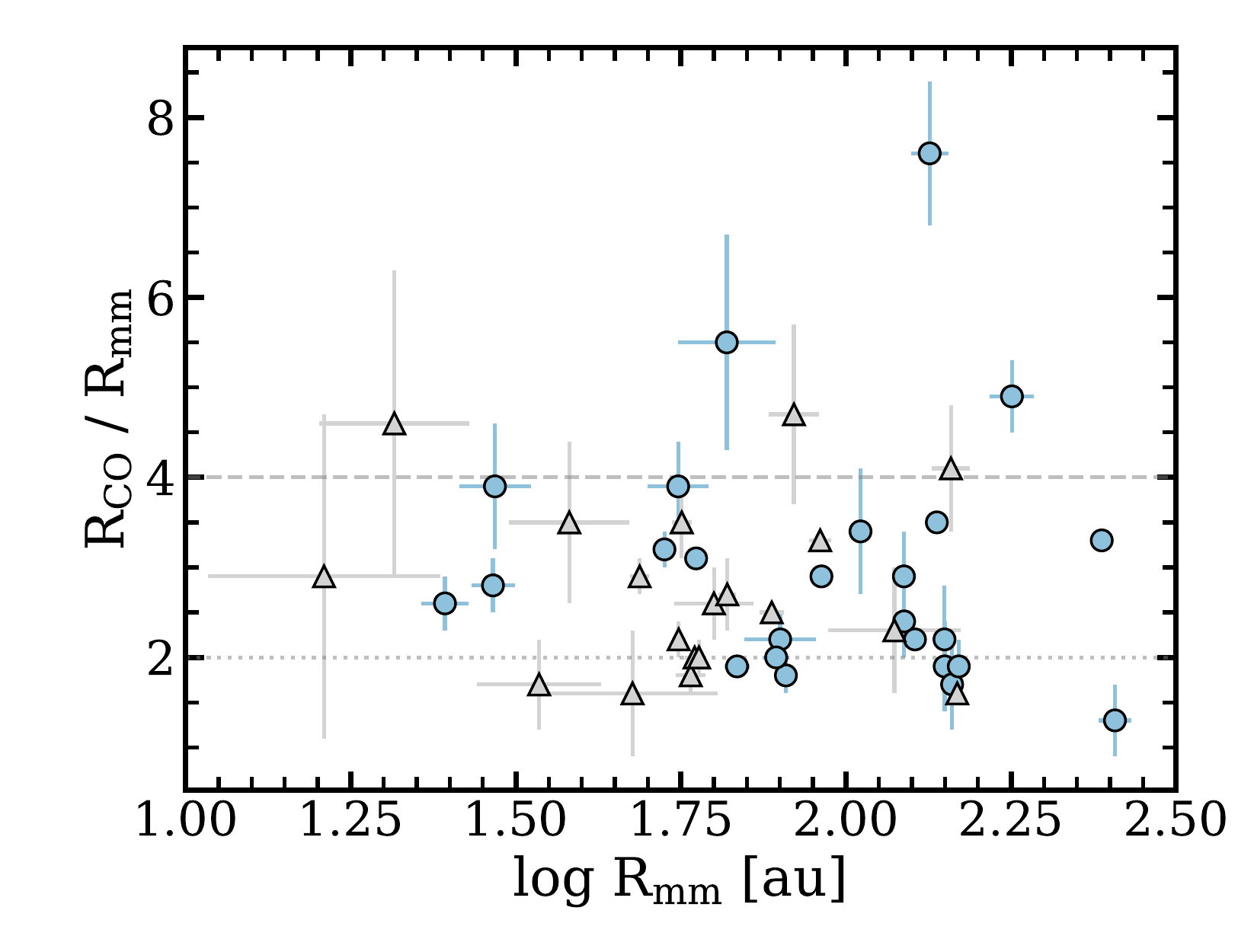} 
    \includegraphics[width=0.45\linewidth]{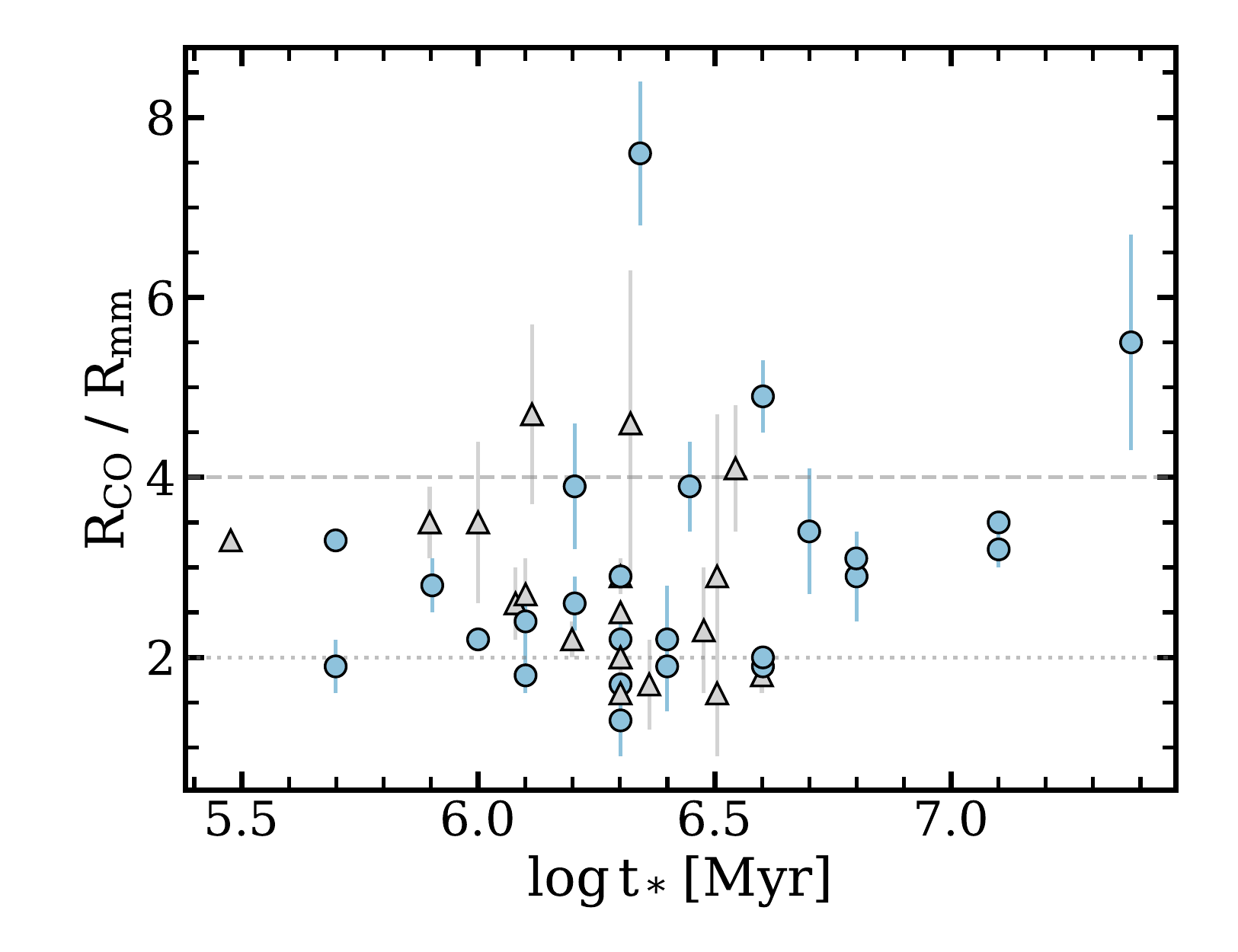} \\
\caption{The distribution of $R_{\rm CO}$/$R_{\rm mm}$ with $R_{\rm mm}$ (left) and stellar age (right). Grey triangles represent disks that either have cloud contamination around systemic velocities or the dust disk is only resolved in less than 2$\times$beam sizes, in both cases $R_{\rm CO}$/$R_{\rm mm}$ is likely underestimated. \label{fig:ratio-more}}
\end{figure*}

Figure~\ref{fig:ratio-more} shows the distribution of $R_{\rm CO}$/$R_{\rm mm}$ with $R_{\rm mm}$ and stellar age. Combined with Figure~\ref{fig:ratio}, We do not find any correlation between the disk size ratios and stellar/disk properties. 

\end{CJK*}
\end{document}